\documentclass{aa520}
\usepackage{epsfig,color,natbib}
\usepackage[dvips]{rotating}
\usepackage{xr2,afterpage}
\def\bibfiles{h3319bib}   

\def\aareferences{\longrefs=0  \bibliographystyle{aa}
            \bibliography{h3319aa,\bibfiles}}

\newcount\longrefs
\def\aap{\ifnum\longrefs=1 {Astron.\ Astrophys.}\else 
                           {A\hbox{\rm \&}A}\fi}
\def\aapr{\ifnum\longrefs=1 {Astron.\ Astrophys.\ Rev.}\else 
                            {A\hbox{\rm \&}AR}\fi}
\def\aaps{\ifnum\longrefs=1 {Astron.\ Astrophys.\ Suppl.}\else 
                            {A\hbox{\rm \&}AS}\fi}
\def\aj{\ifnum\longrefs=1 {Astron.\ J.}\else 
                          {AJ}\fi} 
\def\ao{\ifnum\longrefs=1 {Applied Optics}\else 
                           {Appl.\ Opt.}\fi} 
\def\aspcs{\ifnum\longrefs=1 {Astron.\ Soc.\ Pacific Conf. Series}\else 
                           {ASP Conf.\ Ser.}\fi} 
\def\apj{\ifnum\longrefs=1 {Astrophys.\ J.}\else 
                           {ApJ}\fi} 
\def\apjl{\ifnum\longrefs=1 {Astrophys.\ J. Lett.}\else 
                            {ApJ}\fi} 
\def\aplett{\ifnum\longrefs=1 {Astrophys.\ J. Lett.}\else 
                            {ApJ}\fi} 
\def\apjs{\ifnum\longrefs=1 {Astrophys.\ J. Suppl.}\else 
                            {ApJS}\fi}
\def\apss{\ifnum\longrefs=1 {Astrophys.\ and Space Science}\else 
                            {Ap\hbox{\rm \&}SS}\fi}
\def\araa{\ifnum\longrefs=1 {Ann.\ Rev.\ Astron.\ Astrophys.}\else 
                            {ARA\hbox{\rm \&}A}\fi}
\def\azh{\ifnum\longrefs=1 {Astronomicheskii Zhurnal}\else 
                            {Astron.\ Zhur.}\fi}
\def\baas{\ifnum\longrefs=1 {Bull.\ Am.\ Astron.\ Soc.}\else 
                            {BAAS}\fi}
\def\bain{\ifnum\longrefs=1 {Bull.\ Astronom.\ Institutes Netherlands}\else
                            {Bull.\ Astr.\ Inst.\ Neth.}\fi}
\def\gca{\ifnum\longrefs=1 {Geochim.\ Cosmochim.\ Acta}\else 
                           {Geochim.\ Cosmochim.\ Acta}\fi}
\def\grl{\ifnum\longrefs=1 {Geophys.\ Res.\ Lett.}\else 
                           {Geoph.\ Res.\ Lett.}\fi}
\def\iaucirc{\ifnum\longrefs=1 {IAU Circulars}\else 
                          {IAU Circ.}\fi}
\def\ip{\ifnum\longrefs=1 {in press}\else 
                          {in press}\fi}
\def\jgr{\ifnum\longrefs=1 {J.\ Geophys.\ Res.}\else 
                           {J.\ Geophys.\ Res.}\fi}  
\def\jrasc{\ifnum\longrefs=1 {J.\ Royal Astron.\ Soc.\ Canada}\else 
                           {JRAS Can.}\fi}  
\def\mnras{\ifnum\longrefs=1 {Mon.\ Not.\ Roy.\ Astron.\ Soc.}\else 
                             {MNRAS}\fi} 
\def\nat{\ifnum\longrefs=1 {Nature}\else 
                           {Nat}\fi}
\def\pasj{\ifnum\longrefs=1 {Pub.\ Astron.\ Soc.\ Japan}\else 
                            {PASJ}\fi} 
\def\pasp{\ifnum\longrefs=1 {Pub.\ Astron.\ Soc.\ Pacific}\else 
                            {PASP}\fi} 
\def\physscr{\ifnum\longrefs=1 {Physica Scripta}\else 
                            {Phys.\ Scrip.}\fi} 
\def\planss{\ifnum\longrefs=1 {Planetary \& Space Science}\else 
                            {Plan. \& Space Sci.}\fi} 
\def\procspie{\ifnum\longrefs=1 {Proc.\ SPIE}\else 
                            {Proc.\ SPIE}\fi} 
\def\qjras{\ifnum\longrefs=1 {Quarterly J.\ Royal Astron.\ Soc.}\else 
                            {QJRAS}\fi} 
\def\sa{\ifnum\longrefs=1 {Soviet Astron..}\else 
                               {Sov.\ Astron.}\fi}
\def\skytel{\ifnum\longrefs=1 {Sky \& Telescope}\else 
                            {Sky \& Tel.}\fi} 
\def\solphys{\ifnum\longrefs=1 {Solar Phys.}\else 
                               {Solar Phys.}\fi}
\def\ssr{\ifnum\longrefs=1 {Space Science Rev.}\else 
                               {Space\ Sci.\ Rev.}\fi}

\def\dutch{\def\refname{Referenties}\def\abstractname{Samenvatting}%
  \def\bibname{Bibliografie}\def\chaptername{Hoofdstuk}%
  \def\appendixname{Bijlage}\def\contentsname{Inhoudsopgave}%
  \def\listfigurename{Lijst van figuren}\def\listtablename{Lijst van tabellen}%
  \def\indexname{Index}\def\figurename{Figuur}\def\tablename{Tabel}%
  \def\partname{Deel}\def\enclname{Bijlage(n)}\def\ccname{Ter attentie van}%
  \def\headtoname{Aan}\def\headpagename{Pagina}%
  \def\today{\number\day\space\ifcase\month\or januari\or februari\or maart\or%
     april\or mei\or juni\or juli\or augustus\or september\or oktober\or%
     november\or december\fi \space\number\year}%
  \typeout{
              >>>>> use hlatex209 for Dutch hyphenation <<<<< 
         }}
\newcounter{onefig} \newcounter{fignumber}
\newcount\nocaptions \newcount\nofigures \newcount\figwidth
\newcount\viewgraphs
  \def\paper{}  \def\figlabel{} 
\long\def\nextfig#1{\setcounter{figure}{\value{fignumber}}
  \addtocounter{fignumber}{1}
  \ifnum \viewgraphs=1 \newpage \pagestyle{empty} \fi 
  \ifnum\value{onefig}=0 #1 \fi                 
  \ifnum\value{onefig}=\value{fignumber} #1 \fi}
\def\figwidths#1#2{\ifnum \nocaptions=1 #2mm \else #1mm \fi}  
\def\paper#1{}  
\long\def\plotfig#1#2{\ifnum \nofigures=1 \else #2 \fi}
\long\def\captiontext#1{\ifnum \nofigures=1 \raggedright \fi 
   \ifnum \nocaptions=1 \paper
     \ifnum \viewgraphs=0 
       \newline  \mbox{}\hrulefill\mbox{} \newline 
       \newline label:~\{\figlabel\} 
     \fi 
     \else \ifnum \nofigures=0 \fi 
   #1 \fi}

\newcount\panelwidth \newcount\panelheight 
\newcount\bxmin \newcount\bymin \newcount\bxmax \newcount\bymax
\newcount\tbxmin \newcount\tbymin
\newcount\tpanelwidth \newcount\tpanelheight \newcount\tpdif
\def\panelsize #1,#2;{\panelwidth=#1 \panelheight=#2}  
\def\setbb #1,#2;#3,#4;#5,#6;{
  \tbxmin=#1 \tbymin=#2    
  \bxmin=#3 \bymin=#4      
  \bxmax=#5 \bymax=#6}     
\def\barepanel #1{%
  \ifnum\panelheight=0 
    \tpdif=\bymax \advance\tpdif by -\bymin
    \multiply \tpdif by \panelwidth
    \tpanelheight=\tpdif
    \tpdif=\bxmax \advance\tpdif by -\bxmin
    \divide \tpanelheight by \tpdif
  \else \tpanelheight=\panelheight \fi
  \epsfig{file=#1,%
     bbllx=\bxmin bp,bblly=\bymin bp,bburx=\bxmax bp,bbury=\bymax bp,clip=,%
     width=\panelwidth mm,height=\tpanelheight mm}}
\def\labelypanel #1{
  \ifnum\panelheight=0 
    \tpdif=\bymax \advance\tpdif by -\bymin
    \multiply \tpdif by \panelwidth
    \tpanelheight=\tpdif
    \tpdif=\bxmax \advance\tpdif by -\bxmin
    \divide \tpanelheight by \tpdif
  \else \tpanelheight=\panelheight \fi
  \tpdif=\bxmax \advance\tpdif by -\tbxmin
  \tpanelwidth=\panelwidth \multiply \tpanelwidth by \tpdif
  \tpdif=\bxmax \advance\tpdif by -\bxmin
  \divide \tpanelwidth by \tpdif
  \epsfig{file=#1,%
    bbllx=\tbxmin bp,bblly=\bymin bp,bburx=\bxmax bp,bbury=\bymax bp,%
    clip=,width=\tpanelwidth mm,height=\tpanelheight mm}}
\def\labelxpanel #1{%
  \ifnum\panelheight=0 
    \tpdif=\bymax \advance\tpdif by -\bymin
    \multiply \tpdif by \panelwidth
    \tpanelheight=\tpdif
    \tpdif=\bxmax \advance\tpdif by -\bxmin
    \divide \tpanelheight by \tpdif
  \else \tpanelheight=\panelheight \fi
  \tpdif=\bymax \advance\tpdif by -\tbymin
  \multiply \tpanelheight by \tpdif
  \tpdif=\bymax \advance\tpdif by -\bymin
  \divide \tpanelheight by \tpdif
  \epsfig{file=#1,%
    bbllx=\bxmin bp,bblly=\tbymin bp,bburx=\bxmax bp,bbury=\bymax bp,%
    clip=,width=\panelwidth mm,height=\tpanelheight mm}}
\def\labelxypanel #1{%
  \ifnum\panelheight=0 
    \tpdif=\bymax \advance\tpdif by -\bymin
    \multiply \tpdif by \panelwidth
    \tpanelheight=\tpdif
    \tpdif=\bxmax \advance\tpdif by -\bxmin
    \divide \tpanelheight by \tpdif
  \else \tpanelheight=\panelheight \fi
  \tpdif=\bxmax \advance\tpdif by -\tbxmin
  \tpanelwidth=\panelwidth \multiply \tpanelwidth by \tpdif
  \tpdif=\bxmax \advance\tpdif by -\bxmin
  \divide \tpanelwidth by \tpdif 
  \tpdif=\bymax \advance\tpdif by -\tbymin 
  \multiply \tpanelheight by \tpdif
  \tpdif=\bymax \advance\tpdif by -\bymin
  \divide \tpanelheight by \tpdif
  \epsfig{file=#1,%
    bbllx=\tbxmin bp,bblly=\tbymin bp,bburx=\bxmax bp,bbury=\bymax bp,%
    clip=,width=\tpanelwidth mm,height=\tpanelheight mm}}

\def\CC{\par \vspace*{-2ex} \footnotesize \baselineskip=8pt \begin{verbatim}}

\long\def\startignore #1\stopignore{}   

\def\setlistparams{         
  \topsep=0.7ex                 
  \itemsep=0.7ex                
  \leftmargini=3ex}             
\setlistparams                  

\newcounter{alistindex}       

\newcounter{romenumnr}

\newlength{\minipagewidth}

\newsavebox{\boxcontent}
\newcommand{\ovalhead}[1]{
  \unitlength=1cm
  \sbox{\boxcontent}{\mbox{~~{#1}~~}}
  \begin{center}
    \ifdim\wd\boxcontent>6ex 
    \ifdim\wd\boxcontent<8cm 
    \begin{picture}(8,3) \thicklines     
      \put(4.0,0.8){\oval(8,1.6)} 
      \put(0.0,0.7){\parbox{8cm}{
         \begin{center} \usebox{\boxcontent} \end{center}}}
    \end{picture}
    \else \ifdim\wd\boxcontent<12cm 
    \begin{picture}(12,3) \thicklines     
        \put(6.0,0.8){\oval(12,1.6)} 
        \put(0.0,0.7){\parbox{12cm}{
           \begin{center} \usebox{\boxcontent} \end{center}}}
    \end{picture}
    \else
    \begin{picture}(16,3) \thicklines     
        \put(8.0,0.8){\oval(16,1.6)} 
        \put(0.0,0.7){\parbox{16cm}{
           \begin{center} \usebox{\boxcontent} \end{center}}}
    \end{picture}
    \fi \fi \fi
  \end{center}} 

\newcounter{headnr}            
\newcounter{subheadnr}[headnr]
\newcounter{subsubheadnr}[subheadnr]
\def\head #1\par{
  \stepcounter{headnr}                          
  \vspace{2ex} \noindent                        
  {\bf \theheadnr~~~~#1}\\[1ex] \noindent}      
\def\subhead #1\par{  
  \stepcounter{subheadnr}
  \vspace{1.3ex} \noindent
  {\bf \theheadnr.\arabic{subheadnr}~~~#1}\\[0.3ex] \noindent}
\def\subsubhead #1\par{
  \stepcounter{subsubheadnr}
  \vspace{1.0ex} \noindent
  {\bf \theheadnr.\arabic{subheadnr}.\arabic{subsubheadnr}~~~#1}\\ \noindent}

\font\dropfont= cmr12 scaled \magstep5
\def\dropcap#1#2{{\noindent
    \setbox0\hbox{\dropfont #1}\setbox1\hbox{#2}\setbox2\hbox{(}%
    \count0=\ht0\advance\count0 by\dp0\count1\baselineskip
    \advance\count0 by-\ht1\advance\count0by\ht2
    \dimen1=.5ex\advance\count0by\dimen1\divide\count0 by\count1
    \advance\count0 by1\dimen0\wd0
    \advance\dimen0 by.25em\dimen1=\ht0\advance\dimen1 by-\ht1
    \global\hangindent\dimen0\global\hangafter-\count0
    \hskip-\dimen0\setbox0\hbox to\dimen0{\raise-\dimen1\box0\hss}%
    \dp0=0in\ht0=0in\box0}#2}

\def\level #1 #2#3#4{$#1 \: ^{#2} \mbox{#3} ^{#4}$}   

\def\kms{\hbox{km$\;$s$^{-1}$}}

\def\mathstacksym#1#2#3#4#5{\def#1{\mathrel{\hbox to 0pt{\lower 
    #5\hbox{#3}\hss} \raise #4\hbox{#2}}}}

\mathstacksym\lta{$<$}{$\sim$}{1.5pt}{3.5pt} 
\mathstacksym\gta{$>$}{$\sim$}{1.5pt}{3.5pt} 
\mathstacksym\lrarrow{$\leftarrow$}{$\rightarrow$}{2pt}{1pt} 
\mathstacksym\lessgreat{$>$}{$<$}{3pt}{3pt} 

\bibpunct{(}{)}{;}{a}{}{,}
\newcommand{\ang}{$\rm \AA$}

\newcommand{\Msun}{M$_{\odot}$}
\newcommand{\Rsun}{R$_{\odot}$}

\newcommand{\be}{\begin{equation}}
\newcommand{\ee}{\end{equation}}
\newcommand{\bee}{\begin{eqnarray}}
\newcommand{\ad}{$\theta_d$}
\newcommand{\vt}{$\xi_t$}
\newcommand{\cc}{$\mathrm{^{12}C/^{13}C}$}

\newcommand{\ene}{\end{eqnarray}}
\newcommand{\teff}{T$_{\mathrm{eff}}$}
\newcommand{\mic}{$\mu$m}

\begin{document}

\title{ISO-SWS calibration and the accurate modelling of cool-star
atmospheres \thanks{Based on observations with ISO, an ESA project with
instruments funded by ESA Member States (especially the PI countries France,
Germany, the Netherlands and the United Kingdom) and with the participation of
ISAS and NASA.}}
\subtitle{IV.\ G9 -- M2 stars: APPENDIX}

\author{L.~Decin\inst{1}\thanks{\emph{Postdoctoral Fellow of the Fund for
Scientific Research, Flanders}}  \and
B. Vandenbussche\inst{1} \and
C.~Waelkens\inst{1} \and
G.~Decin\inst{1} \and
K.~Eriksson\inst{2} \and
B.~Gustafsson\inst{2} \and
B.~Plez\inst{3} \and
A.J.~Sauval\inst{4}
}

\offprints{L.\ Decin, e-mail: Leen.Decin@ster.kuleuven.ac.be}

\institute{Instituut voor Sterrenkunde, KULeuven, Celestijnenlaan 200B, B-3001
    Leuven, Belgium
\and
    Institute for Astronomy and Space Physics, Box 515, S-75120 Uppsala, Sweden
\and
    GRAAL - CC72, Universit\'{e} de Montpellier II, F-34095 Montpellier Cedex 5,
France
\and
    Observatoire Royal de Belgique, Avenue Circulaire 3, B-1180 Bruxelles,
    Belgium
}

\date{Received data; accepted date}

\abstract{A detailed spectroscopic study of 11 giants with spectral type from G9
to M2 is presented. The 2.38 -- 4.08\,\mic\ wavelength-range of band 1
of ISO-SWS (Short-Wavelength Spectrometers on board of the Infrared Space
Observatory) in which
many different molecules --- with their own dependence on each of the
stellar parameters --- are absorbing, enables us  to
estimate the effective temperature, the gravity, the microturbulence,
the metallicity, the CNO-abundances, the \cc-ratio and the angular
diameter from the ISO-SWS data. Using the Hipparcos' parallax,
the radius, luminosity and gravity-inferred mass are derived. The
stellar parameters obtained are in good agreement with other published
values, though also some discrepancies with values deduced by other
authors are noted. For a few stars ($\delta$ Dra, $\xi$
Dra, $\alpha$ Tuc, H Sco and $\alpha$ Cet) some parameters --- e.g.\
the CNO-abundances --- are derived for the first time.
By examining the
correspondence between different ISO-SWS observations of the same
object and between the ISO-SWS data and the corresponding synthetic
spectrum, it is shown that the relative accuracy of ISO-SWS in band 1
(2.38 -- 4.08\,\mic) is better than 2\,\% for these high-flux sources.
The high level of correspondence between observations and theoretical
predictions, together with a confrontation of the estimated \teff(ISO)
value with \teff\ values derived from colours --- which demonstrates
the consistency between $V-K$, BC$_K$, \teff\ and \ad\ derived from
optical or IR data --- proves that both the used MARCS models to derive the
stellar quantities and the flux calibration of the ISO-SWS detectors
have reached a high level of reliability.
}

\maketitle
\markboth{L.\ Decin et al.: ISO-SWS and modelling of cool stars}{L.\ Decin et
al.: ISO-SWS and modelling of cool stars}

\appendix
\defcitealias{Decin2000A&A...364..137D}{Paper~I}
\defcitealias{Decin2000b}{Paper~II}
\defcitealias{Decin2000c}{Paper~III}

\section{Calibration accuracy and precision}

In the following subsections, each of the 11 cool giants will be
discussed individually. For each star, calibration details are
specified. These, in conjunction with the general calibration
specifications discussed in Paper~II, gives us an idea on
the calibration accuracy.
If other AOT01 observations {\footnote{Each observation is determined
uniquely by its observation
number, in which the first three digits represent the revolution
number. The observing data can be calculated from the revolution
number which is the number of days after 17 November
1995.}} are available, they are compared with each
other in order to assess the calibration precision of
ISO-SWS. Note that when the ratio of two observations is shown, the
observation with the highest speed number (\,=\,the highest resolving
power) is rebinned to the resolution of the observation with the
lowest speed number.

\subsection{${\delta}$ Dra{\rm: AOT01, speed 4, revolution 206}}

\subsubsection{Some specific calibration details}

When looking to the spectrum, before it was multiplied with the
factors given in Table 3 in \citet[][hereafter referred to as
Paper~II]{Decin2000b}, one could notice that the
subsequent sub-bands of band 1 lie always somewhat lower than the
previous one. The pointing offset in the z-direction is quite
large (dz = $1.48''$).

\subsubsection{Comparison with other observations \rm{(Fig.\ \ref{deldravers})}}

\begin{figure}[h]
\begin{center}
\resizebox{0.5\textwidth}{!}{\rotatebox{90}{\includegraphics{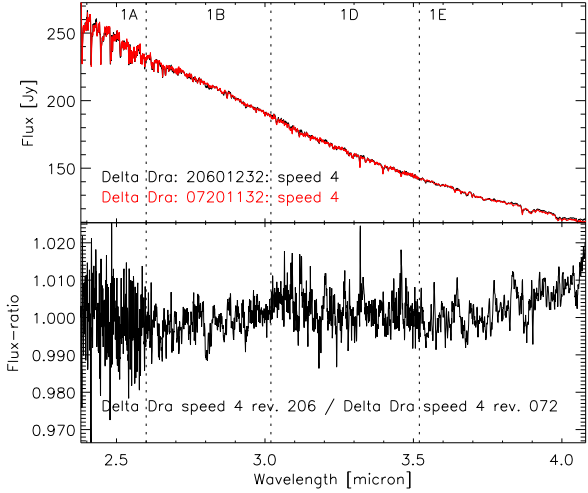}}}
\vspace*{-7ex}
\caption{\label{deldravers}Comparison between the
AOT01 speed-4 observation of \boldmath {\bf{$\delta$ Dra}} \unboldmath
 (revolution 206) and the speed-4
observation (revolution 072). The data of the speed-4 observation,
taken during revolution 072, have been divided by a factor 1.115.}
\end{center}
\end{figure}

During revolution 072, $\delta$ Dra was observed with the AOT01
speed-4 option. Also for this observation, the subsequent
sub-bands lie somewhat lower than the previous one. As for the
other speed-4 observation, the pointing offset in the z-direction
is $> 1''$. The data of band 1A and band 1B are divided by 1.015
and 1.01 respectively, the band-1E flux is multiplied by 1.01. The
spectral features and relative shape agree quite well. Only at the
end of band 1E and around 2.8 \mic, can one see a small difference
in shape probably arising from the pointing difference. The
absolute flux level differs by $\sim 11$\,\%, with observation
07201132 having the highest flux level.


\subsection{${\xi}$ Dra{\rm: AOT01, speed 3, revolution 314}}

\subsubsection{Some specific calibration details}

The AOT01 speed-3 observation of $\xi$ Dra was quite difficult to
calibrate. The larger noise for a speed-3 observation (compared to
a speed-4 observation) together with the large pointing offset
result in a spectrum, after application of the standard
calibration process, being more uncertain than for other
observations. Therefore a speed-1 observation, in conjunction with
a post-helium observation (see next paragraph) were used to obtain
an optimally well calibrated spectrum.

\subsubsection{Comparison with other observations \rm{(Fig.\ \ref{ksidravers})}}

\begin{figure}[h!]
\begin{center}
\resizebox{0.5\textwidth}{!}{\rotatebox{90}{\includegraphics{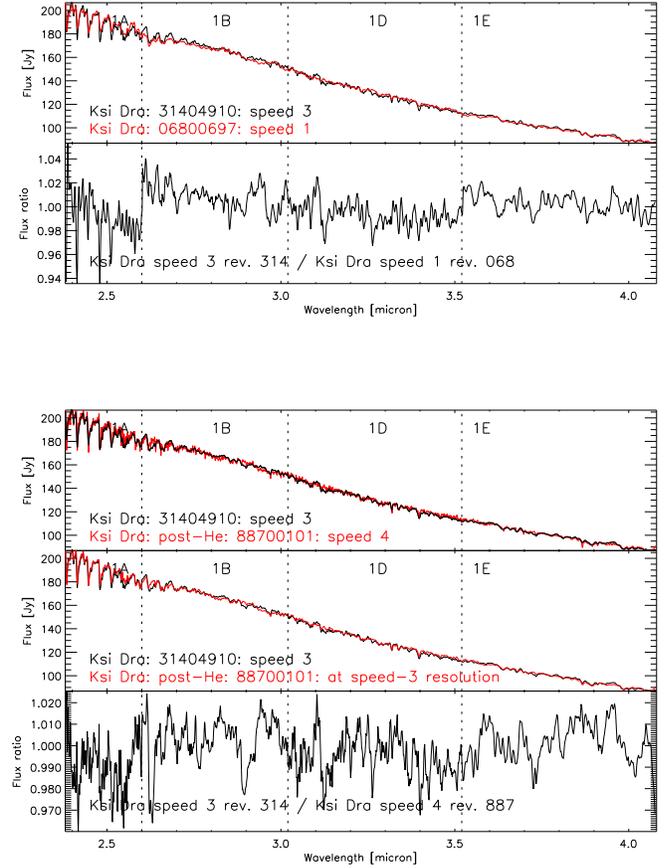}}}
\vspace*{-5ex}
\caption{\label{ksidravers}{\underline{Top:}} Comparison between
the AOT01 speed-3 observation of \boldmath {\bf{$\xi$ Dra}}
\unboldmath
 (revolution 314) and
the speed-1 observation (revolution 068). The data of the speed-1
observation have been divided by a factor 1.09.
{\underline{Bottom:}} Comparison between the AOT01 speed-3
observation of \boldmath {\bf{$\xi$ Dra}} \unboldmath
 (revolution 314) and the post-helium
observation taken during revolution 887. The data of the post-He
observation have been divided by a factor 1.01.}
\end{center}
\end{figure}

An AOT01 speed-1 observation has been taken of $\xi$ Dra.
Unfortunately, the pointing offset of this observation was
approximately as large as for the speed-3 observation (dy =
$-0.560''$, dz = $-1.403''$). The data of bands 1A and 1B were
multiplied with a factor 1.04 and 1.03 respectively. The general
correspondence between the two
observations is not so good at the end of band 1A, in band 1B and band 1D. The
flux of the
speed-1 observation lies $\sim 9$\,\% higher than the flux of the
speed-4 observation (Fig.\ \ref{ksidravers}).

Once ISO had run out off helium, the Short-Wavelength Spectrometer
still has been operated during some 30 days. The observing
programme aimed at obtaining 2.4 -- 4 \mic\ scans at the full
grating resolution of stars spanning the whole MK spectral classification
scheme, i.e.\ to extend this classification to the infrared. $\xi$
Dra was among the targets observed then. The pointing was far more better,
being dy = $0.004''$ and
dz = $-0.073''$. Although the calibration
still has not reached its final stage, these data are already
very useful to study the strength of the spectral
features at a resolving power of $\sim 1500$.
The overall spectrum lies $\sim 1$\,\% higher than
the speed-3 observation. The data of bands 1A and 1B  are
shifted upwards by 5\,\% and 4\,\% respectively. The
comparison between the speed-3 observation (revolution 314) and
the post-He observation (revolution 887) of $\xi$ Dra is shown in
the middle and bottom panel of Fig.\ \ref{ksidravers}. The same
differences as seen in the comparison with the speed-1 observation
can also be seen now. The confrontation with its synthetic
spectrum (see next paragraph), with the speed-1 observation and
with the post-He data indicates clearly that the final calibrated
spectrum of $\xi$ Dra of revolution 314 is still not so reliable.


\subsection{${\alpha}$ Boo{\rm: AOT01, speed 4, revolution 452}}

\subsubsection{Some specific calibration details}

The AOT01 speed-4 observation of $\alpha$ Boo was of very high
quality. The shifts for the different sub-bands of band 1 show the
same trend as most of the observations: the band-1B flux has to be
multiplied by a factor $\le 1.02$ and the band-1E flux by a factor
$\le 1.01$.

\subsubsection{Comparison with other observations \rm{(Fig.\ \ref{aboovers})}}

\begin{figure}[h!]
\begin{center}
\resizebox{0.5\textwidth}{!}{\rotatebox{90}{\includegraphics{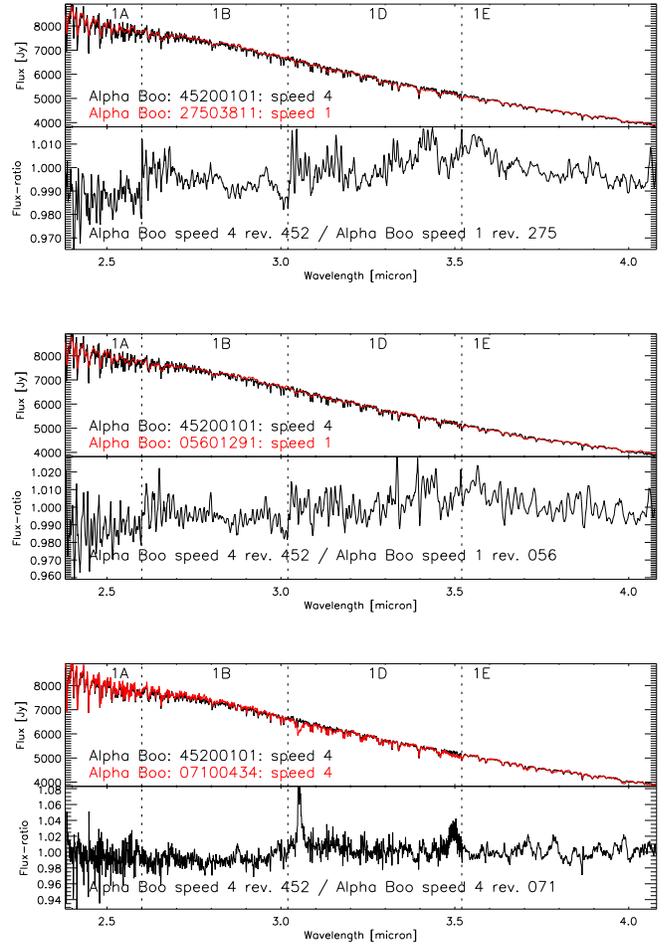}}}
\vspace*{-5ex}
\caption{\label{aboovers}{\underline{Top:}} Comparison between the
AOT01 speed-4 observation of \boldmath {\bf{$\alpha$ Boo}} \unboldmath
 (revolution 452) and the
speed-1 observation (revolution 275). {\underline{Middle:}}
Comparison between the AOT01 speed-4 observation of \boldmath
{\bf{$\alpha$ Boo}} \unboldmath
(revolution 452) and the speed-1 observation (revolution 056).
{\underline{Bottom:}} Comparison between the AOT01 speed-4
observation of \boldmath {\bf{$\alpha$ Boo}} \unboldmath
 (revolution 452) and the speed-4
observation (revolution 071). The data of this latter speed-4
observation have been multiplied by a factor 1.05.}
\end{center}
\end{figure}

$\alpha$ Boo has been observed a few times with SWS. In revolution 056,
244, 275 and 641 a speed-1 observation was taken. In revolution
071, $\alpha$ Boo was observed with the AOT01 speed-4 option.
Observation 24402866 and 64100101 were corrupted. The known
pointing offsets of the two other speed-1 observations are dy =
$0.703''$ and dz = $2.412''$ for observation 05601291, and dy =
$0.667''$ and dz = $-0.197''$ for observation 27503811. In the
upper panel of Fig.\ \ref{aboovers}, the data of band 1A, 1B and 1E
of observation 27503811 are multiplied by 0.965, 0.985 and 1.01
respectively. The data of the sub-bands 1A and 1B of observation
05601291 are multiplied by 0.98 and 1.005 respectively.

The speed-4 observation taken during revolution 071 displays a
scan jump 3415 s after the start of the observation. The spectrum
is therefore (more) suspicious around 3.05 \mic. The pointing
offset of this observation was dy = $0.269''$ and dz = $2.561''$.
The data of band 1A and band 1B are multiplied by 0.95 and 0.97
respectively. The flux of the overall spectrum is $\sim 5\,\%$ lower
than for the observation taken during revolution 452.

The correspondence between the different observations is
satisfactory taking into account the pointing offsets, the lower signal-to-noise
of the speed-1 observations and the less accurate flux calibration at the band
edges.


\subsection{${\alpha}$ Tuc{\rm: AOT01, speed 4, revolution 866}}

\subsubsection{Some specific calibration details}

Just as for $\alpha$ Boo, this speed-4 AOT01 observation was of
very high quality. The data of band 1B were shifted upwards by 2\,\%
and the data of band 1E by 1\,\%. The standard deviations after
rebinning are similar to the ones given for $\alpha$ Tau in Fig.\
6 in \citetalias{Decin2000A&A...364..137D}. Just as for $\alpha$ Cen A
\citepalias{Decin2000c}, $\alpha$ Tuc belongs
to a binary system. Its companion is however too far away to
influence the ISO-SWS spectrum of $\alpha$ Tuc.


\subsection{${\beta}$ UMi{\rm: AOT01, speed 4, revolution 182}}

\subsubsection{Some specific calibration details}

As for most of the stars in the sample, the data of both band 1B
and band 1E are shifted upwards, by 1.5\,\% and 1\,\%
respectively. The pointing offset is $> 1''$ in the y-direction.
No correction is made at this moment, since the beam profiles are
still not available.

\begin{figure}[h!]
\begin{center}
\resizebox{0.5\textwidth}{!}{\rotatebox{90}{\includegraphics{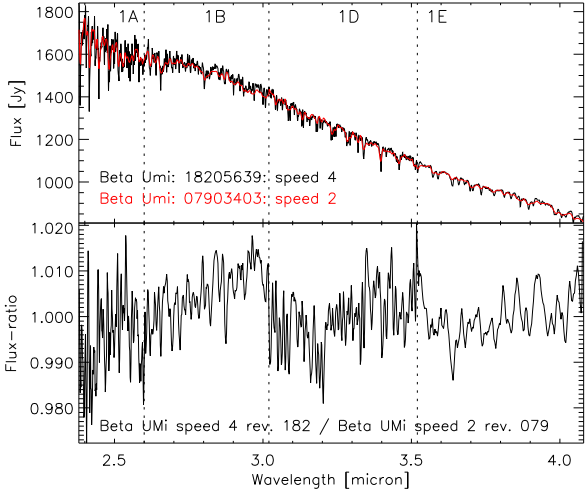}}}
\vspace*{-7ex}
\caption{\label{bumivers}Comparison between the AOT01 speed-4
observation of \boldmath {\bf{$\beta$ UMi}} \unboldmath
 (revolution 182) and the speed-2
observation (revolution 079). The data of the speed-2 observation are divided
by a factor 1.065.}
\end{center}
\end{figure}

\subsubsection{Comparison with other observations \rm{(Fig.\ \ref{bumivers})}}

$\beta$ UMi has also been observed with the speed-2 option during
revolution 079. The pointing was not optimal, with offsets being
dy = $0.780''$ and dz = $-2.360''$. The band-1A and band-1B data
are divided by 1.025 and 1.015 respectively, and the flux level of
the overall spectrum is $\sim 6.5\,\%$ higher than for the speed-4
observation (Fig.\ \ref{bumivers}). Despite the low signal-to-noise
of a speed-2 observation, the two observations agree quite well.


\subsection{${\gamma}$ Dra{\rm: AOT01, speed 4, revolution 377}}

\subsubsection{Some specific calibration details}

$\gamma$ Dra is one of the main calibration sources of the
Short-Wavelength Spectrometer. The pointing of the AOT01 speed-4
observation was not optimal, being dy = $-0.304''$ and dz =
$0.181''$. The shifts of the different sub-bands of band 1 are all
within 0.5\,\%.

\subsubsection{Comparison with other observations \rm{(Fig.\ \ref{gamdravers})}}

\begin{figure}[h!]
\begin{center}
\resizebox{0.5\textwidth}{!}{\rotatebox{90}{\includegraphics{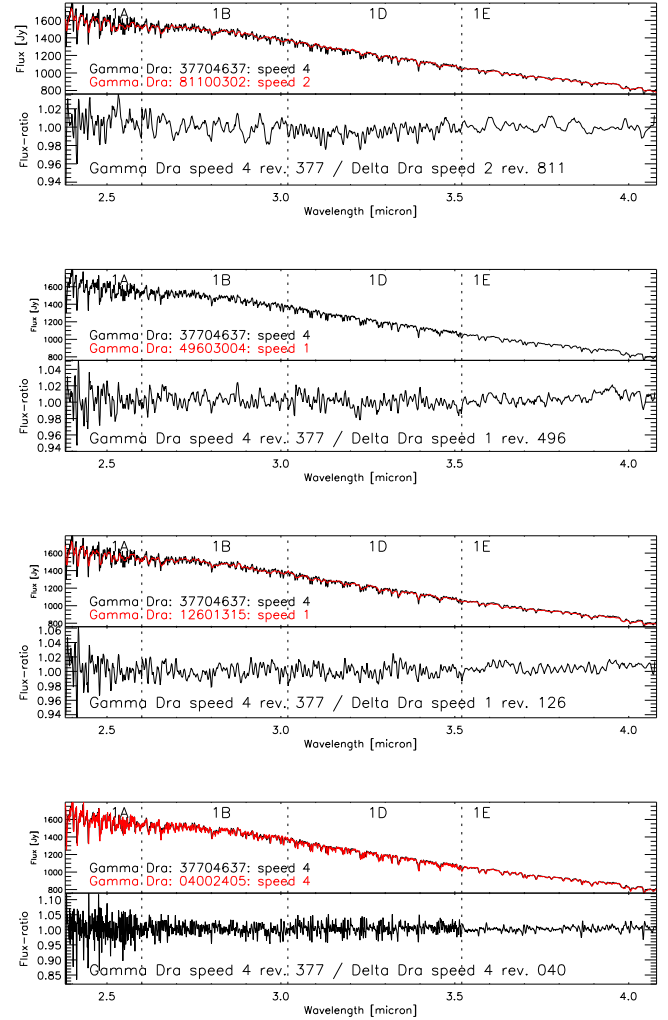}}}
\vspace*{-5ex}
\caption{\label{gamdravers}{\underline{1$^{\mathrm{st}}$ plot:}}
Comparison between the AOT01 speed-4 observation of \boldmath
{\bf{$\gamma$ Dra}} \unboldmath
(revolution 377) and the speed-2 observation (revolution 811). The
data of the speed-2 observation are divided by 1.02.
{\underline{2$^{\mathrm{nd}}$ plot}}: Comparison between the AOT01
speed-4 observation of \boldmath {\bf{$\gamma$ Dra}} \unboldmath 
 (revolution 377) and the speed
1 observation (revolution 496). The data of the speed-1
observation are divided by 1.05. {\underline{3$^{\mathrm{th}}$
plot}}: Comparison between the AOT01 speed-4 observation of
\boldmath {\bf{$\gamma$ Dra}} \unboldmath
 (revolution 377) and the speed-2 observation
(revolution 126). The data of the speed-1 observation are divided
by 1.04. {\underline{4$^{\mathrm{th}}$ plot}}: Comparison between
the AOT01 speed-4 observation of \boldmath {\bf{$\gamma$ Dra}}
\unboldmath  (revolution 377) and
the speed-4 observation (revolution 040). }
\end{center}
\end{figure}

Several observations were taken of $\gamma$ Dra. During
revolution 040 a speed-4 observation was taken, pointing offsets
being dy = $-2.440''$ and dz = $-0.259''$. The data of band 1A and
band 1B are multiplied by 1.015 and 1.01 respectively. Two speed-1
observations were taken during revolution 126 and revolution 496,
the first one with a pointing offset of dy = $-0.586''$ and dz =
$1.320''$ and the latter one with dy = $0.00''$ and dz = $0.00''$.
For observation 12601315 the band-1A and band-1B data are shifted
down by 2\,\%, with the absolute flux level being 4\,\% higher than
for observation 37704637. The band-1A and band-1B data are shifted
upwards by 1\,\% and 1.5\,\% respectively for observation 49603004.
This latter observation lies 5\,\% higher than observation 37704637.


Another (last) AOT01 observation of $\gamma$ Dra was taken during
revolution 811 with the speed-2 option. Pointing offsets were dy
= $0.020''$ and dz = $0.019''$. The band-1A and band-1B data need
however to be multiplied by quite a large factor: 1.05 and 1.065
respectively. The reason for this jump is maybe pointing
jitter when moving from one aperture to another. The shape and the
spectral features of the different observations are in good
agreement.


\subsection{${\alpha}$ Tau{\rm: AOT01, speed 4, revolution 636}}

$\alpha$ Tau has been discussed in \citet{Decin2000A&A...364..137D}.


\subsection{H Sco{\rm: AOT01, speed 4, revolution 847}}

\subsubsection{Some specific calibration details}

Two remarks again have to be made for HD~149447. First, band 1B  is
shifted upwards by 1.5\,\%, secondly, no pointing errors were measured.


\subsection{${\beta}$ And{\rm: AOT01, speed 4, revolution 795}}

\subsubsection{Some specific calibration details}
The speed-4 observation of $\beta$ And was of very high quality.
Only the data of band 1E had to be shifted upwards, by 0.5\,\%. No
pointing offsets were measured in the y and z-direction.

\subsubsection{Comparison with other observations}

\begin{figure}[h!]
\begin{center}
\resizebox{0.5\textwidth}{!}{\rotatebox{90}{\includegraphics{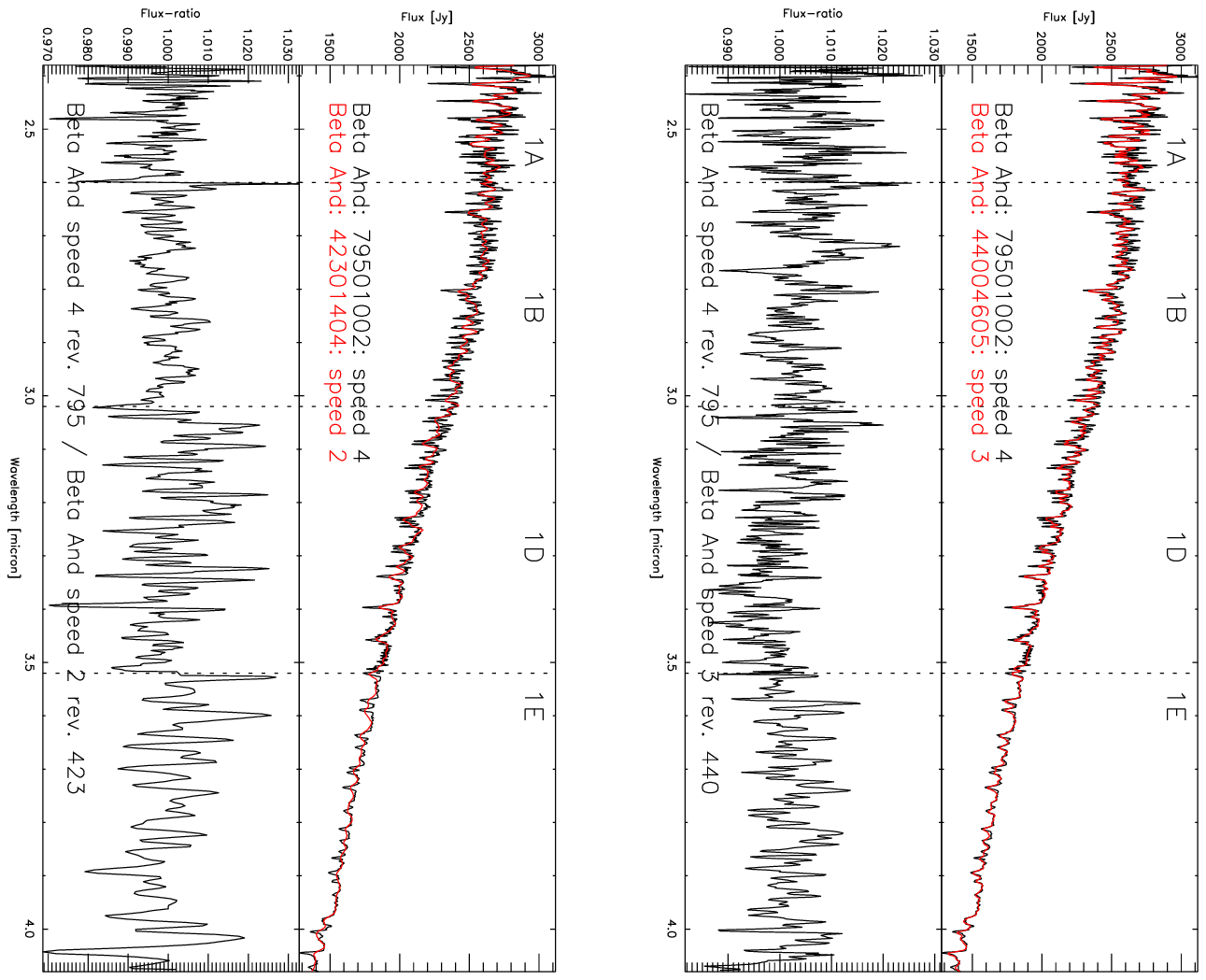}}}
\vspace*{-5ex}
\caption{\label{bandvers}{\underline{Top:}} Comparison between the
AOT01 speed-4 observation of \boldmath {\bf{$\beta$ And}} \unboldmath
 (revolution 795) and the
speed-3 observation (revolution 440). The data of the speed-3
observation have been multiplied by a factor 1.03.
{\underline{Bottom:}} Comparison between the AOT01 speed-4
observation of \boldmath {\bf{$\beta$ And}} \unboldmath
 (revolution 795) and the speed-2
observation (revolution 423). The data of the speed-2 observation
have been multiplied by a factor 1.04.}
\end{center}
\end{figure}

In conjunction with the speed-4 observation, there are also a
speed-3 (revolution 440) and a speed-2 (revolution 423)
observation available. For both observations, the pointing offset
in the z-direction is smaller than in the y-direction. For
observation 44004605, dy = $-0.229''$ and dz = $0.096''$, while
for observation 42301404 dy = $0.038''$ and dz = $-0.15''$. For
the speed-3 observation, the data of band 1A and band 1B are divided
by 1.025 and 1.01 respectively, while the data of
band 1E are multiplied by
1.015, with the total flux being 3\,\% lower than for the speed-4
observation. The flux of band 1A and band 1B of the speed-2
observation was, respectively, 3\,\% and 2\,\% higher than the flux of
band 1D; band
1E is shifted upwards by 1\,\%. The total flux level of this speed-2
observation is 4\,\% lower than for the speed-4 observation. The
shapes of the different sub-bands in band 1 for the three
observations of $\beta$ And do agree very well (Fig.\
\ref{bandvers}).


\subsection{${\alpha}$ Cet{\rm: AOT01, speed 4, revolution 797}}

\subsubsection{Some specific calibration details}
The speed-4 observation of $\alpha$ Cet, taken at the end of the
nominal phase of ISO-SWS, did not suffer from a pointing offset.
As for most of the other observations in the sample, the data of
band 1E were shifted upwards by 0.5\,\%.

\subsubsection{Comparison with other observations}

\begin{figure}[h!]
\begin{center}
\resizebox{0.5\textwidth}{!}{\rotatebox{90}{\includegraphics{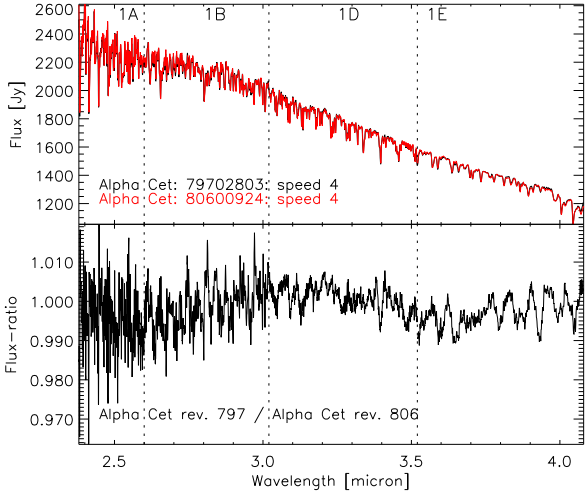}}}
\vspace*{-7ex}
\caption{\label{acetvers}{\underline{Top:}} Comparison between the
AOT01 speed-4 observation of \boldmath {\bf{$\alpha$ Cet}} \unboldmath
 (revolution 797) and the
speed-4 observation (revolution 806). {\underline{Bottom:}} The
AOT01 speed-4 observation of \boldmath {\bf{$\alpha$ Cet}} \unboldmath
 taken during revolution
797 is divided by the AOT01 speed-4 observation of $\alpha$ Cet
taken during revolution 806.}
\end{center}
\end{figure}

A second speed-4 observation of $\alpha$ Cet was made by SWS only nine
revolutions after the first observation. The same calibration remarks
do also apply for
the observation of revolution 806 (almost no pointing offset
and the data of band 1E are multiplied by 1.005). Since the same
time-dependent calibration files are used for these two
observations, they form a good test-vehicle for the stability
during this calibration period and for the flux uncertainty. Good
agreement is found. The ratio of the two spectra is shown in
Fig.\ \ref{acetvers}. The strongest peaks in the lowest plot in
Fig.\ \ref{acetvers} correspond to the strongest CO-peaks in the
uppermost plot. A (very) small difference in pointing causes this
effect.


\subsection{${\beta}$ Peg{\rm: AOT01, speed 4, revolution 551}}

\subsubsection{Some specific calibration details}
$\beta$ Peg has been observed in almost optimal conditions.
The pointing offset was dy = $0.017''$ and dz = $0.095''$. The band-1B data
are multiplied by 1.015 and the band-1E data by 1.005.

\subsubsection{Comparison with other observations}

\begin{figure}[h!]
\begin{center}
\resizebox{0.5\textwidth}{!}{\rotatebox{90}{\includegraphics{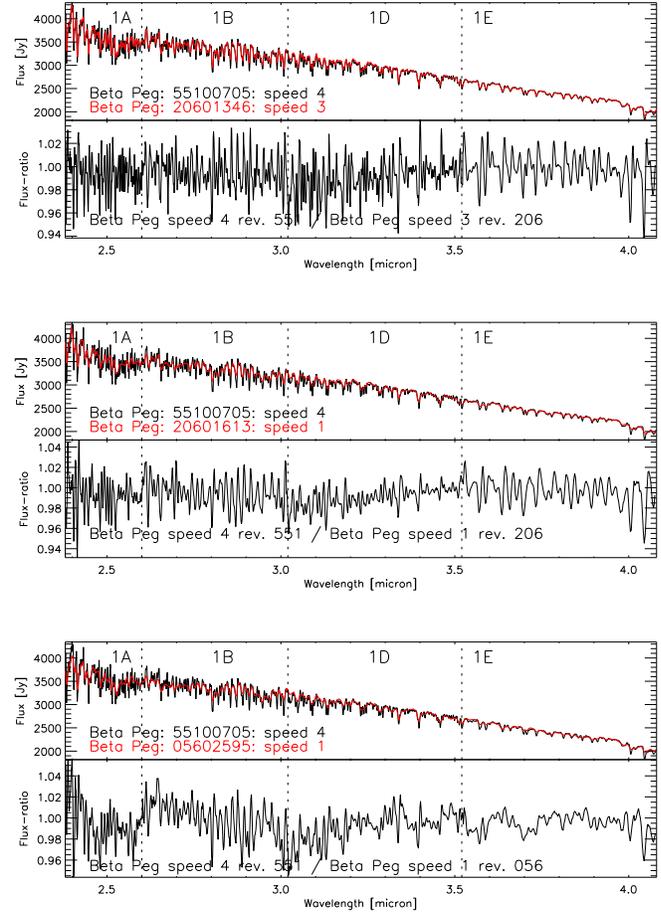}}}
\vspace*{-5ex}
\caption{\label{bpegvers}{\underline{Top:}} Comparison between the
AOT01 speed-4 observation of \boldmath {\bf{$\beta$ Peg}} \unboldmath
 (revolution 551) and the
speed-3 observation (revolution 206). The data of the speed-3
observation are divided by 1.02. {\underline{Middle:}} Comparison
between the AOT01 speed-4 observation of \boldmath {\bf{$\beta$ Peg}}
\unboldmath (revolution
551) and the speed-1 observation (revolution 206). The data of the
speed-1 observation are divided by 1.02. {\underline{Bottom:}}
Comparison between the AOT01 speed-4 observation of \boldmath
{\bf{$\beta$ Peg}} \unboldmath
(revolution 551) and the speed-1 observation (revolution 056). The
data of the speed-1 observation are divided by 1.04.}
\end{center}
\end{figure}

Two speed-1 observations and one speed-3 observation were also
made from $\beta$ Peg. The speed-3 observation was taken during
revolution 206 with pointing offset dy = $0.35''$ and dz =
$-0.76''$. No band shifts were applied and the overall flux level
is $2\,\%$ higher than for the speed-4 observation. The two speed-1
observations were taken during revolution 206 and 056, with
pointing offsets being dy = $0.45''$ and dz = $-0.69''$ and dy =
$0.18''$ and dz = $1.46''$ respectively. The larger pointing offset
for this latter observation results in a shift upwards of the
different sub-bands by $2\,\%$ for the data of band 1A and band 1B
and by $1.5\,\%$ for band 1E. The speed-1 observation from
revolution 206 (056) lies 2\,\% (4\,\%) higher than the speed-4
observation. The overall agreement between the different AOT01
observations is very good (Fig.\ \ref{bpegvers}).

The two observations taken during revolution 206 are once more an indication for
the flux calibration precision. The ratio of the speed-3 observation
(rebinned to the resolution of a speed-1 observation) to the speed-1
observation is displayed in Fig.\ \ref{bpegrev206} and proves a flux
accuracy of $\sim 2\,\%$.

\begin{figure}[h]
\begin{center}
\resizebox{0.5\textwidth}{!}{\rotatebox{90}{\includegraphics{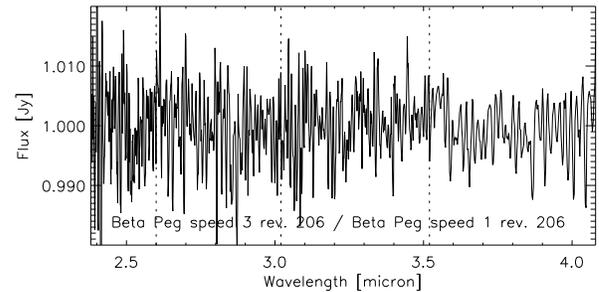}}}
\vspace*{-1truecm}
\caption{\label{bpegrev206}Ratio of a speed-3 observation of \boldmath
{\bf{$\beta$ Peg}} \unboldmath to a
speed-1 observation of $\beta$ Peg. Both observations were taken during
revolution 206.}
\end{center}
\end{figure}

\clearpage
\section{Comparison between the ISO-SWS and synthetic
spectra {\rm{(coloured plots)}}}

In this section, Fig.\ \ref{deldra} -- Fig.\ \ref{bpeg} of the
accompanying article are plotted in colour in order to better
distinguish the ISO-SWS data and the synthetic spectra.


\vspace*{-.5truecm}
\begin{figure}[h]
\begin{center}
\resizebox{0.5\textwidth}{!}{\rotatebox{90}{\includegraphics{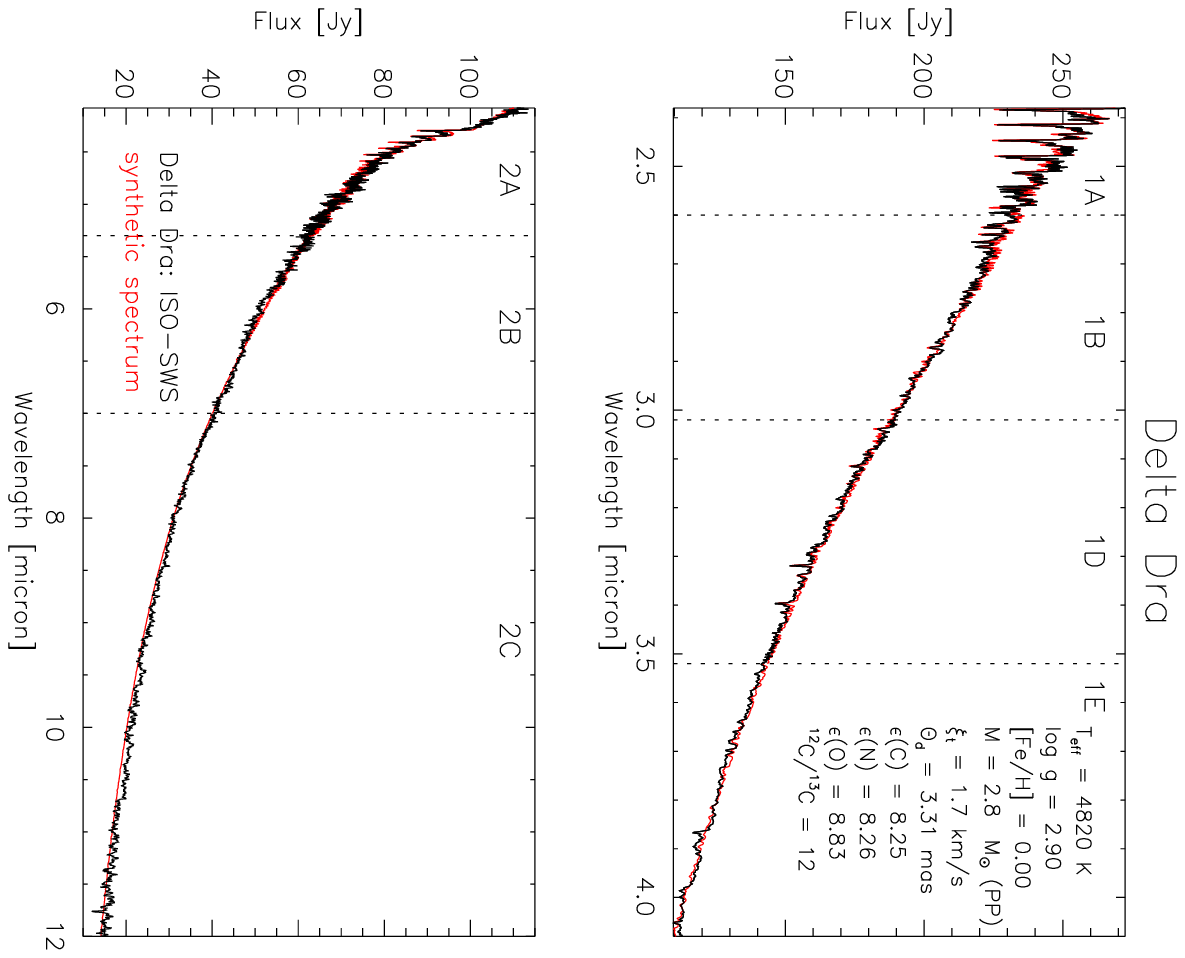}}}
\vspace*{-8ex}
\caption{\label{deldracol} Comparison between band 1 and band 2 of
the ISO-SWS data of \boldmath {\bf{$\mathbf{\delta}$ Dra}} \unboldmath
(black) and the synthetic
spectrum (red) with stellar parameters \teff\ = 4820\,K, $\log$ g =
2.70, M = 2.2\,\Msun, [Fe/H] = 0.00, \vt\ = 2.0\,\kms, \cc\ = 12,
$\varepsilon$(C) = 8.15, $\varepsilon$(N) = 8.26, $\varepsilon$(O)
= 8.93 and \ad\ = 3.30\,mas.}
\end{center}
\end{figure}

\vspace*{-.4truecm}

\begin{figure}[h]
\begin{center}
\resizebox{0.45\textwidth}{!}{\rotatebox{90}{\includegraphics{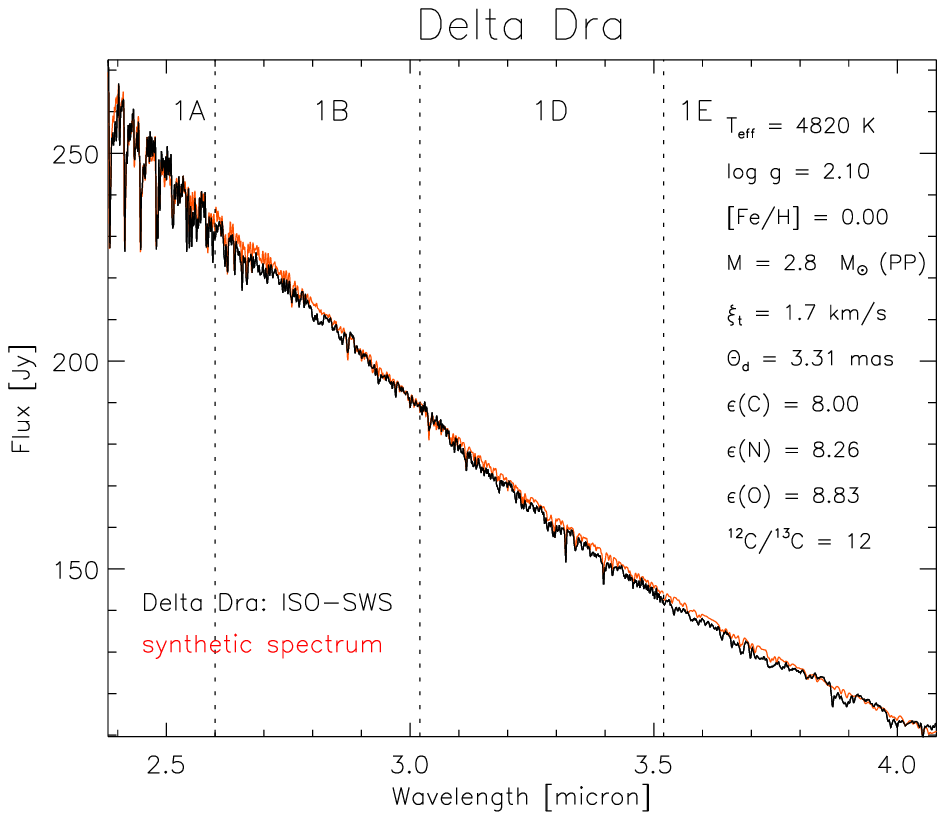}}}
\vspace*{-3ex}
\caption{\label{deldra210col} Comparison between band 1 of
the ISO-SWS data (rev.\ 206) of  \boldmath {\bf{$\mathbf{\delta}$
Dra}} \unboldmath (black) and the synthetic
spectrum (red) with stellar parameters \teff\ = 4820\,K, $\log$ g =
2.10, M = 2.8\,\Msun, [Fe/H] = 0.00, \vt\ = 1.7\,\kms, \cc\ = 12,
$\varepsilon$(C) = 8.00, $\varepsilon$(N) = 8.26, $\varepsilon$(O)
= 8.83 and \ad\ = 3.31\,mas.}
\end{center}
\end{figure}

\begin{figure}[h]
\begin{center}
\resizebox{0.5\textwidth}{!}{\rotatebox{90}{\includegraphics{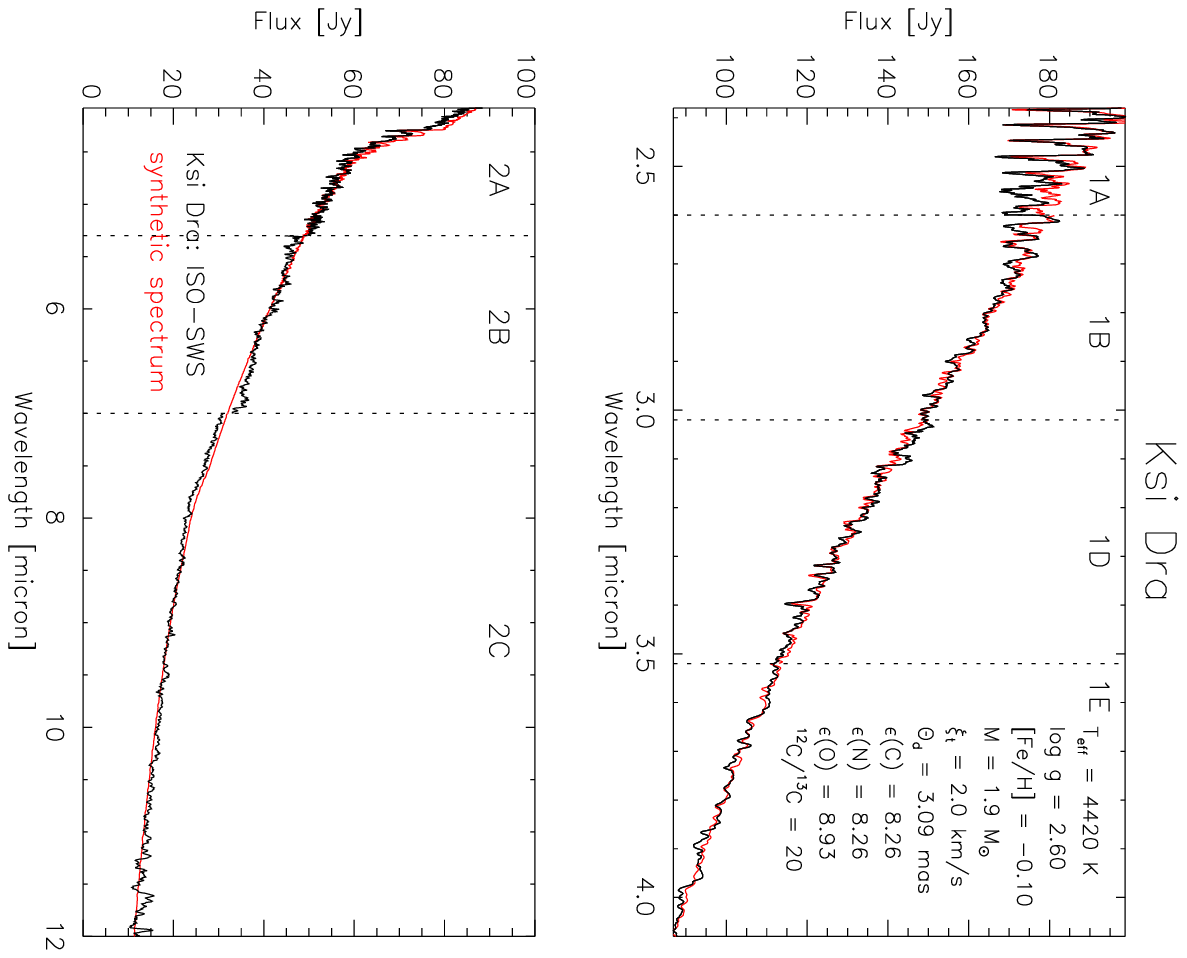}}}
\vspace*{-7ex}
\caption{\label{ksidracol} Comparison between band 1 and band 2 of
the ISO-SWS data (rev.\ 314) of  \boldmath {\bf{$\mathbf{\xi}$ Dra}}
\unboldmath (black) and the synthetic spectrum
(red) with stellar parameters \teff\ = 4440\,K, $\log$ g = 2.40, M
= 1.2\,\Msun, [Fe/H] = $0.10$, \vt\ = 2.0\,\kms, \cc\ = 20,
$\varepsilon$(C) = 8.00, $\varepsilon$(N) = 8.26, $\varepsilon$(O)
= 8.93 and \ad\ = 3.11\,mas.}
\end{center}
\end{figure}

\begin{figure}[h]
\begin{center}
\resizebox{0.5\textwidth}{!}{\rotatebox{90}{\includegraphics{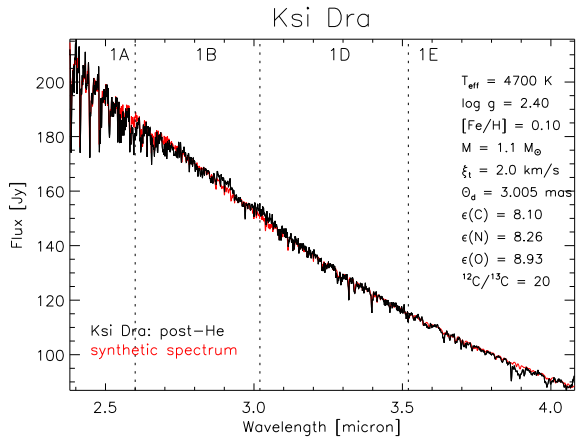}}}
\vspace*{-5ex}
\caption{\label{ksidra2col} Comparison between band 1 of the
post-helium ISO-SWS data of  \boldmath {\bf{$\mathbf{\xi}$ Dra}}
\unboldmath (black) and the synthetic
spectrum (red) with stellar parameters \teff\ = 4440\,K, $\log$ g =
2.40, M = 1.2\,\Msun, [Fe/H] = $0.10$, \vt\ = 2.0\,\kms, \cc\ = 20,
$\varepsilon$(C) = 8.00, $\varepsilon$(N) = 8.26, $\varepsilon$(O)
= 8.93 and \ad\ = 3.125\,mas.}
\end{center}
\end{figure}

\begin{figure}[h]
\begin{center}
\resizebox{0.49\textwidth}{!}{\rotatebox{90}{\includegraphics{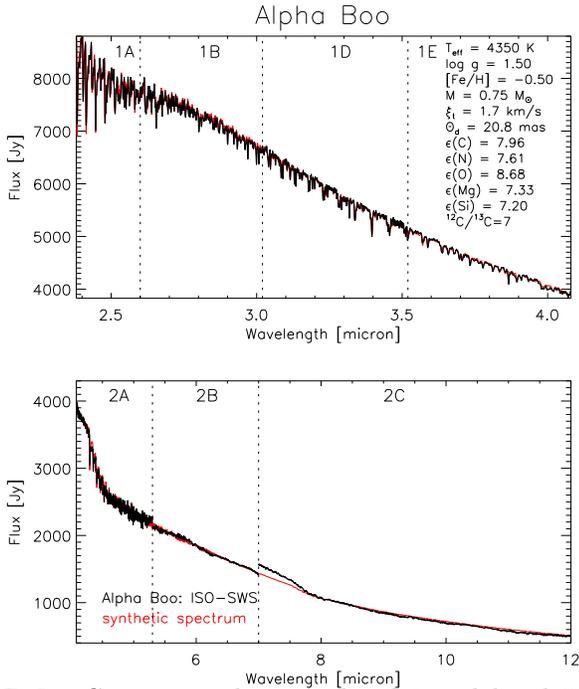}}}
\vspace*{-9ex}
\caption{\label{aboocol} Comparison between band 1 and band 2 of the
ISO-SWS data (rev.\ 452) of \boldmath {\bf{$\mathbf{\alpha}$ Boo}}
\unboldmath (black) and the synthetic spectrum
(red) with stellar parameters \teff\ = 4320\,K, $\log$ g = 1.50, M
= 0.75\,\Msun, [Fe/H] = $-0.50$, \vt\ = 1.7\,\kms, \cc\ = 7,
$\varepsilon$(C) = 7.96, $\varepsilon$(N) = 7.61, $\varepsilon$(O)
= 8.68, $\varepsilon$(Mg) = 7.33, $\varepsilon$(Si) = 7.20 and
\ad\ = 20.72\,mas.}
\end{center}
\end{figure}

\vspace*{-.3truecm}
\begin{figure}[h]
\begin{center}
\resizebox{0.49\textwidth}{!}{\rotatebox{90}{\includegraphics{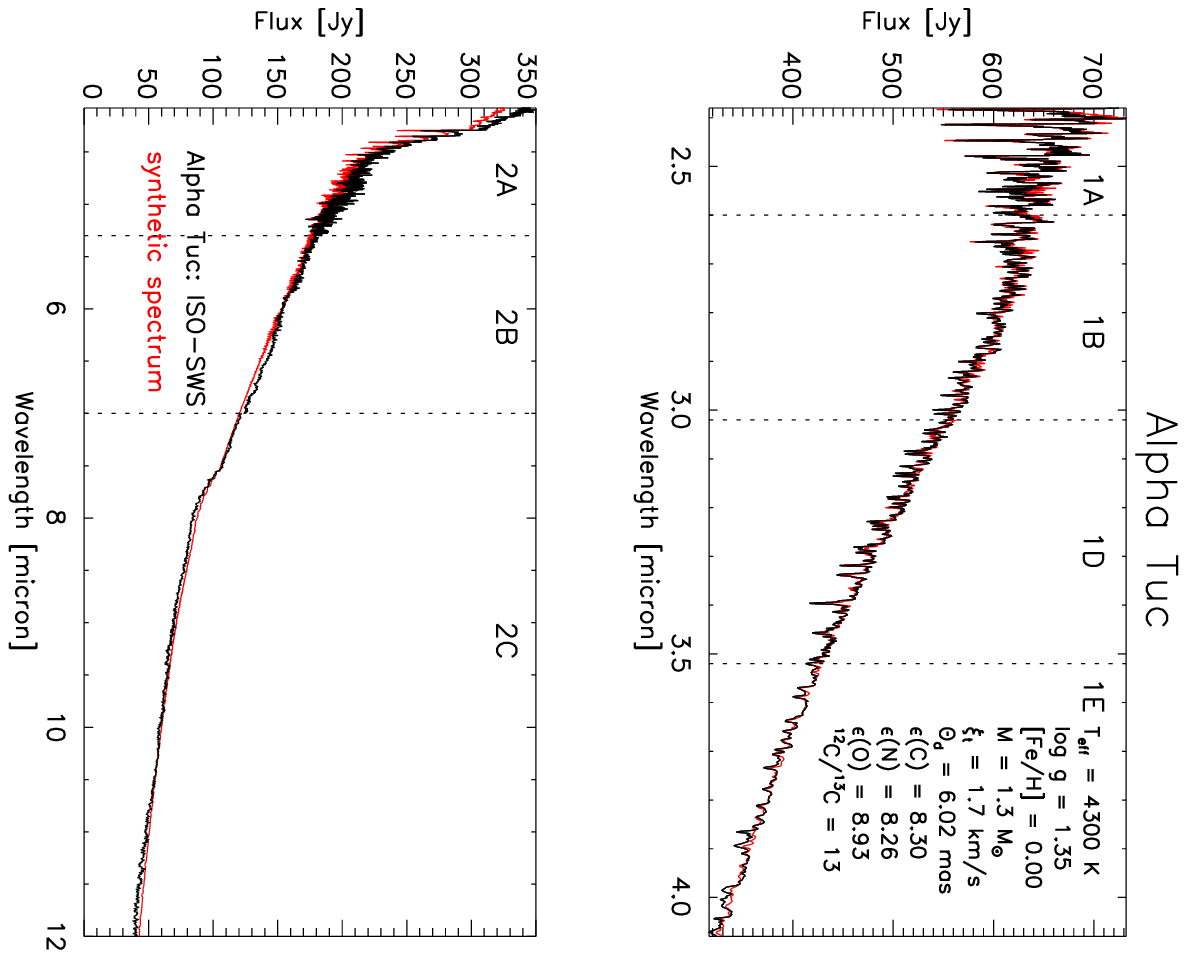}}}
\vspace*{-9ex}
\caption{\label{atuccol} Comparison between band 1 and band 2 of the
ISO-SWS data (rev.\ 866) of \boldmath {\bf{$\mathbf{\alpha}$ Tuc}}
\unboldmath (black) and the synthetic spectrum
(red) with stellar parameters \teff\ = 4300\,K, $\log$ g = 1.35, M
= 1.3\,\Msun, [Fe/H] = 0.00, \vt\ = 1.7\,\kms, \cc\ = 13,
$\varepsilon$(C) = 8.30, $\varepsilon$(N) = 8.26, $\varepsilon$(O)
= 8.93 and \ad\ = 6.02\,mas.}
\end{center}
\end{figure}

\begin{figure}[h]
\begin{center}
\resizebox{0.5\textwidth}{!}{\rotatebox{90}{\includegraphics{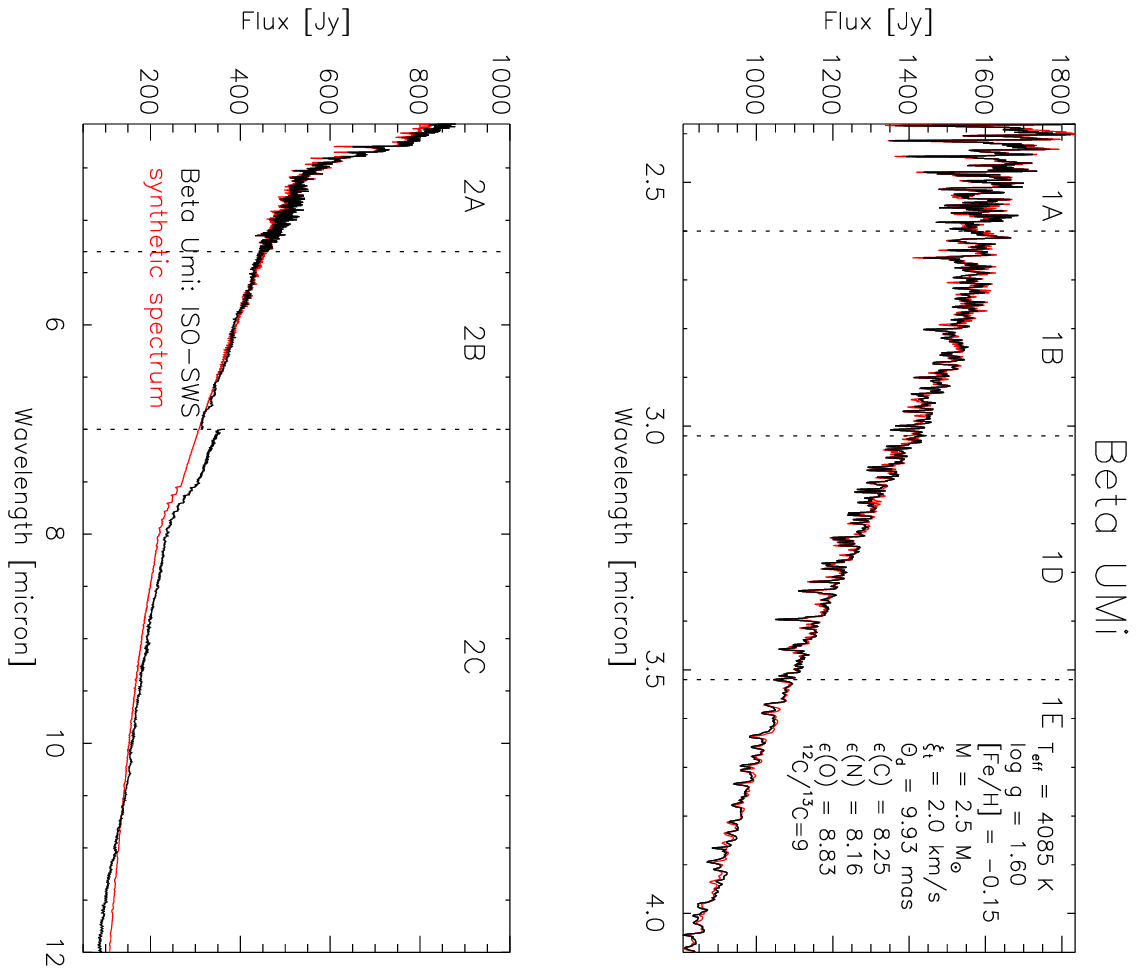}}}
\vspace*{-7ex}
\caption{\label{bumicol} Comparison between band 1 and band 2 of the
ISO-SWS data (rev.\ 182) of \boldmath {\bf{$\mathbf{\beta}$ UMi}}
\unboldmath (black) and the synthetic spectrum
(red) with stellar parameters \teff\ = 4085\,K, $\log$ g = 1.60, M
= 2.50\,\Msun, [Fe/H] = $-0.15$, \vt\ = 2.0\,\kms, \cc\ = 9, $\varepsilon$(C) =
8.25, $\varepsilon$(N) = 8.16, $\varepsilon$(O) = 8.83 and \ad\ =
9.93\,mas.}
\end{center}
\end{figure}

\begin{figure}[h]
\begin{center}
\resizebox{0.5\textwidth}{!}{\rotatebox{90}{\includegraphics{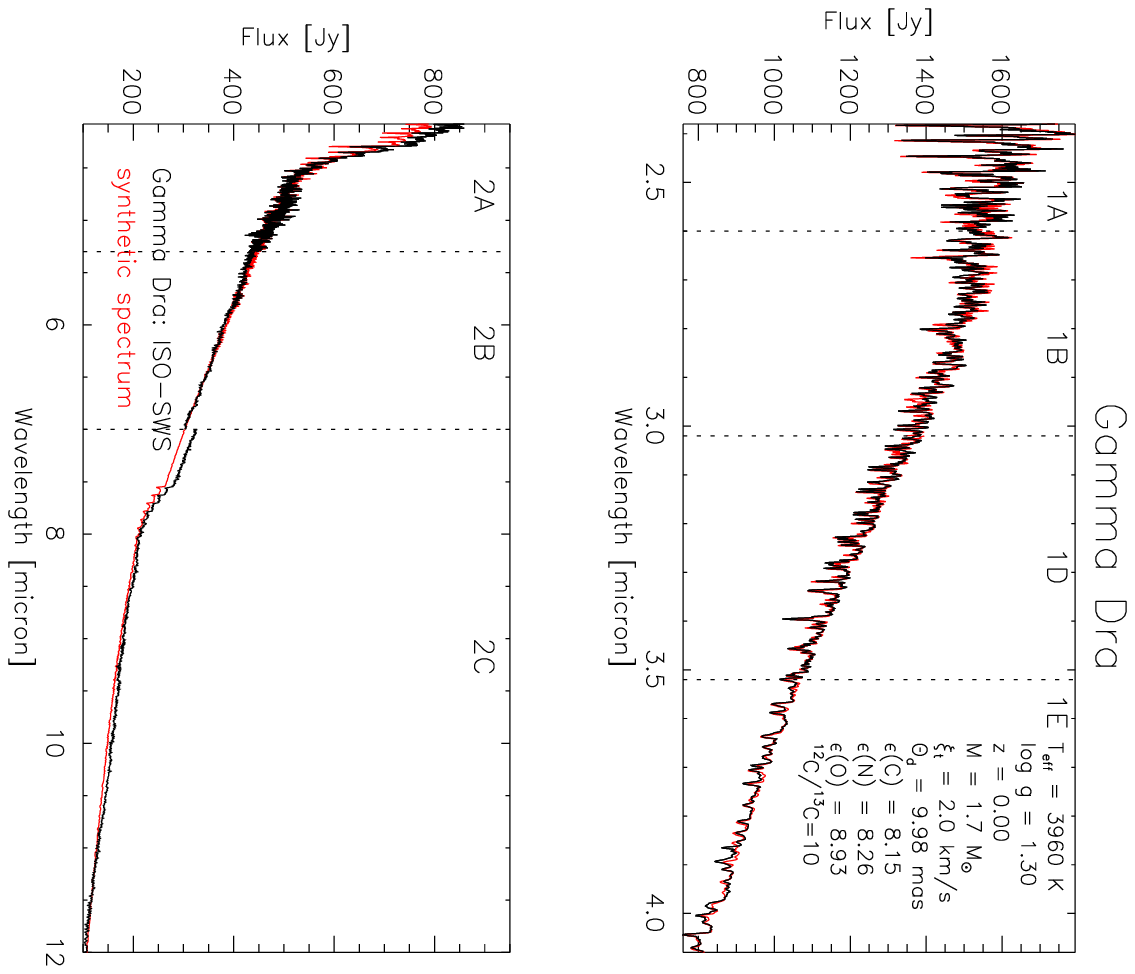}}}
\vspace*{-7ex}
\caption{\label{gamdracol} Comparison between band 1 and band 2 of
the ISO-SWS data (rev.\ 377) of \boldmath {\bf{$\mathbf{\gamma}$ Dra}}
\unboldmath (black) and the synthetic
spectrum (red) with stellar parameters \teff\ = 3960\,K, $\log$ g =
1.30, M = 1.7\,\Msun, [Fe/H] = 0.00, \vt\ = 2.0\,\kms, \cc\ = 10,
$\varepsilon$(C) = 8.15, $\varepsilon$(N) = 8.26, $\varepsilon$(O)
= 8.93 and \ad\ = 9.98\,mas.}
\end{center}
\end{figure}

\begin{figure}[h]
\begin{center}
\resizebox{0.49\textwidth}{!}{\rotatebox{90}{\includegraphics{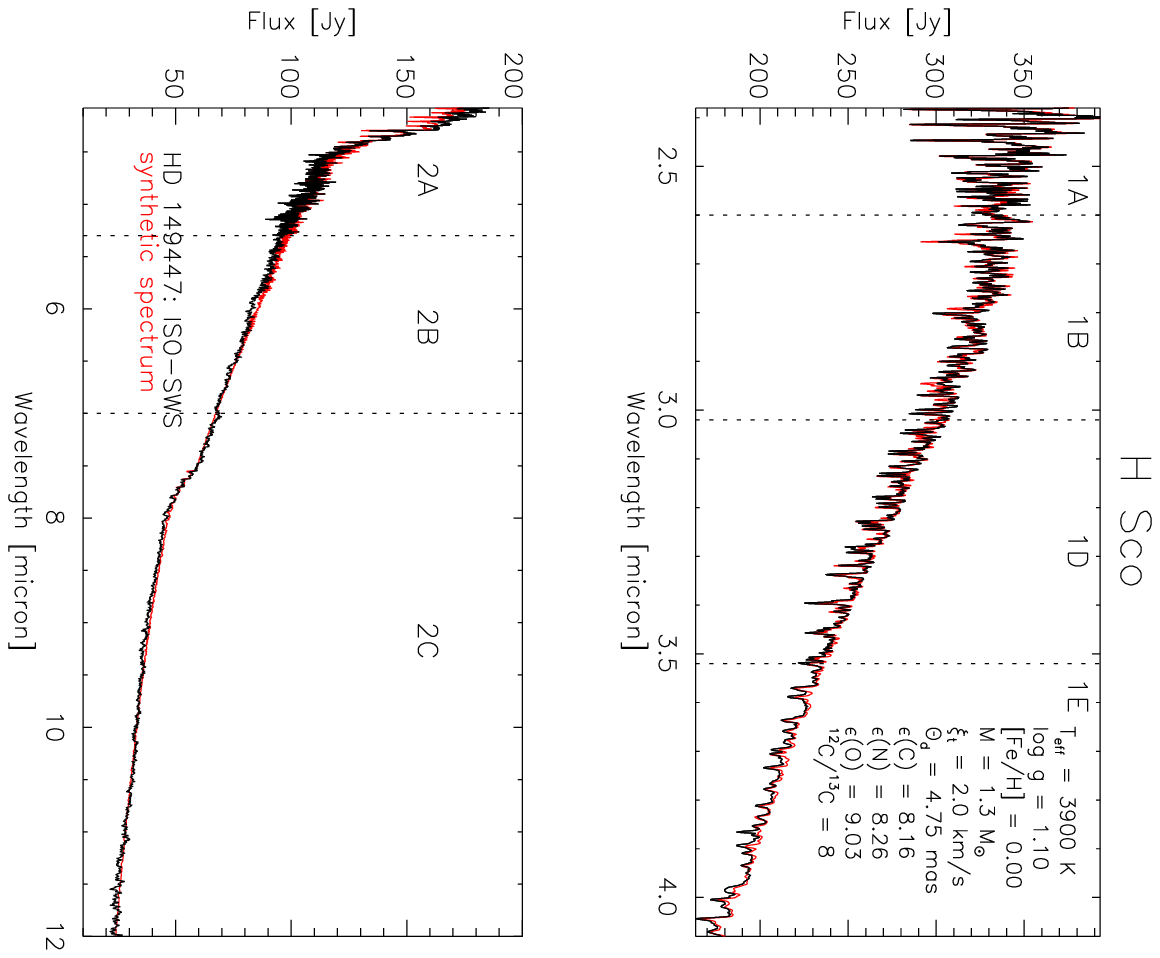}}}
\vspace*{-9ex}
\caption{\label{hd149col} Comparison between band 1 and band 2 of the
ISO-SWS data (rev.\ 847) of {\bf{H Sco}} (black) and the synthetic
spectrum (red)
with stellar parameters \teff\ = 3900\,K, $\log$ g = 1.30, M =
2.0\,\Msun, [Fe/H] = 0.00, \vt\ = 2.0\,\kms, \cc\ = 8, $\varepsilon$(C) =
8.20, $\varepsilon$(N) = 8.26, $\varepsilon$(O) = 8.83 and \ad\ =
4.73\,mas.}
\end{center}
\end{figure}

\begin{figure}[h]
\begin{center}
\resizebox{0.49\textwidth}{!}{\rotatebox{90}{\includegraphics{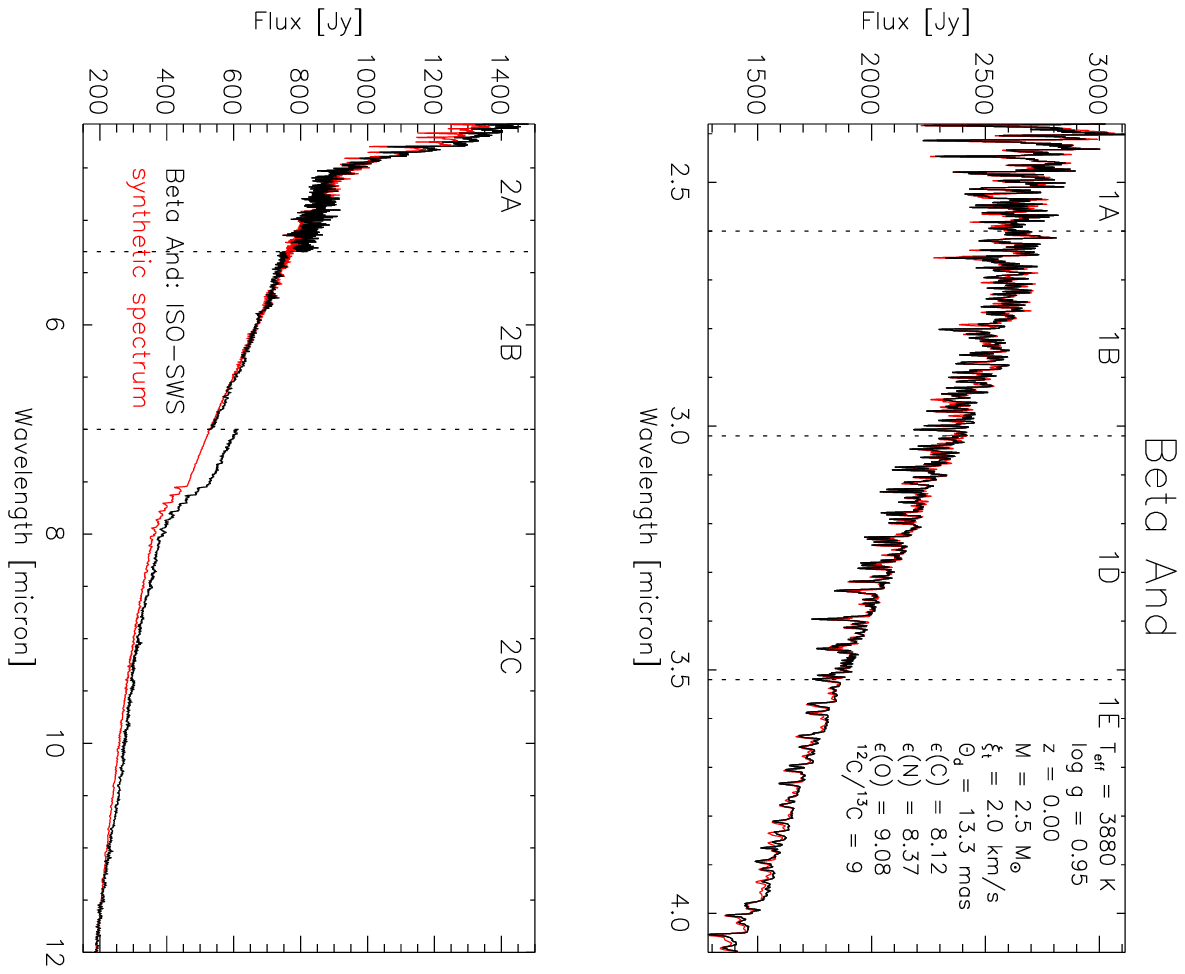}}}
\vspace*{-9ex}
\caption{\label{bandcol} Comparison between band 1 and band 2 of the
ISO-SWS data (rev.\ 795) of \boldmath {\bf{$\mathbf{\beta}$ And}}
\unboldmath (black) and the synthetic spectrum
(red) with stellar parameters \teff\ = 3880\,K, $\log$ g = 0.95, M
= 2.5\,\Msun, [Fe/H] = 0.00, \vt\ = 2.0\,\kms, \cc\ = 9, $\varepsilon$(C)
= 8.12, $\varepsilon$(N) = 8.37, $\varepsilon$(O) = 9.08 and \ad\
= 13.30\,mas.}
\end{center}
\end{figure}

\begin{figure}[h]
\begin{center}
\resizebox{0.5\textwidth}{!}{\rotatebox{90}{\includegraphics{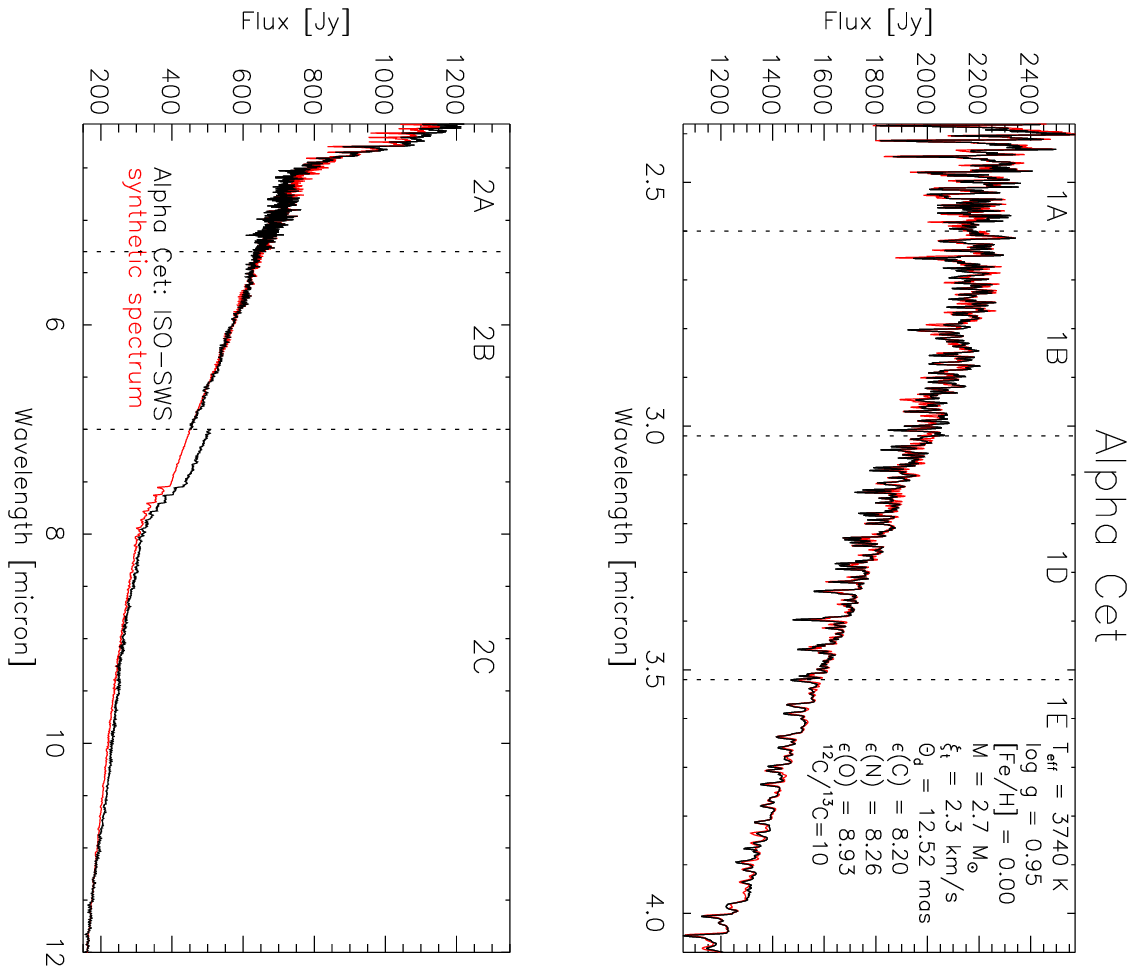}}}
\vspace*{-7ex}
\caption{\label{acetcol} Comparison between band 1 and band 2 of the
ISO-SWS data (rev.\ 797) of \boldmath {\bf{$\mathbf{\alpha}$ Cet}}
\unboldmath (black) and the synthetic spectrum
(red) with stellar parameters \teff\ = 3740\,K, $\log$ g = 0.95, M
= 2.7\,\Msun, [Fe/H] = 0.00, \vt\ = 2.3\,\kms, \cc\ = 10, $\varepsilon$(C)
= 8.20, $\varepsilon$(N) = 8.26, $\varepsilon$(O) = 8.93 and \ad\
= 12.52\,mas.}
\end{center}
\end{figure}

\begin{figure}[h]
\begin{center}
\resizebox{0.5\textwidth}{!}{\rotatebox{90}{\includegraphics{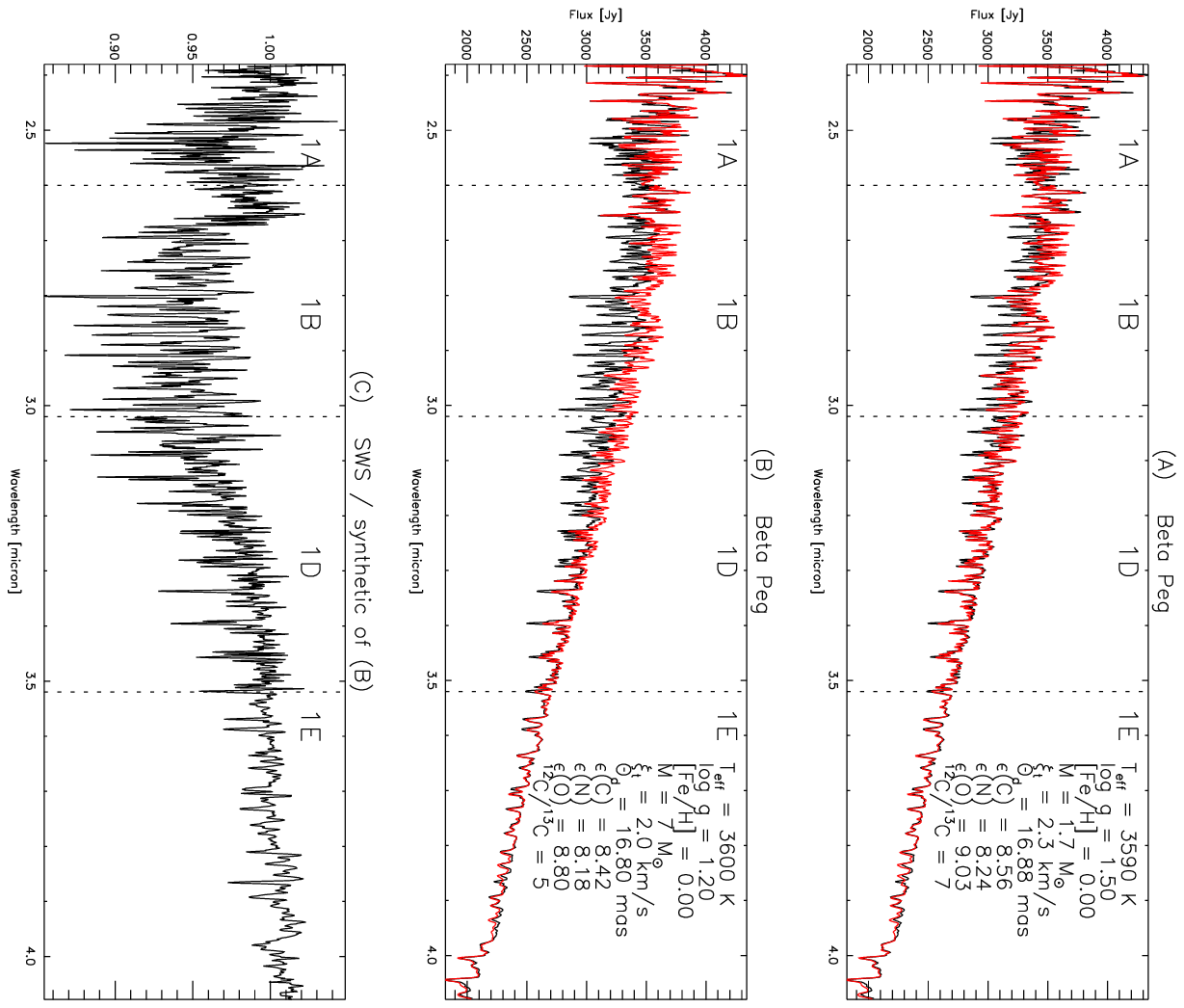}}}
\vspace*{-5ex}
\caption{\label{bpegcol} {\underline{Panel (A):}} Comparison between band 1 of the
ISO-SWS data (rev.\ 551) of \boldmath {\bf{$\mathbf{\beta}$ Peg}}
\unboldmath (black) and the synthetic spectrum
(red) with stellar parameters \teff\ = 3590\,K, $\log$ g = 1.50, M
= 1.7\,\Msun, [Fe/H] = 0.00, \vt\ = 2.3\,\kms, \cc\ = 7, $\varepsilon$(C)
= 8.56, $\varepsilon$(N) = 8.24, $\varepsilon$(O) = 9.03 and \ad\
= 16.88\,mas.
{\underline{Panel (B):}} Comparison between band 1 of the
ISO-SWS data (rev.\ 551) of \boldmath {\bf{$\mathbf{\beta}$ Peg}}
\unboldmath (black) and
the synthetic spectrum
(red) with stellar parameters \teff\ = 3600\,K, $\log$ g = 0.65, M
= 2\,\Msun, [Fe/H] = 0.00, \vt\ = 2.0\,\kms, \cc\ = 5, $\varepsilon$(C)
= 8.20, $\varepsilon$(N) = 8.18, $\varepsilon$(O) = 8.93 and \ad\
= 16.60\,mas.
{\underline{Panel (C):}} ISO-SWS observational data (rev.\ 551) of $\beta$ Peg
divided by the synthetic spectrum with the same parameters as in panel
(B).}
\end{center}
\end{figure}

\newpage \phantom{a}
\newpage\phantom{a}
\newpage
\phantom{a}
\newpage
\boldmath
\section{Temperatures derived from ${V-K}$ colours}
\unboldmath

This section contains the coloured version of Fig.\ \ref{VminK} of the
accompanying paper.

\begin{figure}[h!]
\begin{center}
\resizebox{0.5\textwidth}{!}{\rotatebox{90}{\includegraphics{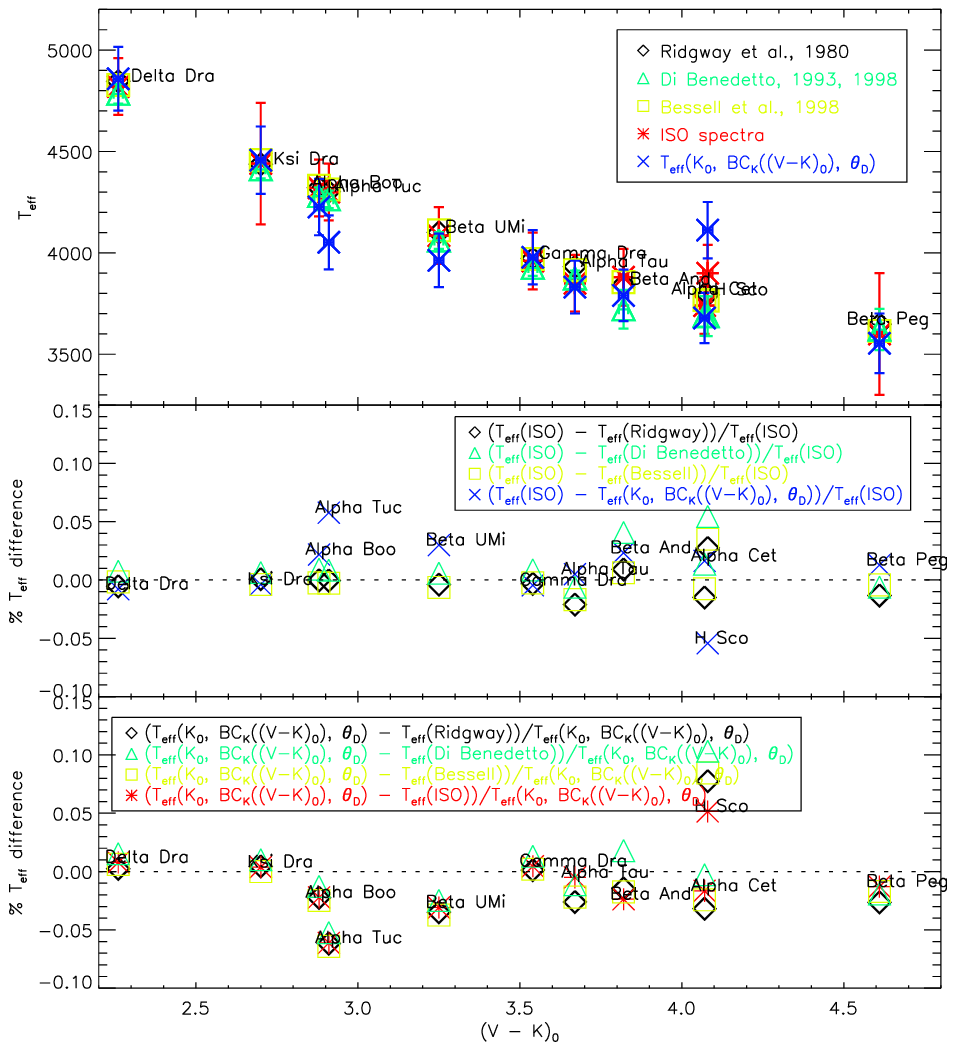}}}
\caption{\label{VminKcol}Comparison between \teff($V-K$) derived from
the relationship given by \citet{Ridgway1980ApJ...235..126R},
\citet{DiBenedetto1993A&A...270..315D},
\citet{DiBenedetto1998A&A...339..858D} or
\citet{Bessell1998A&A...333..231B}, T$_{\rm{eff}}(K_0, {\rm BC}_K, \pi)$
and the effective temperature deduced from the ISO-SWS spectra. }
\end{center}
\end{figure}


\section{Literature study}

\subsection{Summary tables}

For each of the 11 cool stars in our sample, a summary table of the
assumed and deduced stellar parameters retrieved from the various
quoted references in the accompanying paper can be found in this
subsection. The tables have been sorted by spectral type.

\clearpage

\begin{table*}[t!]
\caption{\label{litddra}Literature study of \boldmath
{\bf{$\mathbf{\delta}$ Dra}},  \unboldmath with
the effective
temperature \teff\ given in K, the mass M in \Msun, the microturbulent velocity
$\xi_t$ in km/s, the angular diameter $\theta_d$ in mas, the luminosity L
in L$_{\odot}$ and the radius R in R$_{\odot}$. Angular diameters deduced from
direct measurements (e.g from interferometry) are written in italic, while
others (e.g.\ from spectrophotometric comparisons) are written
upright. Assumed or adopted values are given between parentheses.
The results of this research are mentioned in the last line. A short
description of the methods and/or data used by the
several authors can be found in Sect.\ \ref{literatureapp}.}
\begin{center}
\tiny
\setlength{\tabcolsep}{.9mm}
\begin{tabular}{|c|c|c|c|c|c|c|c|c|c|c|c||l|} \hline
\rule[-3mm]{0mm}{8mm}  \teff & $\log$ g & M & $\xi_t$ &  [Fe/H] &
$\varepsilon$(C) & $\varepsilon$(N) & $\varepsilon$(O) &
$^{12}$C/$^{13}$C & $\theta_d$ & L & R & Ref.\\ \hline
\rule[-0mm]{0mm}{5mm}4940 & 3.00 & & & $-0.09$ & &  & & & & & & 1. \\
$4620 \pm 150$ & 2.7 & & 1.7 & $-0.08$ & &  & & & & & & 2. \\
$4940 \pm 50$ & $2.85 \pm 0.25$ & & & $0.10$ & $8.26 \pm 0.20$ & $8.00 \pm 0.30$
& $8.74 \pm 0.2$ & & & 81 & & 3. \\
4910 & & & & & &  & & & & & & 4. \\
$4670 \pm 170$ & (2.7) & & ($1.7$) & ($-0.08$) &  $8.47 \pm 0.18$
& $8.00 \pm 0.30$ & & & & & & 5. \\
$4910 \pm 60$ & $2.10 \pm 0.15$ & 0.75 & $2.4 \pm 0.1$ & $-0.07$ & &
& & & & & 13 & 6. \\
4830 & 2.4 & & & & &  & & & & & & 7. \\
$4530 \pm 220$ & & & & & &   & & & $\mathit{3.8 \pm 0.3}$ & & $15 \pm 4$ & 8.\\
($4910$) & ($2.10$) & & ($2.4$) & $-0.07$ & 8.55 & 8.36 & 8.70 & & & & & 9. \\
4810 & 2.9 & & & $-0.14$ & &  & & & & & & 10. \\
$4820 \pm 200$ & $2.98 \pm 0.27$ & & $2.0 \pm 0.5$ & $-0.27 \pm 0.12$ & &  &
& & & & & 11. \\
$4793 \pm 50$ & & & & & &  & & & & & & 12. \\
 & & & & $-0.159 \pm 0.047$ & &  & & & & & & 13. \\
\rule[-3mm]{0mm}{3mm}$4820 \pm 140$ & $2.70 \pm 0.20$ & $2.18 \pm 1.04$ & $2.0
\pm 0.5$ & $0.00 \pm 0.25$ & $8.15 \pm 0.25$ & $8.26 \pm 0.25$ & $8.93 \pm 0.25$ & $12 \pm 3$ & $3.30 \pm
 0.20$ & $58 \pm 10$ & $10.90 \pm 0.68$ & 14. \\
\hline
\end{tabular}
\end{center}
{\footnotesize{
1.\ \citet{vanParadijs1973A&A....23..369V};
2.\ \citet{Gustafsson1974A&A....34...99G};
3.\ \citet{vanParadijs1976A&A....53....1V};
4.\ \citet{Linsky1978ApJ...220..619L};
5.\ \citet{Bonnell1982MNRAS.201..253B};
6.\ \citet{Gratton1982MNRAS.201..807G};
7.\ \citet{Burnashev1983IzKry..67...13B};
8.\ \citet{Faucherre1983A&A...120..263F};
9.\ \citet{Gratton1985A&A...148..105G};
10.\ \citet{Brown1989ApJS...71..293B};
11.\ \citet{McWilliam1990ApJS...74.1075M};
12.\ \citet{Blackwell1998A&AS..129..505B};
13.\ \citet{Taylor1999A&AS..134..523T};
14.\ present results}}
\end{table*}


\begin{table*}[t!]
\boldmath
\caption{\label{litkdra}See caption of Table \ref{litddra}, but now
for {\bf{$\mathbf{\xi}$ Dra}}.}
\unboldmath
\begin{center}
\tiny
\setlength{\tabcolsep}{.9mm}
\begin{tabular}{|c|c|c|c|c|c|c|c|c|c|c|c||l|} \hline
\rule[-3mm]{0mm}{8mm}  \teff & $\log$ g & M & $\xi_t$
&  [Fe/H] & $\varepsilon$(C) & $\varepsilon$(N) & $\varepsilon$(O)
& $^{12}$C/$^{13}$C & $\theta_d$ & L & R & Ref.\\ \hline
\rule[-0mm]{0mm}{5mm}$4270 \pm 150$ & $2.5 \pm 0.3$ & & & $0.11 \pm
0.12$ & & & & & & & & 1. \\
4470 & & & & & &  & & $20 \pm 2$ & & 40 & & 2. \\
$4525 \pm 60$ & $2.42 \pm 0.15$ & $1.5 \pm 1.5$ & 2.3 &
$-0.12 \pm 0.14$ & &  & & & & & 14 & 3. \\
4510 & 2.34 & & & $-0.36$ & &  & & & & & & 4. \\
4450 & 2.30 & (1) & (1.7) & $+0.05$ & &  & & & & & & 5. \\
$4493 \pm 45$ & & & & & &  & & & $ 3.076 \pm 0.060$ & & & 6. \\
$4420 \pm 200$ & $2.61 \pm 0.30$ & & $1.9 \pm 0.5$ & $-0.09 \pm 0.09$ & &  & & &
& & &  7.\\
$4525 \pm 30$ & & & & & &  & & & $3.025 \pm 0.060$ & & & 8. \\
$4495 \pm 40$ & & & & & &  & & & & & & 9. \\
 & & & & $-0.043 \pm 0.048$ & &  & & & & & & 10. \\
\rule[-3mm]{0mm}{3mm}$4440 \pm 300$ & $2.40 \pm 0.25$ & $1.20 \pm 0.71$ & $2.0
\pm 1.0$ & $0.10 \pm 0.30$ & $8.00 \pm 0.30$ & $8.26 \pm 0.40$ & $8.93 \pm
0.30$ & $20 \pm 5$&  $3.11 \pm 0.22$ & $46 \pm 14$ & $11.42 \pm 0.83$ & 11. \\
\hline
\end{tabular}
\end{center}
{\footnotesize{
1.\ \citet{Gustafsson1974A&A....34...99G};
2.\ \citet{Dearborn1975ApJ...200..675D};
3.\ \citet{Gratton1982MNRAS.201..807G};
4.\ \citet{Gratton1983MNRAS.202..231G};
5.\ \citet{Brown1989ApJS...71..293B};
6.\ \citet{Blackwell1990A&A...232..396B};
7.\ \citet{McWilliam1990ApJS...74.1075M};
8.\ \citet{Blackwell1991A&A...245..567B};
9.\ \citet{Blackwell1998A&AS..129..505B};
10.\ \citet{Taylor1999A&AS..134..523T};
11.\  present results}}
\end{table*}

\clearpage

\begin{table*}[t!]
\boldmath
\caption{\label{litaboo}See caption of Table \ref{litddra}, but now
for {\bf{$\mathbf{\alpha}$ Boo}}.}
\unboldmath
\begin{center}
\setlength{\tabcolsep}{.1mm}
\tiny
\begin{tabular}{|c|c|c|c|c|c|c|c|c|c|c|c||l|} \hline
\rule[-3mm]{0mm}{8mm}  \teff & $\log$ g & M & $\xi_t$ &  [Fe/H] &
$\varepsilon$(C) & $\varepsilon$(N) & $\varepsilon$(O) &
$^{12}$C/$^{13}$C & $\theta_d$ & L & R & Ref.\\ \hline
\rule[-0mm]{0mm}{5mm} $4350 \pm 50$ & $1.95 \pm 0.25$ & & & $-0.5$ & &
& & & ({\it{23}}) & & & 1. \\
$4260 \pm 50$ & $0.90 \pm 0.35$ & 0.1-0.6 & 1.8 & & $7.81 \pm 0.15$ & $7.03 \pm
0.15$ & $8.23 \pm 0.15$ & & & $229 \pm 100$ & $28 \pm 6$ & 2. \\
$4410 \pm 80$ & & & & & &  & & & $20.1 \pm 1.0$ & & & 3. \\
4240 & & & & & &  & & & & & & 4. \\
$4060 \pm 150$ & & & & & &  & & & $21.6 \pm 1.9$ & & & 5. \\
$4420 \pm 150$ & & & & & &  & & & $19.99 \pm 0.40$ & & & 6. \\
(4260) & (1.6) & & & & &  & & (7) & & & & 7. \\
$4490 \pm 100$ & $2.01 \pm 0.46$ & & 1.8 & $-0.56 \pm 0.07$ & $8.07 \pm 0.06$ &
$7.60 \pm 0.07$ & $8.76 \pm 0.08$ & 7.2 & & 126 & & 8. \\
4350 & & & & & &  & & & $20.5 \pm 1.1$ & & & 9. \\
$4205 \pm 150$ & & & & & &  & & & & & & 10. \\
$4375 \pm 50$ & (1.5) & & & ($-0.5$) & &  & & & & & & 11. \\
4350 & 1.8 & & & ($-0.51$) & $7.89 \pm 0.20$ & $7.61 \pm 0.20$ & $8.68
\pm 0.30$ & & & & &  12. \\
$4490 \pm 200$ & $2.6 \pm 0.3$ & & & $-0.55 \pm 0.30$ & &  & & & & & &
13$^a$. \\
4370 & & & & & &  & & & & & & 13$^b$. \\
(4375) & (1.57) & & (1.6) & & 7.97 &  & & ($7$) & & & & 14. \\
(4375) & $1.6 \pm 0.2$ & 0.42 -- 1.5 & (1.7) & $-0.5$ & &  & & & & & & 15. \\
($4410$) & ($>0.98$) & $> 0.38$ & ($2.5$) &
($-0.50$) & $7.79 \pm 0.10$ & $7.32 \pm 0.10$ & $8.19 \pm 0.10$ & & & & &
16. \\
 & & & & & &  & & & & (24.9) & $27.5 \pm 1.1$ & 17$^a$. \\
& & & & & &  & & & & & 23.4 & 17$^b$. \\
($4225$) & $1.6 \pm 0.2$ & $1.1^{+0.6}_{-0.4}$ & & ($-0.56$) & &
 & & & $21.4 \pm 1$ & ($200$) & ($27$) & 18. \\
4400 & 1.7 & & 2.3 & $-0.6$ & &  & & & & & & 19. \\
(4375) & $1.5 \pm 0.5$ & & $2.22 \pm 0.11$ &  & $7.87 \pm 0.06$ &  & & &
& & & 20. \\
($4300$) & (1.74) & ($1.00$) & (2.5) & & $8.52 \pm 0.52$ & $8.94 \pm 0.06$ &
$8.83 \pm 0.55$ & & & & (26) &  21. \\
$4294 \pm 30$ & & & & & &  & & & ${\mathit{20.95 \pm 0.20}}$ & & & 22. \\
(4375) & $1.97 \pm 0.20$ & & (1.8) & $-0.42$ & &  & & & & & & 23. \\
 & & 1.1 & & & &  & & & & & & 24. \\
4321 & (1.8) & & & ($-0.51$) & &  & & & 21.07 & & & 25. \\
4340 & 1.9 & & (1.7)  & $-0.39$ & &  & & & & & & 26. \\
$4294 \pm 30$ & & & & & &  & & & & ($186$) & $24.6 \pm 1.3$ & 27. \\
4300 & 2.0 & & 1.5 & $-0.69 \pm 0.10$ & &  & & & & & & 28. \\
$4280 \pm 200$ & $2.19 \pm 0.27$ &  & $2.3 \pm 0.5$ & $-0.60 \pm 0.14$ & &  & & & & & & 29. \\
$4362 \pm 45$ & & & & & &  & & & 20.43 & & & 30. \\
$4250 \pm 80$ & $1.6 \pm 0.3$ & 1.1 & & & &  & & & 21.4 & 170 & 27 & 31. \\
($4375$) & $1.5 \pm 0.5$ & & & & $8.06 \pm 0.04$ &  & & & & & & 32. \\
4265 & & & & & &  & & & 21.5 & & & 33. \\
4450 & $1.96|1.98$ & & $1.5|2.0$ & $-0.5$ & &  & $8.97|8.96$ & & & & &
34$^a$. \\
4350 & $1.71|1.73$ & & $1.5|2.0$ & $-0.5$ & &  & $8.86|8.84$ & & & & &
34$^b$. \\
4250 & $1.43|1.44$ & & $1.5|2.0$ & $-0.5$ & &  & $8.72|8.71$ & & & & &
34$^c$. \\
4250 & $1.81|1.82$ & & $1.5|2.0$ & $0.0$ & &  & $9.14|9.12$ & & & & & 34$^d$. \\
$4300 \pm 30$ & $1.5 \pm 0.2$ & $0.70 \pm 0.25$ & $1.7 \pm 0.3$ & $-0.5 \pm 0.1$
& 8.06 & 7.85 & 8.83 & & ${\mathit 20.95}$ & & $24.5 \pm 1.4$ & 35. \\
(4260) & (0.9) & (0.23) & {\tiny{1.68-1.87 $|$ $1.34 \pm 0.07$}} & ($-0.77$) & &
 & & & & & 28 & 36$^a$. \\
(4420) & (1.7) & (1.23) & {\tiny{1.54-1.68 $|$ $1.48 \pm 0.06$}} & ($-0.50$) & &
 & & & & & 28 & 36$^b$. \\
4362 & 2.4 & & & & &  & & & 21.12 & & & 37. \\
$4303 \pm 47$ & & & & & &  & & &  ${\mathit{21.0 \pm 0.2}}$ & & & 38. \\
($4375$) & ($1.5$) & & $1.5 \pm 1.0$ & & ($8.06$) & $7.77 \pm
0.02$ & & & & & &  39. \\
4300 & 1.4 & & & $-0.47$ & 8.03 &  & & $9 \pm 1.5$ & & & & 40. \\
$4291 \pm 48$ & & & & & &   & & & ${\mathit{20.97 \pm 0.20}}$ & & & 41$^a$. \\
4255 & & & & & &  & & &  {\it{20.76}} & & & 41$^b$. \\
$4628 \pm 210$ & & & & & &  & & & ${\mathit 19.1 \pm 1.0}$ & & & 42. \\
4320 & & & & & &  & & & & & & 43. \\
$4321 \pm 44$ & & & & & &  & & &  {\it{20.907}} & & & 44. \\
 & & & & $-0.547 \pm 0.021$ & &  & & & & & & 45. \\
\rule[-3mm]{0mm}{3mm}$4320 \pm 140$ & $1.50 \pm 0.15$ & $0.73 \pm 0.27$ & $1.7
\pm 0.5$ & $-0.50 \pm 0.20$ & $7.96 \pm 0.20$ & $7.61 \pm 0.25$ & $8.68 \pm
0.20$ & $7 \pm 2$ & $20.72 \pm 1.28$ & $197 \pm 36$ & $25.06 \pm 1.56$ & 46.\\
\hline
\end{tabular}
\end{center}
{\footnotesize{
1.\ \citet{vanParadijs1974A&A....35..225V};
2.\ \citet{Mackle1975A&AS...19..303M};
3.\ \citet{Blackwell1977MNRAS.180..177B};
4.\ \citet{Linsky1978ApJ...220..619L};
5.\ \citet{Scargle1979ApJ...228..838S};
6.\ \citet{Blackwell1980A&A....82..249B};
7.\ \citet{Lambert1980ApJ...235..114L};
8.\ \citet{Lambert1981ApJ...248..228L};
9.\ \citet{Manduca1981ApJ...243..883M};
10.\ \citet{Tsuji1981A&A....99...48T};
11.\ \citet{Frisk1982MNRAS.199..471F};
12.\ \citet{Kjaergaard1982A&A...115..145K};
13.\ \citet{Burnashev1983IzKry..67...13B};
14.\ \citet{Harris1984ApJ...285..674H};
15.\ \citet{Bell1985MNRAS.212..497B};
16.\ \citet{Gratton1985A&A...148..105G};
17.\ \citet{Moon1985Ap&SS.117..261M};
18.\ \citet{Judge1986MNRAS.221..119J};
19.\ \citet{Kyrolainen1986A&AS...65...11K};
20.\ \citet{Tsuji1986A&A...156....8T};
21.\ \citet{Altas1987Ap&SS.134...85A};
22.\ \citet{DiBenedetto1987A&A...188..114D};
23.\ \citet{Edvardsson1988A&A...190..148E};
24.\ \citet{Harris1988ApJ...325..768H};
25.\ \citet{Bell1989MNRAS.236..653B};
26.\ \citet{Brown1989ApJS...71..293B};
27.\ \citet{Volk1989AJ.....98.1918V};
28.\ \citet{Fernandez-Villacanas1990AJ.....99.1961F};
29.\ \citet{McWilliam1990ApJS...74.1075M};
30.\ \citet{Blackwell1991A&A...245..567B};
31.\ \citet{Judge1991ApJ...371..357J};
32.\ \citet{Tsuji1991A&A...245..203T};
33.\ \citet{Engelke1992AJ....104.1248E};
34.\ \citet{Bonnell1993MNRAS.264..319B};
35.\ \citet{Peterson1993ApJ...404..333P};
36.\ \citet{Gadun1994AN....315..413G};
37.\ \citet{Cohen1996AJ....112.2274C};
38.\ \citet{Quirrenbach1996A&A...312..160Q};
39.\ \citet{Aoki1997A&A...328..175A};
40.\ \citet{Pilachowski1997AJ....114..819P};
41.\ \citet{DiBenedetto1998A&A...339..858D};
42.\ \citet{Dyck1998AJ....116..981D};
43.\ \citet{Hammersley1998A&AS..128..207H};
44.\ \citet{Perrin1998A&A...331..619P};
45.\ \citet{Taylor1999A&AS..134..523T};
46.\ present results}}
\end{table*}

\clearpage

\begin{table*}[t!]
\boldmath
\caption{\label{litatuc}See caption of Table \ref{litddra}, but now
for {\bf{$\mathbf{\alpha}$ Tuc}}.}
\unboldmath
\vspace{1ex}
\begin{center}
\tiny
\setlength{\tabcolsep}{1.1mm}
\begin{tabular}{|c|c|c|c|c|c|c|c|c|c|c|c||l|} \hline
\rule[-3mm]{0mm}{8mm}  \teff & $\log$ g & M & $\xi_t$ &  [Fe/H] &
$\varepsilon$(C) &  $\varepsilon$(N) & $\varepsilon$(O) &
$^{12}$C/$^{13}$C & $\theta_d$ & L & R & Ref.\\
\hline \rule[-0mm]{0mm}{5mm} 4040 & & & & & & & & & 6.45 & & & 1. \\
(4450) & & & & & &  & & & 4.52 & & & 2. \\
$4040 \pm 70$ & $1.94 \pm 0.50$ & 3.7 & & & &  & &
& & & 34 & 3. \\
 & & & & $-0.36$ & & & & & & & & 4. \\
$3920 \pm 250$ & 1.10 & $1.15 \pm 0.35$ & & & & & & & & & & 5. \\
\rule[-3mm]{0mm}{3mm}$4300 \pm 140$ & $1.35 \pm 0.25$ & $1.27 \pm 0.48$ & $1.7
\pm 0.5$ & $0.00 \pm 0.20$ & $8.30 \pm 0.20$ & $8.26 \pm 0.25$ & $8.93 \pm 0.20$
& $13 \pm 3$ & $6.02 \pm 0.37$ & $479 \pm 93$ & $39.40 \pm 2.82$ & 6. \\
\hline
\end{tabular}
\end{center}
{\footnotesize{
1.\ \citet{Basri1979ApJ...234.1023B};
2.\ \citet{Stencel1980ApJS...44..383S};
3.\ \citet{Glebocki1988A&A...189..199G};
4.\ \citet{Flynn1991A&A...250..400F};
5.\ \citet{Pasquini1992A&A...266..340P};
6.\ present results}}
\end{table*}


\begin{table*}[htb]
\boldmath
\caption{\label{litbumi}See caption of Table \ref{litddra}, but now
for {\bf{$\mathbf{\beta}$ UMi}}.}
\unboldmath
\vspace{1ex}
\begin{center}
\setlength{\tabcolsep}{.8mm}
\tiny
\begin{tabular}{|c|c|c|c|c|c|c|c|c|c|c|c||l|} \hline
\rule[-3mm]{0mm}{8mm}  \teff & $\log$ g & M & $\xi_t$ &  [Fe/H] &
$\varepsilon$(C) &  $\varepsilon$(N) & $\varepsilon$(O)
& $^{12}$C/$^{13}$C & $\theta_d$ & L & R & Ref.\\
\hline \rule[-0mm]{0mm}{5mm}4000 & & & 1.2 & & &  & & $11 \pm 1.65$ & & 224 & & 1. \\
3970 & & & & & &  & & & & & & 2. \\
(4000) & ($2.0$) & & & & &  & & (11) & & & & 3. \\
(3970) & & & & & &   & & & 10.4 & & & 4. \\
$4340 \pm 100$ & $1.86 \pm 0.46$ & & $2.00 \pm 0.25$ & $-0.23 \pm 0.18$ & $8.23
\pm 0.11$ & $8.16 \pm 0.13$ & $8.80 \pm 0.15$ & (11) & & 158 & & 5. \\
4020 & 2.40 & & & & &  & & & & & & 6. \\
$4220 \pm 300$ & & & & & &  & & & ${\mathit{8.9 \pm 1.1}}$ & & $30 \pm 15$ &
7.\\
($4340$) & ($1.86$) & 1.5 & 2.0 & & 8.55 &  & & 12  & & & &
8. \\
$4050 \pm 100$ & (1.6) & & & $-0.14$ & &  & & & & & & 9. \\
4000 & 2.0 & & 1.7 & & &  & & & & & & 10. \\
$4030 \pm 200$ & $1.83 \pm 0.30$ & & $2.4 \pm 0.5$ &
$-0.29 \pm 0.13$ & & & & & & & & 11.\\
$4068 \pm 80$ & 1.6 & $1.5 ^{+1.5}_{-0.5}$ & & & & & & & 9.4 & 220 & 30 & 12. \\
3900 & & & & & &  & & & & & & 13. \\
$4060 \pm 200$ & $2.0 \pm 0.2 $ & & & & &  & & & & & & 14.\\
$4086 \pm 225$ & & & & & &  & & & ${\mathit{9.7 \pm 0.8}}$ & & & 15.\\
 & & & & $-0.132 \pm 0.061$ & &   & & & & & & 16.\\
\rule[-3mm]{0mm}{3mm}$4085 \pm 140$ & $1.60 \pm 0.15$ & $2.49 \pm 0.92$ & $2.0
\pm 0.5$ & $-0.15 \pm 0.20$ & $8.25 \pm 0.20$ & $8.16 \pm 0.25$ & $8.83 \pm
0.20$ & $9 \pm 2$ & $9.93 \pm 0.62$ & $430 \pm 82$ & $41.38 \pm 2.71$ & 17. \\
\hline
\end{tabular}
\end{center}
{\footnotesize{
1.\ \citet{Tomkin1976ApJ...210..694T};
2.\ \citet{Linsky1978ApJ...220..619L};
3.\ \citet{Lambert1980ApJ...235..114L};
4.\ \citet{Stencel1980ApJS...44..383S};
5.\ \citet{Lambert1981ApJ...248..228L};
6.\ \citet{Burnashev1983IzKry..67...13B};
7.\ \citet{Faucherre1983A&A...120..263F};
8.\ \citet{Harris1988ApJ...325..768H};
9.\ \citet{Bell1989MNRAS.236..653B};
10.\ \citet{Brown1989ApJS...71..293B};
11.\ \citet{McWilliam1990ApJS...74.1075M};
12.\ \citet{Judge1991ApJ...371..357J};
13.\ \citet{Cornide1992AJ....103.1374C};
14.\ \citet{Morossi1993A&A...277..173M};
15.\ \citet{Dyck1998AJ....116..981D};
16.\ \citet{Taylor1999A&AS..134..523T};
17.\ present results}}
\end{table*}

\clearpage

\begin{table*}[htb]
\boldmath
\caption{\label{litgamdra}See caption of Table \ref{litddra}, but now
for {\bf{$\mathbf{\gamma}$ Dra}}.}
\unboldmath
\vspace{1ex}
\begin{center}
\setlength{\tabcolsep}{.5mm}
\tiny
\begin{tabular}{|c|c|c|c|c|c|c|c|c|c|c|c||l|} \hline
\rule[-3mm]{0mm}{8mm}  \teff & $\log$ g & M & $\xi_t$
&  [Fe/H] & $\varepsilon$(C) & $\varepsilon$(N) & $\varepsilon$(O)
& $^{12}$C/$^{13}$C & $\theta_d$ & L & R & Ref.\\ \hline
\rule[-0mm]{0mm}{5mm}3780 & 1.5 & & $1.3 \pm 0.5$ & & &  & & $13 \pm 2$ & & 263 & & 1. \\
3820 & & & & & &  & & & $10.2 \pm 0.43$ & & & 2. \\
3950 & 1.6 & & 2.2 & 0.33 | 0.69 & &  & & & & & & 3$^a$. \\
3950 & 0.4 & & 2.2 & 0.12 & & & & & & & & 3$^b$.\\
(3780) & (1.8) & & & & &  & & (13) & & & & 4. \\
$4280 \pm 100$ & $1.55 \pm 0.46$ & & $2.00 \pm 0.25$  & $-0.23 \pm 0.18$  &
$8.29 \pm 0.15$ & $8.06 \pm 0.18$ & $8.81 \pm 0.18$ & 13 & & 316 & & 5. \\
$4300 \pm 230$ & & & & & &  & & & ${\mathit{8.7 \pm 0.8}}$ & & & 6. \\
(3980) & (1.87) & 5 & (2.0) & & 8.42 &  & & ($13$) & & & & 7. \\
$3940 \pm 100$ & & & & & &  & & & $10.2 \pm 0.5$ & $900 \pm 650$ & $64 \pm 23$ &
8. \\
$3981 \pm 62$ & & & & & &  & & & ${\mathit{10.13 \pm 0.24}}$ & & & 9. \\
 & & 2.0 [0.9] & & & &  & & & & & & 10.\\
3955 & (1.55) & & & ($-0.23$) & &  & & & 10.45 & & & 11. \\
3940 & 1.3 & & (1.7) & 0.06 & &  & & & & & & 12. \\
 & & & & & &  & & & ${\mathit{10.2 \pm 0.2}}$ & & & 13. \\
3986 & & & & & &  & & & (9.997) & & & 14. \\
$3930 \pm 200$ & $1.55 \pm 0.30$ & & $2.2 \pm 0.5$ &
$-0.14 \pm 0.16$ & & & & & & & & 15. \\
3960 & $1.20 \pm 0.41$ & 3.3 & $2.0 \pm 0.5$ & $0.00 \pm 0.30$ & &  & & & & & &
16$^a$. \\
3950 & 1.15 [1.20] & & 1.5 [2.0] & 0.00 & &  & 8.86 [8.85] & & & & & 16$^b$. \\
3850 & 0.81 [0.85] & & 1.5 [2.0] & 0.00 & & & 8.70 [8.69] & & & & & 16$^c$. \\
3950 & 0.82 [0.85] & & 1.5 [2.0] & $-0.5$ & &  & 8.51 [8.50] & & & & &
16$^d$. \\
$3934 \pm 200$ & $1.6 \pm 0.2$ & & & $0.00 \pm 0.25$ & &  & & & & & & 17. \\
(3980) & (1.87) & ($3.0$) & & & &  & & ($13$) & & & &  18. \\
3986 & 2.0 & & & & &  & & & $10.17 \pm 0.10$ & & & 19. \\
3941 & & & & & &  & & & {\it{10.17}} & & & 20$^a$. \\
$3964 \pm 55$ & & & & & &  & & & ${\mathit{10.16 \pm 0.20}}$ & & & 20$^b$. \\
$4095 \pm 163$ & & & & & &  & & & ${\mathit{\theta_{\mathrm{UD}}
 = 9.6 \pm 0.3}}$ & & & 21. \\
& & & & & &  & & & ${\mathit{10.16 \pm 0.23}}$ & & & 22. \\
($3985$) & ($1.50$) & $3.0 \pm 0.5$ & ($2.0$) & ($-0.14$) & &  & & ($13$) & (${\mathit{10.17}}$) & ($535$) & (49) &
23. \\
 & & & & $-0.178 \pm 0.046$ & &  & & & & & & 24. \\
\rule[-3mm]{0mm}{3mm}$3960 \pm 140$ & $1.30 \pm 0.25$ & $1.72 \pm 1.02$ & $2.0
\pm 0.5$ & $0.00 \pm 0.20$ & $8.15 \pm 0.25$ & $8.26 \pm 0.25$ & $8.93 \pm 0.20$
& $10 \pm 2$ & $9.98 \pm 0.63$ & $523 \pm 101$ & $48.53 \pm 3.23$ & 25. \\
\hline
\end{tabular}
\end{center}
{\footnotesize{
1.\ \citet{Tomkin1975ApJ...199..436T};
2.\ \citet{Blackwell1977MNRAS.180..177B};
3.\ \citet{Oinas1977A&A....61...17O};
4.\ \citet{Lambert1980ApJ...235..114L};
5.\ \citet{Lambert1981ApJ...248..228L};
6.\ \citet{Faucherre1983A&A...120..263F};
7.\ \citet{Harris1984ApJ...285..674H};
8.\ \citet{Leggett1986A&A...159..217L};
9.\ \citet{DiBenedetto1987A&A...188..114D};
10.\ \citet{Harris1988ApJ...325..768H};
11.\ \citet{Bell1989MNRAS.236..653B};
12.\ \citet{Brown1989ApJS...71..293B};
13.\ \citet{Hutter1989ApJ...340.1103H};
14.\ \citet{Volk1989AJ.....98.1918V};
15.\ \citet{McWilliam1990ApJS...74.1075M};
16.\ \citet{Bonnell1993MNRAS.264..319B};
17.\ \citet{Morossi1993A&A...277..173M};
18.\ \citet{ElEid1994A&A...285..915E};
19.\ \citet{Cohen1996AJ....112.2274C};
20.\ \citet{DiBenedetto1998A&A...339..858D};
21.\ \citet{Dyck1998AJ....116..981D};
22.\ \citet{Perrin1998A&A...331..619P};
23.\ \citet{Robinson1998ApJ...503..396R};
24.\ \citet{Taylor1999A&AS..134..523T};
25.\ present results}}
\end{table*}

\clearpage

\begin{table*}[htb]
\boldmath
\caption{\label{litband}See caption of Table \ref{litddra}, but now
for {\bf{$\mathbf{\beta}$ And}}.}
\unboldmath
\vspace{1ex}
\begin{center}
\tiny
\setlength{\tabcolsep}{.6mm}
\begin{tabular}{|c|c|c|c|c|c|c|c|c|c|c|c||l|} \hline
\rule[-3mm]{0mm}{8mm}  \teff & $\log$ g & M & $\xi_t$
&  [Fe/H] & $\varepsilon$(C) & $\varepsilon$(N) & $\varepsilon$(O)
& $^{12}$C/$^{13}$C & $\theta_d$ & L & R & Ref.\\ \hline
\rule[-0mm]{0mm}{5mm} 3660 & & & $1.6 \pm 0.5$ & & &  & & $11 \pm 1.5$
& & 347 & & 1. \\
3640 & & & & & &  & & & & & & 2. \\
3750 & $1.0 \pm 0.7$ & (1) & & & &  & & & & & & 3. \\
$3520 \pm 150$ & & & & & &  & & & $14.6 \pm 1.3$ & & & 4. \\
$3850 \pm 400$ & & & & & &  & & & $13.5 \pm 0.7$ & & & 5. \\
3675 & & & & & &  & & & & & & 6. \\
$3820 \pm 280$ & & & & & &  & & & ${\mathit{13.2 \pm 1.7}}$ & & $33 \pm 8$ &
7. \\
(3726) & (1.84) & (2) & (1.6) & & 8.37 &  & & ($11$) & & & & 8. \\
 & & & & & &  &  & & ${\mathit{15.45 \pm 1.00}}$ & & & 9. \\
$3800 \pm 100$ & $1.6 \pm 0.3$ & 2.5 & 2.1 & ($-0.10$) & $8.53 \pm 0.08$ & $8.37
\pm 0.08$ & $8.84 \pm 0.06$ & 12 & & & & 10. \\
 & & & & & &  & & & ${\mathit{14.35 \pm 0.19}}$ & & & 11. \\
 & & & & & &  & & & ${\mathit{13.9 \pm 0.2}}$ & & & 12. \\
$3710 \pm 64$ & & & & & &  & & & & ($202$) & $34.4 \pm 4.0$ & 13. \\
($3800$) & ($1.6$) & (2.5) & (2.1) & ($-0.09$) & $8.48 \pm 0.12$ & $8.32 \pm
0.09$ & $8.82 \pm 0.06$ & $11 \pm 3$ & & & & 14. \\
$3839 \pm 10$ & & & & & &  & & & 13.219 & & & 15. \\
$3800 \pm 110$ & $1.6 \pm 0.3$ & 1.8 & & & &  & & & 14.0 & 200 & 34 & 16. \\
3895 & $1.63 \pm 0.30$ & (1.5) & (2.5) & & $8.42 \pm 0.30$ & & & $10 \pm 3$ & & & &
17. \\
 & & & & & &  & & & ${\mathit{13.806 \pm 0.13}}$ & & & 18. \\
 & & & & & &  & & & ${\mathit{\theta_{\mathrm{UD}} = 12.73 \pm 0.18}}$ & & &
19$^a$. \\
 & & & & & &  & & & ${\mathit{\theta_{\mathrm{UD}} = 12.63 \pm 0.22}}$ & & &
19$^b$. \\
$4002 \pm 178$ & & & & & &  & & & ${\mathit{\theta_{\mathrm{UD}} = 12.2 \pm 0.6}}$ & & & 20. \\
\rule[-3mm]{0mm}{3mm}$3880 \pm 140$ & $0.95 \pm 0.25$ & $2.49 \pm 1.48$ & $2.0
\pm 0.5$ & $0.00 \pm 0.20$ & $8.12 \pm 0.30$ & $8.37 \pm 0.40$ & $9.08 \pm 0.30$
& $9 \pm 2$ & $13.30 \pm 0.84$ & $1561 \pm 333$ & $87.37 \pm 6.85 $ & 21. \\
\hline
\end{tabular}
\end{center}
{\footnotesize{
1.\ \citet{Tomkin1976ApJ...210..694T};
2.\ \citet{Linsky1978ApJ...220..619L};
3.\ \citet{Clegg1979ApJ...234..188C};
4.\ \citet{Scargle1979ApJ...228..838S};
5.\ \citet{Manduca1981ApJ...243..883M};
6.\ \citet{Tsuji1981A&A....99...48T};
7.\ \citet{Faucherre1983A&A...120..263F};
8.\ \citet{Harris1984ApJ...285..674H};
9.\ \citet{Koechlin1985A&A...153...91K};
10.\ \citet{Smith1985ApJ...294..326S};
11.\ \citet{DiBenedetto1987A&A...188..114D};
12.\ \citet{Hutter1989ApJ...340.1103H};
13.\ \citet{Volk1989AJ.....98.1918V};
14.\ \citet{Smith1990ApJS...72..387S};
15.\ \citet{Blackwell1991A&A...245..567B};
16.\ \citet{Judge1991ApJ...371..357J};
17.\ \citet{Lazaro1991MNRAS.249...62L};
18.\ \citet{Mozurkewich1991AJ....101.2207M};
19.\ \citet{Quirrenbach1993ApJ...406..215Q};
20.\ \citet{Dyck1998AJ....116..981D};
21.\ present results}}
\end{table*}


\begin{table*}[htb]
\boldmath
\caption{\label{litacet}See caption of Table \ref{litddra}, but now
for {\bf{$\mathbf{\alpha}$ Cet}}.}
\unboldmath
\vspace{1ex}
\begin{center}
\tiny
\setlength{\tabcolsep}{.6mm}
\begin{tabular}{|c|c|c|c|c|c|c|c|c|c|c|c||l|} \hline
\rule[-3mm]{0mm}{8mm}  \teff & $\log$ g & M & $\xi_t$
&  [Fe/H] & $\varepsilon$(C) & $\varepsilon$(N) & $\varepsilon$(O)
& $^{12}$C/$^{13}$C & $\theta_d$ & L & R & Ref.\\ \hline
\rule[-0mm]{0mm}{5mm} 3750 & $1.0 \pm 0.7$ & (1) & & & &  & & & & & & 1. \\
$3560 \pm 150$ & & & & & &  & & & $14.1 \pm 1.3$ & & & 2. \\
$3905 \pm 150$ & & & & & &  & & & & & & 3. \\
3550 & 1.5 & & & $+0.71$ & &  & & & & & & 4$^a$. \\
3680 & 1.8 & & & $+0.26$ & &  & & & & & & 4$^b$. \\
3660 & 0.5 & & & & &  & & & & & & 4$^c$. \\
($3905$) & ($1.5$) & (3) & $3.6 \pm 0.3$ & & $8.06 \pm 0.30$ &  & & &
& & & 5. \\
$3745$ & & & & & &  & & & $12.643 \pm 0.25$ & & & 6. \\
$3767 \pm 100$ & $0.4 \pm 0.3$ & 1.5 & (2.5) & & &  & & & 13.6 & 2000 & 120 & 7. \\
3730 & $1.3 \pm 0.3$ & (1.5) & & & $7.92 \pm 0.30$ &  & & $10 \pm 3$ & & & &
8. \\
 & & & & & &  & & & ${\mathit{13.23 \pm 0.24}}$ & & & 9. \\
($3905$) & ($1.5$) & (3) & $3.3 \pm 1.0$ & & $8.57 \pm 0.04$ &  & & &
& & & 10$^a$. \\
($3905$) & ($1.5$) & (3) & $3.3 \pm 1.0$ & & $8.50 \pm 0.01$ &  & & &
& & & 10$^b$. \\
 & & & & & &  & & & ${\mathit{\theta_{\mathrm{UD}} = 11.95 \pm 0.23}}$ & & &
11$^a$. \\
 & & & & & &  & & & ${\mathit{\theta_{\mathrm{UD}} = 11.66 \pm 0.22}}$ & & &
11$^b$. \\
$3745 \pm 40$ & 1.3 & & & & & & & & $12.77 \pm 0.25$ & & &
12. \\
(3905) & (1.5) & & 3.0 & & (8.57) &  $7.84 \pm 0.05$ & & & & & & 13. \\
$3869 \pm 161$ & & & & & &  & & & ${\mathit{\theta_{\mathrm{UD}} = 11.6 \pm
0.4}}$ & & & 14. \\
 & & & & & &  & & & ${\mathit{\theta_{\mathrm{UD}} = 12.08 \pm 0.60}}$ & & &
15. \\
\rule[-3mm]{0mm}{3mm}$3740 \pm 140$ & $0.95 \pm 0.25$ & $2.69 \pm 1.61$ & $2.3
\pm 0.5$ & $0.00 \pm 0.20$ & $8.20 \pm 0.30$ & $8.26 \pm 0.40$ & $8.93 \pm 0.30$
& $10 \pm 2$ & $12.52 \pm 0.79$ & $1455 \pm 328$ & $90.79 \pm 7.66$ & 16. \\
\hline
\end{tabular}
\end{center}
{\footnotesize{
1.\ \citet{Clegg1979ApJ...234..188C};
2.\ \citet{Scargle1979ApJ...228..838S};
3.\ \citet{Tsuji1981A&A....99...48T};
4.\ \citet{Burnashev1983IzKry..67...13B};
5.\ \citet{Tsuji1986A&A...156....8T};
6.\ \citet{Blackwell1991A&A...245..567B};
7.\ \citet{Judge1991ApJ...371..357J};
8.\ \citet{Lazaro1991MNRAS.249...62L};
9.\ \citet{Mozurkewich1991AJ....101.2207M};
10.\ \citet{Tsuji1991A&A...245..203T};
11.\ \citet{Quirrenbach1993ApJ...406..215Q};
12.\ \citet{Cohen1996AJ....112.2274C};
13.\ \citet{Aoki1997A&A...328..175A};
14.\ \citet{Dyck1998AJ....116..981D};
15.\ \citet{Perrin1998A&A...331..619P};
16.\ present results}}
\end{table*}

\clearpage

\begin{table*}[htb]
\boldmath
\caption{\label{litbpeg}See caption of Table \ref{litddra}, but now
for {\bf{$\mathbf{\beta}$ Peg}}.}
\unboldmath
\vspace{1ex}
\begin{center}
\tiny
\setlength{\tabcolsep}{.4mm}
\begin{tabular}{|c|c|c|c|c|c|c|c|c|c|c|c||l|} \hline
\rule[-3mm]{0mm}{8mm}  \teff & $\log$ g & M & $\xi_t$
&  [Fe/H] & $\varepsilon$(C) & $\varepsilon$(N) & $\varepsilon$(O)
& $^{12}$C/$^{13}$C & $\theta_d$ & L & R & Ref.\\ \hline
\rule[-0mm]{0mm}{5mm} $3650 \pm 250$ & & & & & &  & & & 16 & & & 1. \\
3467 & & & & & &  & & & & & 100 & 2. \\
3600 & & & & & &  & & & $16.3 \pm 1.2$ & & & 3. \\
3280 & & & & & &  & & & & & & 4. \\
3500 & $0.0 \pm 0.7$ & (1) & & & &  & & & & & & 5. \\
$3530 \pm 150$ & & & & & &  & & & $17.8 \pm 1.6$ & & & 6. \\
3600 & & & & & &  & & & $17.00 \pm 0.85$ & & & 7. \\
$3580 \pm 150$ & & & & & &  & & & & & & 8. \\
$3500 \pm 200$ & $2.1 \pm 0.3$ & & $3.0 \pm 0.3$ & & &  & & & & & & 9. \\
(3568) & (1.66) & & (2.0) & & 8.67 &  & & $7 \pm 1$ & & & & 10. \\
$3600 \pm 100$ & $1.2 \pm 0.3$ & & $2.0 \pm 0.3$ & & $8.45 \pm 0.06$ & $8.26 \pm
0.06$ & $8.82 \pm 0.06$ & 8 & & & & 11. \\
(3580) & $1.0 \pm 0.5$ & (3) & $3.12 \pm 0.15$ & & $7.89 \pm 0.12$ & & & & &
& & 12. \\
$3590 \pm 44$  & & & & & &  & & & ${\mathit{16.75 \pm 0.24}}$ & & & 13. \\
(3600) & (1.2) & & 2.3 & & &  & & & & & & 14. \\
 & & 1.7 & & & &  & & & & & & 15. \\
 & & & & & &  & & & ${\mathit{18.4 \pm 0.6}}$ & & & 16. \\
$3547 \pm 35$ & & & & & &  & & & $17.309 \pm 0.519$ & & & 17. \\
(3600) & (1.2) & & (2.0) & & $8.42 \pm 0.07$ & $8.18 \pm 0.08$ & $8.80 \pm 0.06$ & $8 \pm 2$ & & & & 18. \\
3609 & & & & & &  & & & 16.727 & & & 19. \\
$3600 \pm 110$ & $0.6 \pm 0.3$ & 1.7 & & & &  & & & 17.3 & 1700 & 110 & 20. \\
3730 & $1.3 \pm 0.3$ & 1.5 & 2.5 & & $7.92 \pm 0.30$ & & & $10 \pm 3$ & & & &
21. \\
 & & & & & &  & & & ${\mathit{17.98 \pm 0.18}}$ & & & 22. \\
($3580$) & ($1.0$) & 3 & $4.1 \pm 2.8$ & & $8.24 \pm 0.08$
&  & & & & & & 23$^a$. \\
($3580$) & ($1.0$) & 3 & $4.1 \pm 2.8$ & & $8.18 \pm 0.02$
&  & & & & & & 23$^b$. \\
 & & & & & &  & & & ${\mathit{\theta_{\mathrm{UD}} = 17.55 \pm 0.14}}$ & & & 24$^a$. \\
 & & & & & &  & & & ${\mathit{\theta_{\mathrm{UD}} = 16.11 \pm 0.11}}$ & & & 24$^b$. \\
(3580) & (1.2) & & & ($-0.12$) & &  & & & & & & 25. \\
3600 & 1.4 & & & & &  & & & & & & 26. \\
($3580$) & ($1.0$) & 3 & $3.0 \pm 1.0$ & & ($8.24$) & $8.16 \pm 0.04$ & & & & & & 27. \\
3600 & 0.5 & & 3.0 & & &  & & & & & & 28. \\
(3600) & (1.0) & & (1.6) & ($-0.1$) & (8.45) & (8.26) & (8.82) & & & & & 29. \\
$3890 \pm 74$ & & & & & &  & & & ${\mathit{\theta_{\mathrm{UD}} = 14.3 \pm
0.7}}$ & & & 30. \\
 & & & & & &  & & & ${\mathit{16.76 \pm 0.23}}$ & & & 31. \\
\rule[-3mm]{0mm}{3mm}$3600 \pm 300$ & $0.65 \pm 0.40$ & $1.94^{+4.27}_{-1.34}$ & ($2.0$) &
($0.00$) & $8.20 \pm 0.40$ & ($8.18$) & ($8.93$) & $5 \pm 3$ & $16.60
\pm 1.32$ & $1800 \pm 683$ & $108.96 \pm 9.90$ & 32. \\
\hline
\end{tabular}
\end{center}
{\footnotesize{
1.\ \citet{Dyck1974ApJ...189...89D};
2.\ \citet{Sanner1976ApJS...32..115S};
3.\ \citet{Blackwell1977MNRAS.180..177B};
4.\ \citet{Linsky1978ApJ...220..619L};
5.\ \citet{Clegg1979ApJ...234..188C};
6.\ \citet{Scargle1979ApJ...228..838S};
7.\ \citet{Manduca1981ApJ...243..883M};
8.\ \citet{Tsuji1981A&A....99...48T};
9.\ \citet{Burnashev1983IzKry..67...13B};
10.\ \citet{Harris1984ApJ...285..674H};
11.\ \citet{Smith1985ApJ...294..326S};
12.\ \citet{Tsuji1986A&A...156....8T};
13.\ \citet{DiBenedetto1987A&A...188..114D};
14.\ \citet{Lambert1987Ap&SS.133..369L};
15.\ \citet{Harris1988ApJ...325..768H};
16.\ \citet{Hutter1989ApJ...340.1103H};
17.\ \citet{Blackwell1990A&A...232..396B};
18.\ \citet{Smith1990ApJS...72..387S};
19.\ \citet{Blackwell1991A&A...245..567B};
20.\ \citet{Judge1991ApJ...371..357J};
21.\ \citet{Lazaro1991MNRAS.249...62L};
22.\ \citet{Mozurkewich1991AJ....101.2207M};
23.\ \citet{Tsuji1991A&A...245..203T};
24.\ \citet{Quirrenbach1993ApJ...406..215Q};
25.\ \citet{Worthey1994ApJS...95..107W};
26.\ \citet{Cohen1996AJ....112.2274C};
27.\ \citet{Aoki1997A&A...328..175A};
28.\ \citet{Tsuji1997A&A...320L...1T};
29.\ \citet{Abia1998MNRAS.293...89A};
30.\ \citet{Dyck1998AJ....116..981D};
31. \citet{Perrin1998A&A...331..619P};
32.\ present results}}
\end{table*}

\clearpage
\subsection{Comments on published stellar parameters}\label{literatureapp}

In this section of the appendix, a description of the results obtained by
different authors using various methods is given in chronological order.
One either can look to the quoted reference in the accompanying paper
and than search the description in the chronological (and then
alphabetical) listing below
or one can use the cross-reference table (Table \ref{cross}) to find
all the references for one specific star in this numbered listing.

\begin{table*}[htb]
\caption{\label{cross}Cross-reference table in which one can find all
the numbers refering to the papers citing a particular star.}
\begin{center}
\begin{tabular}{l|l} \hline
\rule[-3mm]{0mm}{8mm} name & reference number \\ \hline
\rule[-0mm]{0mm}{5mm}$\delta$ Dra &
\ref{vanParadijs1973AA....23..369V},
\ref{Gustafsson1974AA....34...99G},
\ref{vanParadijs1976AA....53....1V},
\ref{Linsky1978ApJ...220..619L},
\ref{Bonnell1982MNRAS.201..253B},
\ref{Gratton1982MNRAS.201..807G},
\ref{Burnashev1983IzKry..67...13B},
\ref{Faucherre1983AA...120..263F},
\ref{Gratton1985AA...148..105G},
\ref{Brown1989ApJS...71..293B},
\ref{McWilliam1990ApJS...74.1075M},
\ref{Blackwell1998AAS..129..505B},
\ref{Taylor1999AAS..134..523T} \\

$\xi$ Dra &
\ref{Gustafsson1974AA....34...99G},
\ref{Dearborn1975ApJ...200..675D},
\ref{Gratton1982MNRAS.201..807G},
\ref{Gratton1983MNRAS.202..231G},
\ref{Brown1989ApJS...71..293B},
\ref{Blackwell1990AA...232..396B},
\ref{McWilliam1990ApJS...74.1075M},
\ref{Blackwell1991AA...245..567B},
\ref{Blackwell1998AAS..129..505B},
\ref{Taylor1999AAS..134..523T}\\

$\alpha$ Boo &
\ref{vanParadijs1974AA....35..225V},
\ref{Mackle1975AAS...19..303M},
\ref{Blackwell1977MNRAS.180..177B},
\ref{Linsky1978ApJ...220..619L},
\ref{Scargle1979ApJ...228..838S},
\ref{Blackwell1980AA....82..249B},
\ref{Lambert1980ApJ...235..114L},
\ref{Lambert1981ApJ...248..228L},
\ref{Manduca1981ApJ...243..883M},
\ref{Tsuji1981AA....99...48T},
\ref{Frisk1982MNRAS.199..471F},
\ref{Kjaergaard1982AA...115..145K},
\ref{Burnashev1983IzKry..67...13B},
\ref{Harris1984ApJ...285..674H},
\ref{Bell1985MNRAS.212..497B},
\ref{Gratton1985AA...148..105G},
\ref{Moon1985ApSS.117..261M},
\ref{Judge1986MNRAS.221..119J},
\ref{Kyrolainen1986AAS...65...11K},
\ref{Tsuji1986AA...156....8T},
\ref{Altas1987ApSS.134...85A},
\ref{DiBenedetto1987AA...188..114D},
\ref{Edvardsson1988AA...190..148E},
\ref{Harris1988ApJ...325..768H},
\ref{Bell1989MNRAS.236..653B},
\ref{Brown1989ApJS...71..293B},
\ref{Volk1989AJ.....98.1918V},
\ref{Fernandez-Villacanas1990AJ.....99.1961F},
\ref{McWilliam1990ApJS...74.1075M},
\ref{Blackwell1991AA...245..567B},\\
 &
\ref{Judge1991ApJ...371..357J},
\ref{Tsuji1991AA...245..203T},
\ref{Engelke1992AJ....104.1248E},
\ref{Bonnell1993MNRAS.264..319B},
\ref{Peterson1993ApJ...404..333P},
\ref{Gadun1994AN....315..413G},
\ref{Cohen1996AJ....112.2274C},
\ref{Quirrenbach1996AA...312..160Q},
\ref{Aoki1997AA...328..175A},
\ref{Pilachowski1997AJ....114..819P},
\ref{DiBenedetto1998AA...339..858D},
\ref{Dyck1998AJ....116..981D},
\ref{Hammersley1998AAS..128..207H},
\ref{Perrin1998AA...331..619P},
\ref{Taylor1999AAS..134..523T} \\

$\alpha$ Tuc &
\ref{Basri1979ApJ...234.1023B},
\ref{Stencel1980ApJS...44..383S},
\ref{Glebocki1988AA...189..199G},
\ref{Flynn1991AA...250..400F},
\ref{Pasquini1992AA...266..340P} \\

$\beta$ UMi &
\ref{Tomkin1976ApJ...210..694T},
\ref{Linsky1978ApJ...220..619L},
\ref{Lambert1980ApJ...235..114L},
\ref{Stencel1980ApJS...44..383S},
\ref{Lambert1981ApJ...248..228L},
\ref{Burnashev1983IzKry..67...13B},
\ref{Faucherre1983AA...120..263F},
\ref{Harris1988ApJ...325..768H},
\ref{Bell1989MNRAS.236..653B},
\ref{Brown1989ApJS...71..293B},
\ref{McWilliam1990ApJS...74.1075M},
\ref{Judge1991ApJ...371..357J},
\ref{Cornide1992AJ....103.1374C},
\ref{Morossi1993AA...277..173M},
\ref{Dyck1998AJ....116..981D},
\ref{Taylor1999AAS..134..523T} \\

$\gamma$ Dra &
\ref{Tomkin1975ApJ...199..436T},
\ref{Blackwell1977MNRAS.180..177B},
\ref{Oinas1977AA....61...17O},
\ref{Lambert1980ApJ...235..114L},
\ref{Lambert1981ApJ...248..228L},
\ref{Faucherre1983AA...120..263F},
\ref{Harris1984ApJ...285..674H},
\ref{Leggett1986AA...159..217L},
\ref{DiBenedetto1987AA...188..114D},
\ref{Harris1988ApJ...325..768H},
\ref{Bell1989MNRAS.236..653B},
\ref{Brown1989ApJS...71..293B},
\ref{Hutter1989ApJ...340.1103H},
\ref{Volk1989AJ.....98.1918V},
\ref{McWilliam1990ApJS...74.1075M},
\ref{Bonnell1993MNRAS.264..319B},
\ref{Morossi1993AA...277..173M},
\ref{ElEid1994AA...285..915E},
\ref{Cohen1996AJ....112.2274C},
\ref{DiBenedetto1998AA...339..858D},
\ref{Dyck1998AJ....116..981D},
\ref{Perrin1998AA...331..619P},
\ref{Robinson1998ApJ...503..396R},
\ref{Taylor1999AAS..134..523T} \\

$\beta$ And &
\ref{Tomkin1976ApJ...210..694T},
\ref{Linsky1978ApJ...220..619L},
\ref{Clegg1979ApJ...234..188C},
\ref{Scargle1979ApJ...228..838S},
\ref{Manduca1981ApJ...243..883M},
\ref{Tsuji1981AA....99...48T},
\ref{Faucherre1983AA...120..263F},
\ref{Harris1984ApJ...285..674H},
\ref{Koechlin1985AA...153...91K},
\ref{Smith1985ApJ...294..326S},
\ref{DiBenedetto1987AA...188..114D},
\ref{Hutter1989ApJ...340.1103H},
\ref{Volk1989AJ.....98.1918V},
\ref{Smith1990ApJS...72..387S},
\ref{Blackwell1991AA...245..567B},
\ref{Judge1991ApJ...371..357J},
\ref{Lazaro1991MNRAS.249...62L},
\ref{Mozurkewich1991AJ....101.2207M},
\ref{Quirrenbach1993ApJ...406..215Q},
\ref{Dyck1998AJ....116..981D} \\

$\alpha$ Cet &
\ref{Clegg1979ApJ...234..188C},
\ref{Scargle1979ApJ...228..838S},
\ref{Tsuji1981AA....99...48T},
\ref{Burnashev1983IzKry..67...13B},
\ref{Tsuji1986AA...156....8T},
\ref{Blackwell1991AA...245..567B},
\ref{Judge1991ApJ...371..357J},
\ref{Lazaro1991MNRAS.249...62L},
\ref{Mozurkewich1991AJ....101.2207M},
\ref{Tsuji1991AA...245..203T},
\ref{Quirrenbach1993ApJ...406..215Q},
\ref{Cohen1996AJ....112.2274C},
\ref{Aoki1997AA...328..175A},
\ref{Dyck1998AJ....116..981D},
\ref{Perrin1998AA...331..619P} \\

\rule[-3mm]{0mm}{3mm}$\beta$ Peg &
\ref{Dyck1974ApJ...189...89D},
\ref{Sanner1976ApJS...32..115S},
\ref{Blackwell1977MNRAS.180..177B},
\ref{Linsky1978ApJ...220..619L},
\ref{Clegg1979ApJ...234..188C},
\ref{Scargle1979ApJ...228..838S},
\ref{Manduca1981ApJ...243..883M},
\ref{Tsuji1981AA....99...48T},
\ref{Burnashev1983IzKry..67...13B},
\ref{Harris1984ApJ...285..674H},
\ref{Smith1985ApJ...294..326S},
\ref{Tsuji1986AA...156....8T},
\ref{DiBenedetto1987AA...188..114D},
\ref{Lambert1987ApSS.133..369L},
\ref{Harris1988ApJ...325..768H},
\ref{Hutter1989ApJ...340.1103H},
\ref{Blackwell1990AA...232..396B},
\ref{Smith1990ApJS...72..387S},
\ref{Blackwell1991AA...245..567B},
\ref{Judge1991ApJ...371..357J},
\ref{Lazaro1991MNRAS.249...62L},
\ref{Mozurkewich1991AJ....101.2207M},
\ref{Tsuji1991AA...245..203T},
\ref{Quirrenbach1993ApJ...406..215Q},
\ref{Worthey1994ApJS...95..107W},
\ref{Cohen1996AJ....112.2274C},
\ref{Aoki1997AA...328..175A},
\ref{Tsuji1997AA...320L...1T},
\ref{Abia1998MNRAS.293...89A},
\ref{Dyck1998AJ....116..981D},
\ref{Perrin1998AA...331..619P} \\

\hline
\end{tabular}
\end{center}
\end{table*}

\begin{enumerate}

\item{\label{vanParadijs1973AA....23..369V}
\citet{vanParadijs1973A&A....23..369V} has based his analysis on
observations of line strengths on spectrograms of 1.6 and
6.5\,\ang/mm, in the wavelength region between 5000\,\ang\ and
6650\,\ang, and on calculations of weak-line strengths for a grid
of model atmospheres. From the condition of minimum scatter in the
curve of growth, effective temperatures are derived. The gravity
of the stars has been obtained from the requirement that two
ionisation states of one element should give the same abundance.
With the effective temperature and gravity found, curves of growth
have been made for different chemical elements.}

\item{\label{Dyck1974ApJ...189...89D}
\citet{Dyck1974ApJ...189...89D} have derived complete energy
distributions, photometric spectral types and total fluxes from
narrow  and broad-band photometry between 0.55 and 10.2\,\mic. For
stars lacking infrared excesses, an intrinsic relation between the
colour temperature and the spectral type and between the colour
temperature and the effective temperature is found. The error in
the effective temperature may be as large as $\pm 7$\,\%. By
comparing their effective temperature scale with the one of
\citet{Johnson1966ARA&A...4..193J}, they find a good agreement in the
mean relation,
but the individual values of \teff\ do not always agree. Given the
total flux and effective temperature, the angular diameter can be
deduced.}

\item{\label{Gustafsson1974AA....34...99G}
\citet{Gustafsson1974A&A....34...99G} have measured the strength of a
narrow group of weak metal lines. For more than half of the stars
in their sample, the strength of another group with lines on the
flat part of the curve of growth was also measured. The
narrow-band indices, $A(48)$ and $A(58)$, were analyzed as a
function of the fundamental atmospheric parameters by means of
scaled solar model atmospheres. The temperature scale was
established using a (T,\ $R-I$) calibration. The accelerations of
gravity were estimated from absolute magnitudes and evolutionary
tracks. The iron abundances and the microturbulence were
calculated from an iterative process in which the observational
quantities $R-I$, $M_V$, $A(48)$, $A(58)$ and the fundamental
parameters \teff, $\log$ g, [Fe/H] and \vt\ are involved. From the
errors in $R-I$, they estimate $\Delta$ \teff\ to be 150\,K. The
accuracy of the other stellar parameters are $\Delta$[Fe/H] =
0.12, $\Delta$\vt\ = 0.2\,\kms, $\Delta\log$ g = 0.3.}

\item{\label{vanParadijs1974AA....35..225V}
\citet{vanParadijs1974A&A....35..225V} have adopted the angular
diameter from \citet{Currie1974ApJ...187..131C} and
\citet{Gezari1972ApJ...173L...1G} for $\alpha$ Tau and $\alpha$
Boo respectively.  To determine the effective temperature they
used several continuum data, the curve of growth with the line
strengths of Fe~I lines and the surface brightness. Requiring that
the neutral and ionised lines of Fe, Cr, V, Ti and Sc gave the
same abundance, yielded the gravity. Gravity, parallax and angular
diameter resulted in the mass, while the luminosity was determined
from the effective temperature, the parallax and the angular
diameter. \citet{vanParadijs1974A&A....35..225V} quoted that
Wilson (1972) found [Fe/H] = $-0.69$ for $\alpha$ Tau.}

\item{\label{Dearborn1975ApJ...200..675D}
Wavelength intervals around 8000\,\ang\ containing $^{12}$CN 2-0
and $^{13}$CN 2-0 lines, and around 6300\,\ang\ containing weak
$^{12}$CN 4-0 lines were observed by
\citet{Dearborn1975ApJ...200..675D} with
the McDonald 272\,cm telescope and the Tull coud\'{e} scanner. The
effective temperature and bolometric corrections were calculated
from tables by \citet{Johnson1966ARA&A...4..193J} and $UBVRI$
colours. The $^{12}$CN
and $^{13}$CN curves of growth were then used to determine \cc.}

\item{\label{Mackle1975AAS...19..303M}
\citet{Mackle1975A&AS...19..303M} have analysed the Arcturus
spectrum, using the observational material based on the Arcturus
atlas of \citet{Griffin1968}. The effective temperature was
determined from the wings of strong lines (Mgb, NaD, Ca~II IR
triplet) and different photometric observations, the gravity from
requiring that the equivalent widths of the lines of neutral and
ionised atoms yield the same abundance. The microturbulence was
determined from the curve of growth and the mass from the gravity
and the interferometric radius. The luminosity was deduced on the
one hand from \teff\ and R and on the other hand from the observed
parallax, apparent visual magnitude and the bolometric
correction.}

\item{\label{Tomkin1975ApJ...199..436T}
\citet{Tomkin1975ApJ...199..436T} have obtained photoelectric spectral
scans of lines belonging to the red CN system with the McDonald
2.7\,m telescope and the Tull coud\'{e} scanner. Infrared colours
were used to determine \teff. This effective temperature together
with the luminosity and masses - inferred from the locations of the
stars in the H-R diagram - yielded the physical gravity. The
microturbulence was obtained from a curve of growth technique. The
CN lines yielded the isotopic ratio \cc.}

\item{\label{vanParadijs1976AA....53....1V}
The abundances of carbon, nitrogen and oxygen have been
determined in the atmospheres of five G- and K-giants and one
K-subgiant by \citet{vanParadijs1976A&A....53....1V}. Using the method
described in \citet{vanParadijs1973A&A....23..369V} \teff, $\log$ g
and [Fe/H] are determined. The abundance
analysis is based on equivalent widths of the [O~I] line at $\lambda$
6300\,\ang, of rotational lines in the violet bands of CN, of a red
band of CN and of bands of CH.}

\item{\label{Sanner1976ApJS...32..115S}
\citet{Sanner1976ApJS...32..115S} has determined the absolute
visual magnitude, bolometric
correction and effective temperature from a spectral type calibration. The
distance is deduced from the distance modulus and reddening correction. The
radius is then obtained from measured angular diameters found in literature.}

\item{\label{Tomkin1976ApJ...210..694T}
\citet{Tomkin1976ApJ...210..694T} have used infrared colours to determine
\teff. The microturbulence was determined by fitting a theoretical
curve of growth to the $^{12}$CN curve of growth ($^{12}$CN(2,0)
around 8000\,\ang, weak $^{12}$CN(4,0) lines around 6300\,\ang\ and
weak $^{12}$CN(4,2) lines around 8430\,\ang). Together with
$^{13}$CN lines around 8000\,\ang, the isotopic ratio \cc\ was
ascertained. The gravities were estimated from the effective
temperature, the mass and the luminosity (but were not
tabulated). The luminosity has been
computed using different methods, e.g., trigonometric parallax,
narrow-band photometry, K-line luminosity, ...}

\item{\label{Blackwell1977MNRAS.180..177B}
\citet{Blackwell1977MNRAS.180..177B} have described the Infrared
Flux Method (IRFM)
to determine the stellar angular diameters and effective temperatures from
absolute infrared photometry. For 28 stars (including $\alpha$ Car, $\alpha$
Boo, $\alpha$ CMa, $\alpha$ Lyr, $\beta$ Peg, $\alpha$ Cen A, $\alpha$ Tau and
$\gamma$ Dra) the angular diameters are deduced. Only for the first four stars
the corresponding effective temperatures are computed.}

\item{\label{Oinas1977AA....61...17O}
\citet{Oinas1977A&A....61...17O} has obtained observations with
the 60\,inch
telescope on Mt.\ Wilson. Assuming a mass of 2.5\,\Msun\ and taking
the absolute K-line magnitude of \citet{Wilson1976ApJ...205..823W}  in
conjunction with
the bolometric correction of \citet{Johnson1966ARA&A...4..193J}, the gravity was
determined. The effective temperature was obtained from a fit of
the model fluxes to the photoelectric scans, while a curve of
growth technique was used to determine the abundances and the
microturbulence. By using this `physical' gravity a large
discrepancy was found between the iron abundances from the Fe~I
and the Fe~II lines of $\gamma$ Dra. Therefore the surface gravity
was lowered in order  to force equality between the abundances
derived from the neutral and ion lines. This yielded however a very low
gravity-value of $\log$ g = 0.4\,dex for $\gamma$ Dra --- instead of
the physical gravity of $log$ g = 1.6\,dex.}

\item{\label{Linsky1978ApJ...220..619L}
By using a \teff-($V-I$) transformation of
\citet{Johnson1966ARA&A...4..193J},
\citet{Linsky1978ApJ...220..619L} have determined the effective temperature.}

\item{\label{Basri1979ApJ...234.1023B}
\citet{Basri1979ApJ...234.1023B}  have deduced the effective temperature
from the (\teff, $V-R$)-relation from
\citet{Johnson1966ARA&A...4..193J}, while the
angular diameter is computed by using a relation which links the
colour $V-R$ to the apparent angular diameter.}

\item{\label{Clegg1979ApJ...234..188C}
\citet{Clegg1979ApJ...234..188C} have derived both the stellar
effective temperature from a \teff-spectral type relation and a
colour temperature. The surface gravities were derived on the
assumption that all stars have a stellar mass of 1 \Msun\ and a
luminosity estimated from a H-R diagram. In this way, the surface
gravity may be uncertain by a factor 5.}

\item{\label{Scargle1979ApJ...228..838S}
\citet{Scargle1979ApJ...228..838S} have compared the observed
infrared flux curves of cool stars with theoretical predictions in
order to assess the model atmospheres and to derive useful stellar
parameters. This comparison yielded the effective temperature
(determined from flux-curve shape alone) and the angular diameter
(determined from the magnitudes of the fluxes). The overall
uncertainty in \teff\ is probably about 150\,K, which translates
into about a 9\,\% error in the angular diameter.}

\item{\label{Blackwell1980AA....82..249B}
\citet{Blackwell1980A&A....82..249B} have determined the
effective temperature and the angular diameter for 28 stars using the
IRFM method.}

\item{\label{Lambert1980ApJ...235..114L}
\citet{Lambert1980ApJ...235..114L} took model parameters from published
papers which constrained the \cc\ ratio, e.g.\
\citet{Tomkin1976ApJ...210..694T} and
\citet{Lambert1976MSRSL...9..405L} to derive the lithium abundance for
about 50 G and K giants.}

\item{\label{Stencel1980ApJS...44..383S}
\citet{Stencel1980ApJS...44..383S} have taken \teff, $\log$ g and [Fe/H]
from sources referenced by \citet{Linsky1979ApJS...41...47L},
\citet{Johnson1966ARA&A...4..193J} and \citet{Gustafsson1975A&A....42..407G}.
The apparent stellar diameter is
related with the ($V-R$) photometric colour. Uncertainties in \teff\
and the bolometric correction can be significant in the
determination of \ad. Especially the values given for $\alpha$ Tuc
seems to be quite unreliable.}

\item{\label{Lambert1981ApJ...248..228L}
\citet{Lambert1981ApJ...248..228L} have used high-resolution
low-noise spectra. The parameters were ascertained by demanding
that the spectroscopic requirements (ionisation balance,
independence of the abundance of an ion versus the excitation
potential and equivalent width) should be fulfilled.
 The effective temperature
was found from the Fe~I excitation temperature and the model
atmosphere calibration of the excitation temperature as a function
of \teff. As quoted by \citet{Ries1981},
\citet{Harris1988ApJ...325..768H} and \citet{Luck1995AJ....110.2968L}
their \teff\ and $\log$ g are too high and
should be lowered by 240\,K and 0.40\,dex respectively. The isotopic
ratio \cc\ was taken from \citet{Tomkin1975ApJ...199..436T}, while the
luminosity was estimated from the K-line visual magnitude $M_V(K)$ given by
\citet{Wilson1976ApJ...205..823W} and the bolometric correction BC by
\citet{Gustafsson1979A&A....74..313G}. The abundances of carbon,
nitrogen and oxygen were
based on C$_2$, [O~I] and the red system CN lines respectively.
\citet{Luck1995AJ....110.2968L} wondered whether the nitrogen
abundance for $\alpha$ Tau [N/Fe]=$-0.20$ reflects a typographic error
and should rather
be [N/Fe]=$+0.20$, resulting in $\varepsilon$(N)=7.86, which is more in
agreement with being a red giant branch star.}

\item{\label{Manduca1981ApJ...243..883M}
\citet{Manduca1981ApJ...243..883M} have compared absolute flux
measurements in the
2.5 -- 5.5\,\mic\ region with fluxes computed for model stellar atmospheres. The
stellar angular diameters obtained from fitting the fluxes at 3.5\,\mic\ are in
good agreement with observational values and with angular diameters deduced from
the relation between visual surface brightness and ($V-R$) colour. The
temperatures obtained from the shape of the flux curves are in satisfactory
agreement with other temperature estimates. Since the average error is expected
to be well within 10\,\%, the error for the angular diameter is estimated to be
in the order of 5\,\%.}

\item{\label{Tsuji1981AA....99...48T}
\citet{Tsuji1981A&A....99...48T} has used the IRFM method to determine the
effective temperature. A new calibration of effective temperatures
against spectral type is given on the basis of direct analyses of
several objects in each sub-class.}

\item{\label{Bonnell1982MNRAS.201..253B}
\citet{Bonnell1982MNRAS.201..253B} have used synthetic DDO photometric
indices with precise DDO photometric observations of Population I
G and K giants to determine carbon and nitrogen abundances and
temperatures for these stars. The adopted gravity, metallicity and
microturbulence were taken from \citet{Gustafsson1974A&A....34...99G}. The
obtained accuracy is $\Delta$\teff\ = 170\,K,
$\Delta\varepsilon$(C) = 0.18\,dex and $\Delta\varepsilon$(N) =
0.30\,dex. To
convert [C/Fe] and [N/Fe] to $\varepsilon$(C) and
$\varepsilon$(N) we have used the solar
abundances given by \citet{Anders1989GeCoA..53..197A}.}

\item{\label{Frisk1982MNRAS.199..471F}
\citet{Frisk1982MNRAS.199..471F} have used photometry in the
wavelength range from
0.4 -- 2.2\,\mic\ in order to determine the effective temperature of Arcturus. A
comparison with detailed model atmospheres (with $\log$ g = 1.5, mixing-length
parameter $l/H_p =
1.5$ and chemical composition [A/H] = $-0.5$) leads to an effective
temperature of 4375\,K with an estimated maximum uncertainty of about 50\,K.}

\item{\label{Gratton1982MNRAS.201..807G}
\citet{Gratton1982MNRAS.201..807G} have analyzed blue-violet
high-dispersion spectra of 26 K giants by a semi-automatic
procedure for the determination of \teff, \vt, g,
$\varepsilon$(Fe) and $\varepsilon$(Ti). The procedure consists in
a comparison of observed curves of growth with those obtained by
model atmospheres. They mainly have used weak and medium strong
lines. The standard errors of these stellar parameters are
$\Delta$\teff\ = 60\,K, $\Delta$\vt\ = 0.1\,\kms\ and $\Delta \log$ g =
0.15\,dex. Due to the uncertain excitation temperature of $\xi$
Dra, the temperature was estimated from a colour-temperature
relation. The radius was derived from the visual surface
brightness and was used to obtain the mass and the bolometric
absolute magnitude. They found an average mass of the field
K giants being $1.1 \pm 0.2$\,\Msun. The accuracy on the radius is
$\Delta \log$ R = 0.06, on the mass $\Delta \log$ M = 0.19 and on
the iron abundance $\Delta \varepsilon$(Fe) = 0.14.}

\item{\label{Kjaergaard1982AA...115..145K}
\citet{Kjaergaard1982A&A...115..145K} have derived the effective
temperatures for the programme stars from observed ($R-I$) and
($72-110$) values and preliminary guesses for $\log$ g and [Fe/H].
They noted that their effective temperature derived from this
method could be systematically too high by about 50\,K, due to the
lack of an opacity source in the violet-ultraviolet spectral
region of the used models. The gravity was assigned by using an
empirical relation which relates the width of the Ca~II line
emission feature, the effective temperature, the metallicity and
the gravity. The metal abundances were taken from
\citet{Gustafsson1974A&A....34...99G} and a general value of \vt\ =
1.7\,\kms\ was adopted for
all the programme stars. The C, N and O abundances were then
derived from high-resolution scans of the [O~I] 6300\,\ang line and of
the lines of the red CN (2,0) band and very narrow band index
measurements of the C$_2$ Swan (0,1) band head.}

\item{\label{Burnashev1983IzKry..67...13B}
\citet{Burnashev1983IzKry..67...13B} has determined \teff,
$\log$ g and [Fe/H]
from narrow-band photometric colours in the visible part of the
electromagnetic spectrum, obtained from different
spectrophotometric catalogues.}

\item{\label{Faucherre1983AA...120..263F}
\citet{Faucherre1983A&A...120..263F} have used the two-telescope
stellar interferometer
(I2T) at CERGA to measure angular diameters. The eleven colours of
\citet{Johnson1966ARA&A...4..193J} were used for the bolometric flux
to yield the effective temperature.}

\item{\label{Gratton1983MNRAS.202..231G}
\citet{Gratton1983MNRAS.202..231G} has used an analogous method as
in 1982, but now very strong iron lines are used. Therefore, he
has first analysed the solar lines to derive a damping enhancement
in order to match model atmosphere predictions and observational
data. This yielded the quite low solar iron abundance
$\varepsilon$(Fe) = 7.40. Using these damping enhancements, the
stellar parameters are determined. This results in an iron
abundance which is $\sim 0.15$\,dex lower than the value deduced
by \citet{Gratton1982MNRAS.201..807G} with an accuracy of $\Delta
\varepsilon$(Fe) = 0.12. This lower iron abundance may be
explained by observational uncertainties.}

\item{\label{Harris1984ApJ...285..674H}
\citet{Harris1984ApJ...285..674H} have taken \teff, $\log$ g and \vt\
determined by Dominy, Hinkle and Lambert (1984), a reference which
we could not trace back. The isotopic ratio \cc\ was adopted from
\citet{Tomkin1975ApJ...199..436T}. The carbon abundance was found by fitting
weak $^{12}$C$^{16}$O lines at 1.6\,\mic, 2.3\,\mic\ and 5\,\mic.}

\item{\label{Bell1985MNRAS.212..497B}
\citet{Bell1985MNRAS.212..497B} have estimated the surface gravity
of Arcturus using the strengths of MgH features, strong metal
lines and the Fe~I - Fe~II ionisation equilibrium. The MgH lines
give $\log$ g = $1.8 \pm 0.5$, for an effective temperature of
4375\,K \citep{Frisk1982MNRAS.199..471F} and \vt\ = 1.7\,\kms. The
uncertainty in \teff\ leads to one of the most important
uncertainties of the gravity when estimated from MgH lines: a
reduction of 50\,K in temperature leads to $\log$ g = 1.6. Using
the wings of pressure-broadened strong lines, a gravity $\log$ g =
$1.6 \pm 0.19$ is obtained, while a gravity $\log$ g = $1.5 \pm
0.5$ is found from the Fe~I - Fe~II ionisation equilibrium. A
reason for the discrepancy between this last result and the result
of other determinations could be the departure from LTE for Fe~I
lines or could be situated in the ionisation equilibrium.
Combining the results of these three methods yields $\log$ g =
$1.6 \pm 0.2$, corresponding to an Arcturus mass of $0.42 \le
\mathrm{M} \le 1.5$\,\Msun.}

\item{\label{Gratton1985AA...148..105G}
\citet{Gratton1985A&A...148..105G} has derived C, N and O
abundances by means of a
model-atmosphere analysis of blends of the CN and C~II bands
and of literature data about the 6300\,\ang\ [O~I] line. The
atmospheric parameters (\teff, $\log$ g, \vt, [Fe/H]) were taken
from \citet{Gratton1982MNRAS.201..807G}. The stellar abundances were computed
relative to the solar CNO abundances (which were not given). To
convert [C/H], [N/H] and [O/H] to $\varepsilon$(C),
$\varepsilon$(N) and $\varepsilon$(O) we have used the solar
abundances given by \citet{Anders1989GeCoA..53..197A}.}

\item{\label{Koechlin1985AA...153...91K}
\citet{Koechlin1985A&A...153...91K} have used the interferometer at
CER\-GA (I2T) to determine the angular diameter.}

\item{\label{Moon1985ApSS.117..261M}
\citet{Moon1985Ap&SS.117..261M} has found a linear relation
between the visual surface
brightness parameter $F_{\nu}$ and the ($b-y$)$_0$ colour index of $ubvy\beta$
photometry for spectral types later than G0. Using this relation, tables of
intrinsic colours and indices, absolute magnitude and stellar radius are given
for the ZAMS and luminosity classes Ia - V over a wide range of spectral
types. Using this $uvby\beta$ photometry, he found a radius of 23.4\,\Rsun\ for
$\alpha$ Boo, which is in good agreement with the value 27.5\,\Rsun\ obtained
from the parallax (97\,mas) and the angular diameter (24.9\,mas). The estimated
accuracy of the radius is 1.1\,\Rsun.}

\item{\label{Smith1985ApJ...294..326S}
High-resolution spectra of OH ($\Delta v = 2$), CO ($\Delta v =
3$), CO ($\Delta v = 2$) and  CN ($\Delta v = 2$) were obtained by
\citet{Smith1985ApJ...294..326S}. They have used ($V-K$) colours
and the calibration provided by \citet{Ridgway1980ApJ...235..126R}
to determine \teff. Using the spectroscopic requirement that
$\varepsilon$(Fe~I) = $\varepsilon$(Fe~II) yields, e.g., $\log$ g
= 0.8 for $\alpha$ Tau, which is too low for a K5~III giant. They
suggested that the reason for this low value is the
over-ionisation of iron relative to the LTE situation. They then
computed the surface gravity starting from a mass estimated from
evolutionary tracks in the H-R diagram. The metallicity was taken
from \citet{Kovacs1983A&A...120...21K} and for the microturbulence
they used Fe~I, Ni~I and Ti~I lines, demanding that the abundances
are independent of the equivalent width. Using the molecular
lines, they determined $\varepsilon$(C), $\varepsilon$(N),
$\varepsilon$(O) and \cc.}

\item{\label{Judge1986MNRAS.221..119J}
\citet{Ayres1982ApJ...263..791A} have determined \teff, L and R
from the distance, the
angular diameter and the bolometric luminosity found in literature. These
parameters were then used by \citet{Judge1986MNRAS.221..119J} to
determine the gravity and the
mass. The metallicity was taken from \citet{Lambert1981ApJ...248..228L}.}

\item{\label{Kyrolainen1986AAS...65...11K}
\citet{Kyrolainen1986A&AS...65...11K} have observed $\alpha$ Boo
with the 2.6\,m
telescope of the Crimean Astrophysical Observatory during the years
1974 -- 1980. The discussion of the results and methods used would be published
later, which does not seem to be the case.}

\item{\label{Leggett1986AA...159..217L}
The IRFM method is used by \citet{Leggett1986A&A...159..217L} to determine
the effective temperature and the angular diameter. The IRFM
procedures true angular diameters, so that no limb darkening
correction needs to be applied to these values. The parallax is
combined with the angular diameter to give the linear radius and
with the absolute integrated flux to give the luminosity.}

\item{\label{Tsuji1986AA...156....8T}
The stellar parameters quoted by
\citet{Tsuji1986A&A...156....8T} were
based on the results of \citet{Tsuji1981A&A....99...48T}, in which the
temperature was determined by the IRFM method. Only for $\alpha$ Boo,
the temperature determined by \citet{Frisk1982MNRAS.199..471F} was
used. A mass of 3\,\Msun\ was
assumed to ascertain the gravity. \citet{Tsuji1986A&A...156....8T} has
used high-resolution FTS
spectra of the CO first-overtone lines to determine
$\varepsilon$(C) and \vt\ by assuming that the abundance should be
independent of the equivalent width of the lines.}

\item{\label{Altas1987ApSS.134...85A}
\citet{Altas1987Ap&SS.134...85A} has done the analysis of
$\alpha$ Boo on base of
the Arcturus atlas of \citet{Griffin1968}. The atmospheric parameters
were taken from different literature sources.}

\item{\label{DiBenedetto1987AA...188..114D}
\citet{DiBenedetto1987A&A...188..114D} used Michelson interferometry by
the two-telescope baseline located at CERGA. Combining this
angular diameter with the bolometric flux F$_{\mathrm{bol}}$
(resulting from a direct integration using the trapezoidal rule
over the flux distribution curves, after taking  interstellar
absorption into account) they found the effective temperature,
which was in good agreement with results obtained from the lunar
occultation technique.}

\item{\label{Lambert1987ApSS.133..369L}
\citet{Lambert1987Ap&SS.133..369L} have presented the first
determinations of the Si isotopic ratios in M, MS and S stars.
Therefore they have obtained observations with the KPNO \,m
telescope and Fourier Transform Spectrometer. They however have
used the not so accurate $gf$-values of the SiO lines from a
recipe of \citet{Tipping1981} \citep[see, e.g., ][]{Langhoff1993,
Drira1997A&A...319..720D, Tsuji1994A&A...289..469T}.  The
temperatures and gravities were taken from the analysis of
\citet{Smith1985ApJ...294..326S} and
\citet{Smith1986A&A...165..126S}. With the total equivalent width
of the weak and strong lines, abundances were determined using the
LTE spectrum synthesis program MOOG \citep{Sneden1974}. The
microturbulent velocity was set to the value which gave the same
abundance of $^{28}$SiO for both weak and strong lines. This
yielded a silicon abundance of 7.83 for $\beta$ Peg. As
\citet{Lambert1987Ap&SS.133..369L} quoted,
\citet{Smith1985ApJ...294..326S, Smith1986A&A...165..126S} did not
include Si, but the expected Si abundance may be estimated from
their determinations of the iron and nickel abundance. This method
yielded a silicon abundance of 7.40 for $\beta$ Peg. The
discrepancy between these two silicon abundances was attributed
either to an inaccurate dissociation energy of SiO, a too high
effective temperature, problems with the model atmosphere or with
the SiO $gf$-values. This latter possibility is likely to be one
of the most important reasons: using the electric dipole moment
function of \citet{Langhoff1993} would increase the Einstein
coefficient for the 2-0 overtone band by a factor of about two and
consequently, the derived silicon abundance from the SiO lines
should decrease by about 0.3\,dex
\citep{Tsuji1994A&A...289..469T}. }

\item{\label{Edvardsson1988AA...190..148E}
\citet{Edvardsson1988A&A...190..148E} estimated logarithmic
surface gravities from the analysis of pressure broadened wings of
strong metal lines. Comparisons with trigonometrically determined
surface gravities give support to the spectroscopic results.
Surface gravities determined from the ionisation equilibria of Fe
and Si are found to be systematically lower than the strong line
gravities, which may be an effect of errors in the model
atmospheres, or departures from LTE in the ionisation equilibria.}

\item{\label{Glebocki1988AA...189..199G}
\citet{Glebocki1988A&A...189..199G} have obtained spectra of
binaries from the IUE archive. The masses of the primaries, M, the
mass ratios, $q$, and the inclinations of the orbits, $i$, have been
evaluated from the radial velocity amplitude, from the
mass-function and from the position of the primary on the
evolutionary track calculations. The effective temperature of the
primary is derived from infrared broad-band photometry, the error
being smaller than $70$\,K. The radius R is calculated from the
bolometric magnitudes, with the error being 20 -- 30\,\%. The error of
the logarithm of the ratio of the tidal acceleration at surface to
the surface gravity is estimate to be 0.5\,dex.}

\item{\label{Harris1988ApJ...325..768H}
\citet{Harris1988ApJ...325..768H} have analysed 5 stars,
containing $\beta$ Umi. They have taken \teff and $\log$ g-values
from literature, slightly adjusted to give the best fit.
Microturbulent velocity, $\varepsilon$(C) and \cc-ratio were
obtained by requiring an optimal fit with the observed spectra.
Different methods were used to estimate the stellar mass for 12
targets.}

\item{\label{Bell1989MNRAS.236..653B}
\citet{Bell1989MNRAS.236..653B} first determined the temperature
from the Johnson K band at 2.2\,\mic\ using the IRFM method. By
comparison with temperatures deduced from the colours Glass $J-H$,
$H-K$, $K-L$ and $K$; Cohen, Frogel and Persson $J-H$, $H-K$,
$K-L$ and $K$; Johnson $V-J$, $V-K$, $V-L$ and $K$; Cousins $V-R$,
$R-I$; Johnson and Mitchell 13-colour and Wing's near-infrared
eight-colour photometry they found that \teff(IRFM) was $\sim 80$K
too high, by which they corrected the temperature. The gravity and
[Fe/H] were adopted from literature values or were selected as
being characteristic for the MK class in question (e.g.\ $\beta$
UMi).}

\item{\label{Brown1989ApJS...71..293B}
\citet{Brown1989ApJS...71..293B} have used two different methods to
determine the effective temperature. In the first method published
photometry was used. The observed DDO $C(45-48)$ and $C(42-45)$
indices were used together with a theoretical model-atmosphere DDO
temperature calibration. In the second method, the ($R-I$) colour
index and a colour-temperature relation for giants was used. Equal
weights were assigned to each of these temperature values. The
gravity was ascertained by using the K-line absolute magnitude
$M_V(K)$ of \citet{Wilson1976ApJ...205..823W} scaled to a Hyades
distance modulus of
3.30, the effective temperature and an assumed stellar mass of
1\,\Msun. A microturbulent velocity of 1.7\,\kms\ was adopted
\citep{Gustafsson1974A&A....34...99G}. Using the observations from the 2.1\,m
reflector and coud\'{e} spectrograph at Mc Donald Observatory,
together with up to four different literature sources, a mean iron
abundance was determined.}

\item{\label{Hutter1989ApJ...340.1103H}
\citet{Hutter1989ApJ...340.1103H} have presented stellar diameter
measurements of 24 stars made with the MarkIII optical
interferometer. This gives the uniform disk angular diameter,
which was then corrected for limb darkening on the base of
appropriate model atmospheres. Their obtained angular diameters
are in good agreement with the angular diameters of stars in
similar ranges of spectral type measured through lunar
occultations for \ad\ $> 5$\,mas.}

\item{\label{Volk1989AJ.....98.1918V}
\citet{Volk1989AJ.....98.1918V} determined the effective temperature
directly from the literature values of angular-diameter
measurements and total-flux observations (also from literature).
The distance was taken from the Catalog of Nearby Stars
\citep{Gliese1969VeARI..22....1G} or from the Bright Star Catalogue
\citep{Hoffleit1982}.}

\item{\label{Blackwell1990AA...232..396B}
The effective temperature \teff\ and the angular diameter \ad\
given by \citet{Blackwell1990A&A...232..396B} were determined by the infrared
flux method (IRFM), a semi-empirical method which relies upon a
theoretical calibration of infrared bolometric corrections with
effective temperature. The infrared flux method uses a quantity
$R$ which is the ratio of the total flux to the flux at an
infrared wavelength. The temperature is obtained by comparing the
observed value of $R$ to theoretical values, based on model
atmosphere calculations. Since the infrared wavelength is in the
Rayleigh-Jeans tail of the flux distribution, $R$ is proportional
to the cube of the temperature. Thus, the temperature uncertainty
is one third of the measured flux error. One expects that the IRFM
should yield results better than 1\,\% for the effective temperature
and 2 -- 3\,\% for the angular diameter. The final effective
temperature is a weighted mean of T(J$_n$), T(K$_n$) and T(L$_n$),
with J$_n$ at 1.2467\,\mic, K$_n$ at 2.2135\,\mic\ and L$_n$ at
3.7825\,\mic.}

\item{\label{Fernandez-Villacanas1990AJ.....99.1961F}
By taking the mean value of the different temperatures delivered by
calibrations with photometric indices ($U-B$, $B-V$, $V-R$, $V-I$, $V-J$, $V-K$,
$V-L$), \citet{Fernandez-Villacanas1990AJ.....99.1961F} have fixed the effective
temperature.  For the gravity, the DDO photometry indices $C(45-48)$ and
$C(42-45)$ were used. The Fe~I lines served for the determination of \vt.}

\item{\label{McWilliam1990ApJS...74.1075M}
\citet{McWilliam1990ApJS...74.1075M} based his results on
high-resolution spectroscopic observations with resolving power
40000. The effective temperature was determined from empirical and
semi-empirical results found in the literature and from broad-band
Johnson colours. The gravity was ascertained by using the
well-known relation between g, \teff, the mass M and the
luminosity L, where the mass was determined by locating the stars
on theoretical evolutionary tracks. So, the computed gravity is
fairly insensitive to errors in the adopted L. High-excitation
iron lines were used for the metallicity [Fe/H], in order that the
results are less spoiled by non-LTE effects. The author refrained
from determining the gravity in a spectroscopic way (i.e.\ by
requiring that the abundance of neutral and ionised species yields
the same abundance) because {\it{`A gravity adopted by demanding
that neutral and ionised lines give the same abundance, is known
to yield temperatures which are $\sim 200$ K higher than found by
other methods. This difference is thought to be due to non-LTE
effects in Fe~I lines.'}}. By requiring that the derived iron
abundance, relative to the standard 72 Cyg, were independent of
the equivalent width of the iron lines, the microturbulent
velocity $\xi_t$ was found. }

\item{\label{Smith1990ApJS...72..387S}
\citet{Smith1990ApJS...72..387S} have determined the chemical
composition of a sample of M, MS and S giants. The use of a slightly
different set of lines for the molecular vibration-rotation lines of
CO, OH and CN, along with improved $gf$-values for CN and NH, results
in a small difference in the carbon, nitrogen and oxygen abundance
with respect to the values deduced by
\citet{Smith1985ApJ...294..326S}.}

\item{\label{Blackwell1991AA...245..567B}
\citet{Blackwell1991A&A...245..567B} is a revision of
\citet{Blackwell1990A&A...232..396B} where the H$^-$ opacity has been
improved. They
investigated the effect of the improved H$^-$ opacity on the IRFM
temperature scale and derived angular diameters. Also here, the
mean temperature is a weighted mean of the temperatures for $J_n$,
$K_n$ and $L_n$. Relative to Blackwell et al.\ (1990) there was a
change of temperature up to 1.4\,\% and a decrease by 3.5\,\% in \ad.}

\item{\label{Flynn1991AA...250..400F}
\citet{Flynn1991A&A...250..400F} have derived the metallicity from
DDO photometry.}

\item{\label{Judge1991ApJ...371..357J}
\citet{Judge1991ApJ...371..357J} have used literature values for the
effective temperatures of K stars. If these were not available the
($V-K$) versus \teff\ calibration of
\citet{Ridgway1980ApJ...235..126R} was used. Direct measurements
available in literature for ($V-R$) versus
visual surface flux corrections were used for \ad. In conjunction
with the distance --- obtained from the parallax --- this yielded the
radii. Stellar gravities from photospheric line studies, which
should be reliable up to $\sim 0.3$ dex were used.
Finally the mass was obtained from the surface gravity or from
stellar masses for groups of stars. On the basis of initial mass
function and time scales of evolution, \citet{Scalo1978ApJ...225..523S}
argue that most O-rich giants must have masses around $1$\,\Msun.}

\item{\label{Lazaro1991MNRAS.249...62L}
\citet{Lazaro1991MNRAS.249...62L} have obtained
intermediate-re\-so\-lu\-tion ($\lambda / \Delta\lambda \approx
600$) scans of 70 bright late-type giants of luminosity class III
(K5-M6) with a cooled grating spectrometer (CGS) at the Cassegrain
focus of the 1.5\,m Infrared Flux Collector of the Observatorio del
Teide. The wavelength range was at least 2.20 -- 2.45\,\mic. Since
the infrared scans have insufficient information to allow the
determination of \teff\ and $\log$ g, the MK spectral types are
used with the calibration of \citet{Ridgway1980ApJ...235..126R} to determine
\teff. A physical gravity was deduced, with absolute magnitudes
taken from \citet{Wilson1976ApJ...205..823W} and masses inferred from
the H-R diagram.
An advantage of using spectral types rather than a colour index
such as ($V-K$) to class the stars in temperature is that the former
is independent of the (uncertain or unknown) reddening. Their use
of the composite H-R diagram for red giants corresponds roughly to
the adoption of a mean giant mass of 1.5\,\Msun. This is in good
accord with estimates for the mean mass of Scalo \citep[][and
references therein]{Lazaro1991MNRAS.249...62L}. Lazaro denoted that the adoption
by \citet{Tsuji1986A&A...156....8T}of a mean mass of 3.0\,\Msun\ seems
too high.  Adopting a microturbulence velocity of 2.5\,\kms\ then yielded
$\varepsilon$(C) and \cc\ from the first-overtone lines of CO.}

\item{\label{Mozurkewich1991AJ....101.2207M}
\citet{Mozurkewich1991AJ....101.2207M} have used the MarkIII Optical
Interferometer. The uniform disk angular diameter $\theta_{\mathrm{UD}}$ has a
residual of 1\,\% for the 800\,nm observations and less than 3\,\% for
the 450\,nm observations. The limb darkened
diameter was then obtained by multiplying the uniform disk angular diameter with
a correction factor \citep[using the quadratic limb darkening
coefficient of][]{Manduca1979A&AS...36..411M}.}

\item{\label{Tsuji1991AA...245..203T}
The stellar parameters quoted by \citet{Tsuji1991A&A...245..203T} were
based on the results of \citet{Tsuji1981A&A....99...48T}, in which the
temperature was
determined by the IRFM method. A mass of 3\,\Msun\ was assumed to
ascertain the gravity. \citet{Tsuji1991A&A...245..203T} used
CO lines of the second overtone band. A standard analysis of
these CO lines could, however, yield different results than a
linear analysis of the weak lines.}

\item{\label{Cornide1992AJ....103.1374C}
Effective temperatures have been computed by
\citet{Cornide1992AJ....103.1374C}  with
the temperature calibration of \citet{BohmVitense1981ARA&A..19..295B}.}

\item{\label{Engelke1992AJ....104.1248E}
\citet{Engelke1992AJ....104.1248E} has derived a two-parameter
analytical expression approximating the long-wavelength (2 --
60\,\mic) infrared continuum of stellar calibration standards.
This generalised result is written in the form of a Planck
function with a brightness temperature that is a function of both
observing wavelength and effective temperature. This function is
then fitted to the best empirical flux data available, providing
thus the effective temperature and the angular diameter.}

\item{\label{Pasquini1992AA...266..340P}
\citet{Pasquini1992A&A...266..340P} have used the effective
temperature given by \citet{Pasquini1990A&A...234..277P}, who have used the
spectral classification and the $(V-R)_0$ colour index to
construct a colour-temperature relation. The accuracy on \teff\ is
250\,K. Together with the bolometric magnitude and a mass inferred
from the comparison with evolutionary tracks, they have deduced
the gravity. Due to a raw division of the stellar masses into four
sub-groups, their gravity is the parameter containing the major
uncertainties.}

\item{\label{Bonnell1993MNRAS.264..319B}
\citet{Bonnell1993MNRAS.264..319B} constructed a grid based on the
effective temperature \teff\ of \citet{Manduca1981ApJ...243..883M}. They used
ground-based high-resolution FTS spectra of OH and [O~I] lines.
The requirement that the oxygen abundances determined from the
[O~I] and OH line widths are in agreement amounts to finding the
intersection of the loci of points defined by the measured widths
in the ([O/H], $\log$ g) plane. The determination of \vt\ was
based on OH-lines, but was hampered by a lack of weak lines from
this radical for some stars. A least-square fit was then
performed, which yielded \vt. A lower \vt\ for the OH ($\Delta v
= 2$) lines than for the OH($\Delta v = 1$) lines was attributed
to a greater average depth of formation for the OH ($\Delta v =
2$) sequence lines.}

\item{\label{Morossi1993AA...277..173M}
Using the catalogue of \citet{Cayrel1992A&AS...95..273C},
\citet{Morossi1993A&A...277..173M} have selected programme stars having a solar
chemical composition. Effective temperature values have been
derived from calibrations of broad-band Johnson colours, by
adopting the polynomial relationships of
\citet{McWilliam1990ApJS...74.1075M}. The
surface gravity has been obtained from DDO photometry.}

\item{\label{Peterson1993ApJ...404..333P}
Using opacity distribution functions based on a newly expanded
atomic and molecular line list, \citet{Peterson1993ApJ...404..333P} have
calculated a model atmosphere for Arcturus which reproduces the
observed flux distribution of the Griffin atlas \citep{Griffin1968}. Individual
line parameters in the list were adjusted to match the solar
spectrum. Using spectral lines for which the solar $gf$-values are
well determined in the region 5000 -- 5500\,\ang, 6000 -- 6500\,\ang\ and
7500 -- 8875\,\ang, the effective temperature, surface gravity and
microturbulence are determined.}

\item{\label{Quirrenbach1993ApJ...406..215Q}
\citet{Quirrenbach1993ApJ...406..215Q} have determined the uniform disk
angular diameter in the strong TiO band at 712\,nm and in a
continuum band at 754\,nm with the MarkIII stellar interferometer
on Mount Wilson. Because limb darkening is expected to be
substantially larger in the visible than in the infrared, the
measured uniform disk diameters should be larger in the visible
than in the infrared. This seems however not always to be the
case. Using the same factor as \citet{Mozurkewich1991AJ....101.2207M} we have
converted their continuum uniform disk value for $\beta$ And,
$\alpha$ Cet, $\alpha$ Tau and $\beta$ Peg into a limb darkened
angular diameter, yielding a value of 15.23\,mas, 13.62\,mas,
22.73\,mas and 18.11\,mas respectively.  There is a systematic uncertainty
in the limb-darkened angular diameter of the order of 1\,\% in
addition to the measurement uncertainty of the uniform disk
angular diameter \citep{Davis1998IAUS..189...31D}.}

\item{\label{ElEid1994AA...285..915E}
The effective temperature, surface gravity and mass of
\citet{Harris1984ApJ...285..674H} were used by
\citet{ElEid1994A&A...285..915E}. He noted a correlation between the
$^{16}$O/$^{17}$O ratio and the stellar mass and the \cc\ ratio and the stellar
mass for evolved stars. Using this ratio
in conjunction with evolutionary tracks, El Eid has determined the mass for
$\gamma$ Dra.}

\item{\label{Gadun1994AN....315..413G}
\citet{Gadun1994AN....315..413G} has used model parameters and
equivalent widths of Fe~I and Fe~II lines for $\alpha$ Cen,
$\alpha$ Boo and $\alpha$ Car found in literature. It turned out
that the Fe~I lines were very sensitive to the temperature
structure of the model and that iron was over-ionised relative to
the LTE approximation due to the near-ultraviolet excess
$J_{\nu}-B_{\nu}$. Since the concentration of the Fe~II ions is
significantly higher than the concentration of the neutral iron
atoms, the iron abundance was finally determined using these Fe~II
lines. It is demonstrated that there is a significant difference
in behaviour of \vt\ from the Fe~I lines for solar-type stars,
giants and supergiants. The microturbulent velocity decreases in
the upper photospheric layers of solar-type stars, in the
photosphere of giants (like Arcturus) \vt\ has the tendency to
increase and in Canopus, a supergiant, a drastic growth of \vt\ is
seen. This is due to the combined effect of convective motions and
waves which form the base of the small-scale velocity field. The
velocity of convective motions decreases in the photospheric
layers of dwarfs and giants, while the velocity of waves increases
due to the decreasing density. In solar-type stars the convective
motion penetrates in the line-forming region, while the behaviour
of \vt\ in Canopus may be explained by the influence of gravity
waves. The characteristics of the microturbulence determined from
the Fe~II lines differ from that found with Fe~I lines. These
results can be explained by 3D numerical modeling of the
convective motions in stellar atmospheres, where it is shown that
the effect of the lower gravity is noticeable in the growth of
horizontal velocities above the stellar `surface' (in the region
of Fe~I line formation). But in the Fe~II line-forming layers the
velocity fields are approximately equal in 3D model atmospheres
with a different surface gravity and same \teff. Both values for
\vt\ derived from the Fe~I and Fe~II lines are listed. If the
microturbulence varies, the values of \vt\ are given going outward
in the photosphere.}

\item{\label{Worthey1994ApJS...95..107W}
\citet{Worthey1994ApJS...95..107W} has used \teff, $\log$ g and
[Fe/H] from \citet{Worthey1994ApJS...94..687W} for the
construction of detailed models for intermediate and old stellar
populations. Most temperatures come from a transformation of
($V-K$) colours. \citet{Worthey1994ApJS...94..687W} have
determined indices from 21 optical absorption features. These
indices are summarised in fitting functions which give index
strengths as function of stellar temperature, gravity and [Fe/H].
$\beta$ Peg was, however, not mentioned in this project.}

\item{\label{Cohen1996AJ....112.2274C}
\citet{Cohen1996AJ....112.2274C} have derived the effective
temperature and angular
diameter from the composite spectra of $\alpha$ Boo, $\gamma$ Dra, $\alpha$ Cet
and $\gamma$ Cru using the \citet{Engelke1992AJ....104.1248E}. Using a --- not
published --- statistical relation between the spectral type and the
gravity, the gravities of these stars were assigned. A gravity of
2.0 was adopted for $\alpha$ Tau, though in
\citet{Cohen1992AJ....104.2030C} a value of 1.5
--- taken from \citet{Smith1985ApJ...294..326S} --- was used.}

\item{\label{Quirrenbach1996AA...312..160Q}
The diameter of Arcturus has been measured by
\citet{Quirrenbach1996A&A...312..160Q} with the MarkIII interferometer
at five wavelengths
between 450\,nm and 800\,nm. By using the limb-darkening coefficient
of \citet{Manduca1979A&AS...36..411M} and
\citet{Manduca1981ApJ...243..883M} they have computed the
true limb-darkened diameter of Arcturus as being $21.0 \pm
0.2$\,mas. By combining this value with the bolometric flux, the
effective temperature is determined, being $4303 \pm 47$\,K.}

\item{\label{Aoki1997AA...328..175A}
\citet{Aoki1997A&A...328..175A} have taken the same stellar parameters
as \citet{Tsuji1991A&A...245..203T} (taken from
\citet{Tsuji1981A&A....99...48T} or from
\citet{Frisk1982MNRAS.199..471F} for $\alpha$ Boo), but they now used
CN lines to determine $\varepsilon$(N) and $\xi_t$.}

\item{\label{Pilachowski1997AJ....114..819P}
\citet{Pilachowski1997AJ....114..819P} reported on the
determination of the
carbon isotope ratios using the first-overtone $^{12}$CO and
$^{13}$CO high-resolution observations. The analysis of the
spectra proceeded via synthetic spectrum calculations. This
provided both the carbon abundance and the isotopic ratio \cc.}

\item{\label{Tsuji1997AA...320L...1T}
\citet{Tsuji1997A&A...320L...1T} have used a model photosphere with
parameters \teff\ = 3600\,K, $\log$ g = 0.5, \vt\ = 3.0\,\kms,
without explaining the used parameters.}

\item{\label{Abia1998MNRAS.293...89A}
\citet{Abia1998MNRAS.293...89A} have derived the abundances of a
variety of
elements from Sr to Eu relative to iron. For $\beta$ Peg, they quoted that the
stellar parameters were taken from
\citet{McWilliam1990ApJS...74.1075M}.
\citet{McWilliam1990ApJS...74.1075M} did, however, not mention the
target $\beta$ Peg.}

\item{\label{Blackwell1998AAS..129..505B}
The IRFM method has been used by
\citet{Blackwell1998A&AS..129..505B} to obtain the effective
temperature of a sample of stars selected for the flux calibration
of ISO. The accuracy estimates are based on considerations
concerning the accuracy of absolute fluxes, the agreement between
temperatures derived using the various filters and on the relation
between temperatures and various photometric indices. Because no
account has been taken of uncertainties due to difficulties in
theoretical modelling of stellar atmospheres, these accuracies are
a useful indication of the relative accuracy of the determined
stellar temperature by \citet{Blackwell1998A&AS..129..505B}.}

\item{\label{DiBenedetto1998AA...339..858D}
\citet{DiBenedetto1998A&A...339..858D} calibrated the surface
brightness-colour correlation using a set of high-precision
angular diameters measured by modern interferometric techniques.
The stellar sizes predicted by this correlation were then combined
with bolometric-flux measurements, in order to determine
one-dimensional (T, $V-K$) temperature scales of dwarfs and giants.
Both measured and predicted values for the angular diameter are
listed.}

\item{\label{Dyck1998AJ....116..981D}
\citet{Dyck1998AJ....116..981D} have obtained observations at
2.2\,\mic\ at the Infrared
Optical Telescope Array (IOTA) interferometer. Comparisons
with lunar occultations at 1.65 and 2.2\,\mic, interferometry at 2.2\,\mic\ at
CERGA and at IOTA with the FLUOR beam combination system can be seen to be
good. Using broad-band photometry for the value of the bolometric flux, the
effective temperature was determined.}

\item{\label{Hammersley1998AAS..128..207H}
\citet{Hammersley1998A&AS..128..207H} have determined the
stellar effective
temperature using on the one hand the infrared flux method (IRFM)
and on the other hand the ($V-K$) versus \teff\ relationship of
\citet{DiBenedetto1993A&A...270..315D}.}

\item{\label{Perrin1998AA...331..619P}
\citet{Perrin1998A&A...331..619P} have derived the effective temperature
for nine giant stars from diameter determinations at 2.2\,\mic\
with the FLUOR beam combiner on the IOTA interferometer. This
yielded the uniform disk angular diameter of $\alpha$ Boo and
$\alpha$ Tau. The averaging effect of a uniform model leads to an
underestimation of the diameter of the star. Therefore, they have
fitted their data with limb-darkened disk models published in the
literature. The average result is a ratio between the uniform and
the limb-darkened disk diameters of 1.035 with a dispersion of
0.01. This ratio could then also be used for the uniform disk
angular diameters of $\gamma$ Dra and $\beta$ Peg, listed by
\citet{DiBenedetto1987A&A...188..114D}, and $\alpha$ Cet, which were
based on a
photometric estimate. Several photometric sources were used to
determine the bolometric flux, which then, in conjunction with the
limb-darkened diameter, yielded the effective temperature.}

\item{\label{Robinson1998ApJ...503..396R}
\citet{Robinson1998ApJ...503..396R} have used different sources
of scientific literature. Their quoted mass-value was computed from
the adopted radius and gravity value.}

\item{\label{Taylor1999AAS..134..523T}
\citet{Taylor1999A&AS..134..523T} prepared a catalogue of
temperatures and [Fe/H] averages for evolved G and K giants. This
catalogue is available at CDS via anonymous ftp to cdsarc.u-strasbg.fr.}

\end{enumerate}
\aareferences
\end{document}


\newcommand{\ang}{$\rm \AA$}
\newcommand{\tauross}{$\tau_{\mathrm{ross}}$}
\newcommand{\Msun}{M$_{\odot}$}
\newcommand{\Mbol}{M$_{\rm{bol}}$}
\newcommand{\Rsun}{R$_{\odot}$}
\newcommand{\Lsun}{L$_{\odot}$}
\newcommand{\be}{\begin{equation}}
\newcommand{\ee}{\end{equation}}
\newcommand{\bee}{\begin{eqnarray}}
\newcommand{\ad}{$\theta_d$}
\newcommand{\vt}{$\xi_t$}
\newcommand{\cc}{$\mathrm{^{12}C/^{13}C}$}
\newcommand{\kn}{$\kappa_{\nu}$}
\newcommand{\lnu}{$l_{\nu}$}
\newcommand{\ha}{H$_{\alpha}$}
\newcommand{\ea}{et al.}
\newcommand{\ene}{\end{eqnarray}}
\newcommand{\teff}{T$_{\mathrm{eff}}$}
\newcommand{\mic}{$\mu$m}
\newcommand{\hoogte}[1]{\rule{0pt}{#1}}
\newcommand{\hminff}{H$^-_{\rm{ff}}$}
\newcommand{\teffo}{T$_{\mathrm{eff,0}}$}
\hyphenation{Kol-mo-go-rov}
\hyphenation{Kol-mo-go-rov--Smir-nov}

\def\offinterlineskip{\baselineskip=-1000pt \lineskip=1pt
\lineskiplimit=\maxdimen}
\def\pra{\mathrel{\mathchoice {\vcenter{\offinterlineskip\halign{\hfil$\displaystyle##$\hfil\cr\propto\cr\sim\cr}}}
{\vcenter{\offinterlineskip\halign{\hfil$\textstyle##$\hfil\cr\propto\cr\sim\cr}}}
{\vcenter{\offinterlineskip\halign{\hfil$\scriptstyle##$\hfil\cr\propto\cr\sim\cr}}}
{\vcenter{\offinterlineskip\halign{\hfil$\scriptscriptstyle##$\hfil\cr\propto\cr\sim\cr}}}}}

\def\ga{\mathrel{\mathchoice {\vcenter{\offinterlineskip\halign{\hfil$\displaystyle##$\hfil\cr>\cr\sim\cr}}}
{\vcenter{\offinterlineskip\halign{\hfil$\textstyle##$\hfil\cr>\cr\sim\cr}}}
{\vcenter{\offinterlineskip\halign{\hfil$\scriptstyle##$\hfil\cr
>\cr\sim\cr}}}
{\vcenter{\offinterlineskip\halign{\hfil$\scriptscriptstyle##$\hfil\cr>\cr\sim\cr}}}}}

\def\la{\mathrel{\mathchoice {\vcenter{\offinterlineskip\halign{\hfil$\displaystyle##$\hfil\cr<\cr\sim\cr}}}
{\vcenter{\offinterlineskip\halign{\hfil$\textstyle##$\hfil\cr<\cr\sim\cr}}}
{\vcenter{\offinterlineskip\halign{\hfil$\scriptstyle##$\hfil\cr
<\cr\sim\cr}}}
{\vcenter{\offinterlineskip\halign{\hfil$\scriptscriptstyle##$\hfil\cr><cr\sim\cr}}}}}


\title{ISO-SWS calibration and the accurate modelling of cool-star atmospheres
\thanks{Based on observations with ISO, an ESA project with
instruments funded by ESA Member States (especially the PI countries France,
Germany, the Netherlands and the United Kingdom) and with the participation of
ISAS and NASA.}}
\subtitle{IV.\ G9 to M2 stars}


\author{L.\ Decin\inst{1}\thanks{\emph{Postdoctoral Fellow of the Fund for
Scientific Research, Flanders}}
\and B.\ Vandenbussche\inst{1}
\and C.\ Waelkens\inst{1}
\and G.\ Decin\inst{1}
\and K.\ Eriksson\inst{2}
\and B.\ Gustafsson\inst{2}
\and B.\ Plez\inst{3}
\and A.J.\ Sauval\inst{4}}

\offprints{L.\ Decin, e-mail: Leen.Decin@ster.kuleuven.ac.be}

\institute{Instituut voor Sterrenkunde, KULeuven, Celestijnenlaan 200B, B-3001
    Leuven, Belgium
\and
    Institute for Astronomy and Space Physics, Box 515, S-75120 Uppsala, Sweden
\and
    GRAAL - CC72, Universit\'{e} de Montpellier II, F-34095 Montpellier Cedex 5,
France
\and
    Observatoire Royal de Belgique, Avenue Circulaire 3, B-1180 Bruxelles,
    Belgium
}

\date{Received data; accepted date}

\abstract{
A detailed spectroscopic study of 11 giants with spectral type from G9
to M2 is presented. The 2.38 -- 4.08\,\mic\ wavelength-range of band 1
of ISO-SWS (Short-Wavelength Spectrometers on board of the Infrared Space
Observatory) in which
many different molecules --- with their own dependence on each of the
stellar parameters --- are absorbing, enables us  to
estimate the effective temperature, the gravity, the microturbulence,
the metallicity, the CNO-abundances, the \cc-ratio and the angular
diameter from the ISO-SWS data. Using the Hipparcos' parallax,
the radius, luminosity and gravity-inferred mass are derived. The
stellar parameters obtained are in good agreement with other published
values, though also some discrepancies with values deduced by other
authors are noted. For a few stars ($\delta$ Dra, $\xi$
Dra, $\alpha$ Tuc, H Sco and $\alpha$ Cet) some parameters --- e.g.\
the CNO-abundances --- are derived for the first time.
By examining the
correspondence between different ISO-SWS observations of the same
object and between the ISO-SWS data and the corresponding synthetic
spectrum, it is shown that the relative accuracy of ISO-SWS in band 1
(2.38 -- 4.08\,\mic) is better than 2\,\% for these high-flux sources.
The high level of correspondence between observations and theoretical
predictions, together with a confrontation of the estimated \teff(ISO)
value with \teff\ values derived from colours --- which demonstrates
the consistency between $V-K$, BC$_K$, \teff\ and \ad\ derived from
optical or IR data --- proves that both the used MARCS models to derive the
stellar quantities and the flux calibration of the ISO-SWS detectors
have reached a high level of reliability.
}

\maketitle
\markboth{L.\ Decin et al.: ISO-SWS and modelling of cool stars}{L.\ Decin et
    al.: ISO-SWS and modelling of cool stars}

\defcitealias{Decin2000A&A...364..137D}{Paper~I}
\defcitealias{Decin2000b}{Paper~II}
\defcitealias{Decin2000c}{Paper~III}

\section{Introduction}\label{introduction}

In the fourth article of this series, we discuss the ISO-SWS
(Short-Wavelength Spectrometers on board of the Infrared Space
Observatory) spectra of 11 giants cooler than the Sun. These
giants are part of a larger sample selected as stellar standard
candles for the calibration of the detectors of ISO-SWS. In
\citet[][hereafter referred to as
Paper~I]{Decin2000A&A...364..137D} a method was developed to
analyse the infrared data (2.38 -- 12\,\mic) of cool stars: the
specific dependence of the absorption pattern by the different
molecules gives the possibility to estimate fundamental stellar
parameters like the effective temperature (\teff), the gravity
($\log$ g), the microturbulence (\vt), the metallicity, the
CNO-abundances, the \cc-ratio and the angular diameter (\ad). We
now apply this method of analysis to the ISO-SWS data of these 11
giants. Keeping the general calibration problems and the accuracy
of the theoretical models and synthetic spectra  (based on the
MARCS and Turbospectrum code \citep{Gustafsson1975A&A....42..407G,
Plez1992A&A...256..551P, Plez1993ApJ...418..812P}; version May
1998) in mind --- as discussed in \citet[][hereafter referred to
as Paper~II]{Decin2000b}
--- the error bars on the different stellar parameters are determined
and the resultant stellar parameters are confronted with other
published parameters.

This paper is organised in the following way: in Sect.\
\ref{summary} a summary of the general discrepancies between the
ISO-SWS spectra of the cool giants and the theoretical spectra, as
described in Paper~II, is given. We will go more deeply into the
determination of the different stellar parameters and their error
bar in Sect.\ \ref{stelparameters}. After an overview of the
deduced stellar parameters, each star is discussed in more detail.
Not only are the accuracy and precision of the ISO-SWS data
assessed, but the deduced stellar parameters are also evaluated
with respect to other published values. A discussion on some of
the stellar parameters obtained is held in Sect.\
\ref{discussion}, where we focus on (i) the effective temperature
and (ii) the \cc\ ratio.  The conclusions are given in Sect.\
\ref{conclusions}.

The appendix of this article is published electronically. Most of
the grey-scale plots in the article are printed in colour in the
appendix, in order to better distinguish the different spectra or
symbols.


\section{Summary of general discrepancies (Paper~II)}
\label{summary}

The origin(s) of different types of discrepancies seen in the
spectra of the cool stars could be settled from a detailed
comparison between ISO-SWS data and theoretical predictions. As far as the
calibration is concerned, we demonstrated in
\citetalias{Decin2000b} that problems with the relative spectral
response functions in band 1A (2.38 -- 2.60\,\mic) and band 2 (4.08 --
12\,\mic), fringes at the end of
band 1D and memory-effects in band 2 destroy the flux accuracy in
this wavelength range. Problems with the CO lines underline
that one needs to know the instrumental profile and resolution
accurately. Concerning the modelling part, the main problems were
situated in the incomplete knowledge of the atomic oscillator
strengths in the infrared. The fact that the low-excitation
first-overtone CO lines were predicted as being too strong and the
fundamental OH lines as being to weak, indicated that some of the
assumptions the models are based on --- e.g. hydrostatic equilibrium,
homogeneity ---- are questionable for these cool stars.


\section{Stellar parameters}\label{stelparameters}

As was demonstrated in Paper~I, it is possible to estimate different
stellar parameters for cool giants from their ISO-SWS spectrum and
this due to the presence of many molecular absorbers in the 2.38 --
4.08\,\mic\ wavelength range. E.g.\ the slope of the continuum between
3.1\,\mic\ and 4.0\,\mic\ is indicative for the effective temperature (\teff),
while the slope in band 1B and band 1D makes it possible to pin down
the gravity ($\log$ g). The microturbulent velocity (\vt), the
abundance of carbon, nitrogen and oxygen and the \cc-ratio can be
deduced from the relative strength of the molecular features. For the
determination of the metallicity ([Fe/H]), both the slope of the
continuum and the strength of the features are required (see Fig. 2 in
Paper~I). Finally, the angular diameter (\ad) can be extracted directly from
the absolute flux-level of the ISO-SWS data.

Although, these IR ISO-SWS data give us a lot of information
concerning the stellar parameters, an inspection of a large grid
with stellar parameters typical for these stars, learned us that
there exists an actual danger to be trapped in a local minimum in
this huge parameter space. Moreover, the preliminary results of a
statistical study (Shkedy et al., {\it in prep}; Decin et al.,
{\it in prep}) indicate that the {\it maximum} error bar on the
logaritm of the gravity (as determined from the ISO-SWS spectra)
can be quite large ($\ga 0.40$\,dex) inducing significant error
bars on e.g.\ the stellar abundances. Hence, we decided to
determine first the effective temperature from ($V-K$) colours
(see Sect. \ref{firstguess}). A first guess for the gravity was
found from $K$-band photometry, the bolometric correction BC$_K$,
the Hipparcos' parallax $\pi$, \teff($V-K$) and a mass M$_{hr}$
estimated from evolutionary tracks (see Sect. \ref{firstguess}).
Using then the ISO-SWS spectra, the values of these two parameters
were refined if necessary. In this way, the {\it maximum}
{\footnote{This maximum error is {\it not} the error as obtained
from varying one parameter at a time, since this only gives us an
underestimate of the uncertainties. An error in one parameter may
couple to the error in another parameter `constructively'. The
effect of this coupling will be different for different type of
stars. Considering here the G9 to M2 giants in our sample, a large
grid with typical stellar parameters have been inspected in order
to estimate this coupling and to determine the {\it maximum}
error}} error on \teff, $\log$ g, \vt, [Fe/H], $\varepsilon$(C),
$\varepsilon$(N), $\varepsilon$(O) and \cc-ratio could be
restricted to 200\,K, 0.20\,dex, 1\,\kms, 0.30\,dex, 0.30\,dex,
0.40\,dex, 0.30\,dex and 3 respectively. A {\it mean} error is
then $\sim 2/3$ of this maximum error, depending on the quality of
the observation. As will be discussed in Sect.\ \ref{disbpeg}, it
was impossible to deduce \teff, $\log$ g, \vt, [Fe/H] and the
CNO-abundances from the ISO-SWS spectrum of $\beta$ Peg. The error
bars are estimated from a comparison with other quoted literature
values.

\subsection{First guess for \teff\ and $\log$ g} \label{firstguess}

One can calculate \teff\ directly from $V-K$ using (semi)-empirical
(\teff, $V-K$) temperature scales. Such a calibration is e.g.\ given
by \citet{Bessell1998A&A...333..231B}, who determined a polynomial fit
between \teff\ and $V-K$ based on interferometer and IRFM data
\citep[Infrared Flux Method][]{Blackwell1977MNRAS.180..177B} for
giants.
In Table \ref{firsttemp}, the $V$ and $K$ magnitude of Simbad are listed. For
$\beta$ UMi, we list the $K$ magnitude of
\citet{Faucherre1983A&A...120..263F} ; for $\alpha$ Tuc and H
Sco the $K$ magnitudes found in the atlas of
\citet{Gezari1999yCat.2225....0G} are
given. The ($V-K$) colour has been corrected for interstellar
extinction using $A_V = 0.8$\,mag/kpc
\citep{Blackwell1990A&A...232..396B} and $E(V-K) = A_V/1.1$
\citep{Mathis1990ARA&A..28...37M}, with the distance calculated from the
Hipparcos' parallax (see Table \ref{parameters}).
The values for the effective
temperature calculated from formula abcd given in Table 7 in
\citet{Bessell1998A&A...333..231B} are listed in Table
\ref{firsttemp}. A comparison with other
(\teff, $V-K$) temperature scales will be given in Sect.\ \ref{colours}. A
typical error bar on the derived \teff($(V-K)_0$) values is 50\,K.

\begin{table*}
\caption{\label{firsttemp}Magnitude measurements for the Johnson $V$
and $K$ pass-bands are tabulated. The effective temperature (in K) derived
from the (\teff, $(V-K)_0$) relationship as given by
\citet{Bessell1998A&A...333..231B} is listed in the last column. }
\begin{center}
\begin{tabular}[htb]{llrrccc}\hline
\rule[-3mm]{0mm}{8mm} name & HD-number & \multicolumn{1}{c}{$V$} &
\multicolumn{1}{c}{$K$} & $V-K$ & $(V-K)_0$ & \teff($(V-K)_0$) \\
\hline \hline \rule[0mm]{0mm}{5mm}$\delta$ Dra & HD~180711 &
3.08$^a$ & 0.80$^b$ & 2.28 & 2.26 & 4828 \\
$\xi$ Dra & HD~163588 & 3.75$^a$ & 1.03$^b$ & 2.72 & 2.70 & 4457 \\
$\alpha$ Boo & HD~124897 & $-0.06^a$ & $-2.95^c$ & 2.89 & 2.88 & 4331 \\
$\alpha$ Tuc & HD~211416 & $2.86^g$ & $-0.09^f$ & 2.95 & 2.91 & 4312 \\
$\beta$ UMi & HD~131873 & 2.06$^a$ & $-1.22^d$ & 3.28 & 3.25 & 4111 \\
$\gamma$ Dra & HD~164058 & 2.23$^a$ & $-1.34^b$ & 3.57 & 3.54 & 3970 \\
$\alpha$ Tau & HD~29139 & 0.87$^a$ & $-2.81^{b,c}$ & 3.68 & 3.67 & 3915 \\
H Sco & HD~149447 & $4.16^b$ & $0.00^h$ & 4.16 & 4.08 & 3764\\
$\beta$ And & HD~6860 & 2.03$^a$ & $-1.83^c$ & 3.86 & 3.82 & 3856 \\
$\alpha$ Cet & HD~18884 & 2.49$^a$ & $-1.63^e$ & 4.12 & 4.07 & 3768 \\
\rule[-3mm]{0mm}{3mm}$\beta$ Peg & HD~217906 & 2.45$^a$ & $-2.20^c$ &
4.65 & 4.61 & 3615 \\ \hline
\end{tabular}
\end{center}
{\footnotesize{
$a$: \citet{Mermilliod1986EgUBV.........M};
$b$: \citet{Johnson1966CoLPL...4...99J};
$c$: \citet{Lee1970ApJ...162..217L};
$d$: \citet{Faucherre1983A&A...120..263F};
$e$: \citet{Glass1974MNSSA..33...53G};
$f$: \citet{Wamsteker1981A&A....97..329W};
$g$: \citet{Nicolet1978A&AS...34....1N};
$h$: \citet{Price1968AJ.....73..431P}}}
\end{table*}

In order to determine the gravity, we need the radius R and the
mass. The first parameter is assessed from $K$, BC$_K$, $\pi$ (from
Hipparcos) and
\teff($(V-K)_0$). We used BC$_K$($(V-K)_0$)\,=\,m$_{\rm{bol}} - K$ from
\citet{Bessell1998A&A...333..231B}. The work of
\citet{Bessell1998A&A...333..231B} is based on current SOSMARCS-models
and synthetic colours, and ensures consistency between derived
parameters and model spectra.
The error on the bolometric correction
is assumed to be 0.05; M$_{{\rm{bol}}.\odot}$ is assumed to be 4.74
\citep{Bessell1998A&A...333..231B}. The radii derived in this way are
tabulated in Table \ref{firstlogg}.

Mass values for the stars in our sample are estimated from
evolutionary tracks with appropriate metallicity as calculated by
\citet{Girardi2000A&AS..141..371G}, with the only exception being
the more massive star $\alpha$ Car for which the evolutionary tracks of
\citet{Lejeune2001A&A...366..538L} have been used. The two
sets of evolutionary tracks from the zero age main sequence (ZAMS)
to the red giant branch (RGB) agree well with each other. Only
\citet{Girardi2000A&AS..141..371G} have given the evolutionary tracks
from the zero
age horizontal branch (ZAHB) to the thermally pulsing
asymptotic giant branch (TP-AGB). By plotting (L(\Mbol) $\pm \sigma$(L(\Mbol)),
\teff($(V-K)_0$) $\pm \sigma$(\teff($(V-K)_0$))) for each star on the HR-diagram
the mass values M$_{hr}$ are obtained
(see Table \ref{firstlogg}). Note that most of the
giants lie above the horizontal branch and hence could burn
hydrogen in a shell surrounding the helium core or could be
post-He-flash stars. From our observations no diagnostic tools
exist to decide upon the evolutionary status, with the only
exception being $\beta$ Peg, since its colours do not match any RGB
evolutionary tracks. In Table \ref{firstlogg} the mass values are
given for a giant on the RGB (except for $\beta$ Peg), with
a typical error bar of 0.3\,\Msun\ when M$_{hr} \le
2.5\,$\Msun. When a star would be on the AGB, this value
should be lowered by $\sim 0.2$\,\Msun.

These radius- and mass values were then used to find first-guess
values for the gravity (last column in Table \ref{firstlogg}). Also
the mean error for all these parameters is listed. 

\begin{table*}
\caption{\label{firstlogg}Magnitude measurements for the Johnson $K$ band
are tabulated. The absolute bolometric magnitude
\Mbol\ and the luminosity L(\Mbol) (in \Lsun)  are derived from
the $(V-K)_0$ colour index, the bolometric correction BC$_K$($(V-K)_0$) and the
Hipparcos' parallax $\pi$. The radius R is calculated from L(\Mbol) and
\teff($(V-K)_0$). Mass values estimated from evolutionary tracks are given
in \Msun\ for all stars in our sample. A typical error bar for M$_{hr}$
is 0.3\,\Msun\ when M$_{hr} \le 2.5$\,\Msun. The gravity is derived from
R and M$_{hr}$. More details are given in the text.}
\begin{center}
\begin{tabular}[htb]{lrcrrcccc}\hline
\rule[-3mm]{0mm}{8mm} name & \multicolumn{1}{c}{$K_0$} & BC$_K$ & \multicolumn{1}{c}{m$_{\rm{bol}}$} &
\multicolumn{1}{c}{\Mbol(m$_{\rm{bol}},\pi$)} & L(\Mbol) &
R(L(\Mbol), \teff($(V-K)_0$)) & M$_{hr}$ & $\log$ g \\
\hline \hline \rule[0mm]{0mm}{5mm}$\delta$ Dra &
0.80 & 1.94 & $2.74 \pm 0.05$ & $0.30 \pm 0.06$ &
$59 \pm 3$ & $11.01 \pm 0.37$ & $2.2$ & $2.70 \pm 0.07$ \\
$\xi$ Dra & 1.03  & 2.22 & $3.25 \pm 0.05$ & $0.58 \pm
0.06$ & $46 \pm 3$ & $11.40 \pm 0.41$ & $1.1$ & $2.36 \pm 0.12$ \\
$\alpha$ Boo & $-2.95$ & 2.31 & $-0.64
\pm 0.05$ & $-0.90 \pm 0.05$ & $180 \pm 9$ & $23.81 \pm 0.80$ & $1.1$
& $1.72 \pm 0.12$ \\
$\alpha$ Tuc & $-0.10$ & 2.33 & $2.23 \pm 0.05$  & $-1.70 \pm 0.05$ &
$376 \pm 32$ & $34.72 \pm 1.69$ & 2.5 & $1.75 \pm 0.07$ \\
$\beta$ UMi & $-1.22$ & 2.46 & $1.24 \pm 0.05$ & $-1.71
\pm 0.07$ & $379 \pm 23$ & $38.36 \pm 1.50$ & $2.2$ & $1.61 \pm 0.07$\\
$\gamma$ Dra & $-1.34$ & 2.55 & $1.21 \pm 0.05$ &
$-2.07 \pm 0.07$ & $530 \pm 33$ & $48.61 \pm 1.94$ & $1.80$ & $1.31
\pm 0.08$ \\
$\alpha$ Tau & $-2.81$ & 2.59 & $-0.22 \pm 0.05$ &
$-1.72 \pm 0.06$ & $395 \pm 23$ & $42.62 \pm 1.67$ & $1.2$ & $1.26 \pm
0.11$ \\
H Sco & $-0.01$ & 2.70 & $2.69 \pm 0.05$ & $-2.40 \pm 0.19$ & $716 \pm
122$ & $62.87 \pm 5.61$ & 1.4 & $0.99 \pm 0.12$ \\
$\beta$ And & $-1.84$ & 2.63 & $0.79 \pm 0.05$ & $-3.14
\pm 0.11$ & $1415 \pm 147$ & $84.23 \pm 4.88$ & $2.2$ & $0.93 \pm
0.08$ \\
$\alpha$ Cet & $-1.64$ & 2.70 & $1.06 \pm 0.05$ & $-3.09
\pm 0.13$ & $1356 \pm 164$ & $86.35 \pm 5.71$ & $2.5$ & $0.96 \pm
0.08$ \\
\rule[-3mm]{0mm}{3mm}$\beta$ Peg & $-2.21$  &
2.80 & $0.59 \pm 0.05$ & $-3.34 \pm 0.11$ & $1702 \pm 169$ & $105.10
\pm 5.97$ & $1.85$ & $0.66 \pm 0.09$ \\ \hline
\end{tabular}
\end{center}
{\footnotesize{
$a$: \citet{Johnson1966CoLPL...4...99J};
$b$: \citet{Lee1970ApJ...162..217L};
$c$: \citet{Faucherre1983A&A...120..263F};
$d$: \citet{Glass1974MNSSA..33...53G}}}
\end{table*}

\subsection{Results}

Using the ISO-SWS spectra, these first guesses were further refined
and the [Fe/H], $\varepsilon$(C,N,O), \cc-ratio and \ad\ were
derived. The results from the detailed spectroscopic study and the
final error bars are presented in Table \ref{parameters}. The distance
D (in pc) is calculated from the parallax measurements (in mas) of
Hipparcos. R(D, \ad) in combination with the gravity then yielded the
gravity-inferred mass M$_g$ (in \Msun). The luminosity L (in \Lsun),
extracted from the radius R(D, \ad) and \teff(ISO), is the last
quantity listed in this table.
The objects have been sorted by spectral type, but since the
spectral type does not necessarily provide a good indicator of the
temperature \citep{Tsuji1981A&A....99...48T}, this ordering is not
always by decreasing effective temperature.
The theoretical models and synthetic spectra for stars
with spectral type earlier than K0 were calculated using a
plane-parallel geometry, for the other ones a spherically
symmetric geometry was assumed \citep[see Fig.\ 1
in][]{Decin1997fiso.work..185D}.  In order to judge upon the
goodness-of-fit of the observed and theoretical data, the deviation
estimating parameter $\beta$ derived from the Kolmogorov-Smirnov test
\citepalias[see][]{Decin2000A&A...364..137D} has been calculated (see
Table \ref{beta}). This parameter $\beta$ checks the {\it{global}} accordance
between the two spectra. The lower the $\beta$-value, the better the
accordance. The maximum acceptable $\beta$-values were derived from
$\alpha$ Tau in \citepalias{Decin2000A&A...364..137D}. These
$\beta_{\rm{max}}$ values do not depend on the flux level of the
observed target, since the $\beta$ parameter is a {\it{global}}
goodness-of-fit parameter and these $\beta_{\rm{max}}$ values may thus
be used for all observations.

\begin{table*}[htb]
\caption{\label{parameters}
Final fundamental stellar parameters
for the selected stars in the sample. The effective temperature
\teff\ is given in K, the logarithm of the gravity in c.g.s.
units, the microturbulent velocity \vt\ in \kms, the angular
diameter in mas, the parallax $\pi$ in mas, the distance D in
parsec, the radius R(D, \ad) in \Rsun, the gravity-inferred mass M$_g$ in
\Msun\ and the luminosity L in \Lsun.}
\begin{center}
\leavevmode
\footnotesize
\begin{tabular}{lcccccc}  \hline
\rule[-3mm]{0mm}{8mm}  & \object{$\delta$ Dra} & \object{$\xi$ Dra} &
\object{$\alpha$ Boo} & \object{$\alpha$ Tuc} & \object{$\beta$ UMi} &
\object{$\gamma$ Dra} \\
\hline \hline
\rule[0mm]{0mm}{5mm}Sp. Type & G9~III & K2~III & K2~IIIp & K3~III &
K4~III & K5~III \\
\teff & $4820 \pm 140$ & $4440 \pm 300$ & $4320 \pm 140$ & $4300 \pm
140$ & $4085 \pm 140$ & $3960 \pm 140$ \\
$\log$ g & $2.70 \pm 0.20$ & $2.40 \pm 0.25$ & $1.50 \pm 0.15$ & $1.35
\pm 0.15$ & $ 1.60 \pm 0.15$ & $1.30 \pm 0.25$ \\
\vt & $2.0 \pm 0.5$ & $2.0 \pm 1.0$ & $1.7 \pm 0.5$ & $1.7 \pm 0.5$ &
$2.0 \pm 0.5$ & $2.0 \pm 0.5$ \\
${\mathrm{[Fe/H]}}$ & $0.00 \pm 0.25$ & $0.10 \pm 0.30$ & $-0.50 \pm
0.20$ & $0.00 \pm 0.20$ & $-0.15 \pm 0.20$ & $0.00 \pm 0.20$ \\
$\varepsilon$(C) & $8.15 \pm 0.25$ & $8.00 \pm 0.30$ & $7.96 \pm 0.20$
& $8.30 \pm 0.20$ & $8.25 \pm 0.20$ & $8.15 \pm 0.25$ \\
$\varepsilon$(N) & $8.26 \pm 0.25$ & $8.26 \pm 0.40$ & $7.61 \pm 0.25$
& $8.26 \pm 0.25$ & $8.16 \pm 0.25$ & $8.26 \pm 0.25$ \\
$\varepsilon$(O) &  $8.93 \pm 0.25$ & $8.93 \pm 0.30$ & $8.68 \pm
0.20$ & $8.93 \pm 0.20$ & $8.83 \pm 0.20$ & $8.93 \pm 0.20$ \\
\cc & $12 \pm 3$ & $20 \pm 5$ & $7 \pm 2$ & $13 \pm 3$ & $9 \pm 2$ &
$10 \pm 2$ \\
\ad & $3.30 \pm 0.20$ & $3.11 \pm 0.22$ & $20.72 \pm 1.28$ & $6.02 \pm
0.37$ & $9.93 \pm 0.62$ & $9.98 \pm 0.63$ \\
$\pi$ & $32.54 \pm 0.46$ & $29.26 \pm 0.49$ & $88.85 \pm 0.74$ &
$16.42 \pm 0.59$ & $25.79 \pm 0.52$ & $22.10 \pm 0.46$ \\
D & $30.73 \pm 0.43$ & $34.34 \pm 0.57$ & $11.26 \pm 0.09$ & $60.90
\pm 2.19$ & $38.78 \pm 0.78$ & $45.25 \pm 0.94$ \\
R & $10.90 \pm 0.68$ & $11.42 \pm 0.83$ & $25.06 \pm 1.56$ & $39.40
\pm 2.82$ & $41.38 \pm 2.71$ & $48.53 \pm 3.23$ \\
M$_g$ & $2.18 \pm 1.04$ & $1.20 \pm 0.71$ & $0.73 \pm 0.27$ & $1.27
\pm 0.48$ & $2.49 \pm 0.92$ & $1.72 \pm 1.02$ \\
\rule[-3mm]{0mm}{3mm}L & $58 \pm 10$ & $46 \pm 14$ & $197 \pm 36$ &
$479 \pm 93$ & $430 \pm 82$ & $523 \pm 101$ \\ \hline
\end{tabular}

\hspace*{.1cm}\begin{tabular}{lccccc}  \hline
\rule[-3mm]{0mm}{8mm} & \object{$\alpha$ Tau} & \object{H Sco} &
\object{$\beta$ And} & \object{$\alpha$ Cet} & \object{$\beta$ Peg}
\\ \hline \rule[0mm]{0mm}{5mm}Sp. Type & K5~III & K6~III & M0~III & M2~III & M2.5~III\\
\teff & $3850 \pm 140$ & $3900 \pm 140$ & $3880 \pm 140$ & $3740 \pm 140$
& $3600 \pm 300$ \\
$\log$ g & $1.50 \pm 0.15$ & $1.30 \pm 0.15$ & $0.95 \pm 0.25$ & $0.95
\pm 0.25$ & $0.65 \pm 0.40$ \\
\vt & $1.7 \pm 0.5$ & $2.0 \pm 0.5$ & $2.0 \pm 0.5$ & $2.3 \pm 0.5$ & $2.0 \pm
 0.3$ \\
${\mathrm{[Fe/H]}}$ & $-0.15 \pm 0.20$ & $0.00 \pm 0.20$ & $0.00 \pm 0.30$ &
 $0.00 \pm 0.30$ & $0.00$ \\
$\varepsilon$(C) & $8.35 \pm 0.20$ & $8.25 \pm 0.20$ & $8.12 \pm 0.30$ &
$8.20 \pm 0.30$ & $8.20 \pm 0.40$ \\
$\varepsilon$(N) & $8.35 \pm 0.25$ & $8.26 \pm 0.25$ & $8.37 \pm 0.40$
& $8.26 \pm 0.40$ & $8.18 \pm 0.40$ \\
$\varepsilon$(O) & $8.93 \pm 0.20$ & $8.83 \pm 0.20$ & $9.08 \pm 0.30$ &
$8.93 \pm 0.30$ & $8.93 \pm 0.40$ \\
\cc & $10 \pm 2 $ & $8 \pm 2$ & $ 9 \pm 2$ & $10 \pm 2$ & $5 \pm 3$ \\
\ad & $20.77 \pm 1.31$ & $4.73 \pm 0.30$ & $13.30 \pm 0.84$ & $12.52 \pm 0.79$ &
$16.60 \pm 1.32$ \\
$\pi$ & $50.09 \pm 0.95$ & $9.63 \pm 0.79$ & $16.36 \pm 0.76$ & $14.82
\pm 0.83 $ & $16.37 \pm 0.72$ \\
D & $19.96 \pm 0.38$ & $103.84 \pm 8.52$ & $61.12 \pm 2.84$ & $67.48 \pm
3.78$ & $61.08 \pm 2.69$ \\
R & $44.56 \pm 2.94$ & $52.79 \pm 5.47$ & $87.37 \pm 6.85$ &
$90.79 \pm 7.66$ & $108.98 \pm 9.90$ \\
M$_g$ & $2.30 \pm 0.85$ & $2.03 \pm 0.82$ & $2.49 \pm   1.48$ &
$2.69 \pm 1.61$ & $1.94^{+4.27}_{-1.34}$ \\
\rule[-3mm]{0mm}{3mm}L & $394 \pm 77$ & $582 \pm 147$ & $1561 \pm 332$
& $1455 \pm 328$ & $1800 \pm 683$\\ \hline
\end{tabular}
\end{center}
\end{table*}

\begin{table*}[htb]
\caption{\label{beta} The $\beta$-values from the Kolmogorov-Smirnov
test are given not only for the different sub-bands, but also for the
different molecular
 absorbers the $\beta$-values. In the second column, one can find the
 maximum acceptable $\beta$-value for that specific wavelength-range as
 estimated for a high-quality AOT01 speed-4 observation \citepalias{Decin2000A&A...364..137D}.}
\begin{center}
\leavevmode
\footnotesize
\setlength{\tabcolsep}{1.8mm}
\begin{tabular}{llccccccccccc}  \hline
\rule[-3mm]{0mm}{8mm} & $\beta_{\rm{max}}$ & \object{$\delta$ Dra} & \object{$\xi$ Dra} &
\object{$\alpha$ Boo} & \object{$\alpha$ Tuc} & \object{$\beta$ UMi} &
\object{$\gamma$ Dra} & \object{$\alpha$ Tau} & \object{H Sco} &
\object{$\beta$ And} & \object{$\alpha$ Cet} & \object{$\beta$ Peg}\\
\hline \hline
\rule[0mm]{0mm}{5mm}$\beta_{\mathrm{1A}}$ & 0.06 &
0.051 & 0.074 & 0.015 & 0.036  & 0.050 & 0.058 & 0.040 &
0.033 & 0.049 & 0.057 & 0.157\\
$\beta_{\mathrm{1B}}$  & 0.05 & 0.046 & 0.072 & 0.041 & 0.013 & 0.029 & 0.048 &
0.034 & 0.026 & 0.046 & 0.041 & 0.150\\
$\beta_{\mathrm{1D}}$ & 0.04 & 0.040 & 0.069 & 0.036 & 0.021 & 0.025 & 0.011 &
0.036 & 0.019 & 0.019 & 0.024 & 0.218 \\
\rule[-3mm]{0mm}{3mm}$\beta_{\mathrm{1E}}$ & 0.04 & 0.051 &
0.022 & 0.043 & 0.035 & 0.040 & 0.014 & 0.027 &
0.033 & 0.044 & 0.044 & 0.084 \\ \hline
\rule[0mm]{0mm}{5mm}$\beta_{\mathrm{CO(1A)}}$ & 0.06 & 0.051 &
0.074 & 0.015 & 0.036 & 0.050 & 0.058 &
0.040 & 0.033 & 0.040 & 0.057 & 0.157 \\
$\beta_{\mathrm{CO(1B)}}$ & 0.05 & 0.029 & 0.044 & 0.038 & 0.017 & 0.017 & 0.026
& 0.039 & 0.034 & 0.056 & 0.038 & 0.167\\
$\beta_{\mathrm{OH/H_2O(1B)}}$ & 0.04 & 0.010 & 0.010 & 0.010 & 0.027 & 0.020 &
0.023 & 0.024 & 0.021 & 0.032 & 0.019 & 0.027 \\
$\beta_{\mathrm{OH/H_2O(1D)}}$ & 0.04 & 0.040 & 0.069 & 0.036 & 0.021 & 0.025 &
0.011 & 0.036 & 0.019 & 0.020 & 0.024 & 0.218 \\
$\beta_{\mathrm{OH(1E)}}$ & 0.03 & 0.038 & 0.019 & 0.041 & 0.023 & 0.042 & 0.017
& 0.025 & 0.017 & 0.031 & 0.036 &  0.047\\
\rule[-3mm]{0mm}{3mm}$\beta_{\mathrm{SiO(1E)}}$ & 0.03 &
0.024 & 0.015 & 0.014 & 0.021 & 0.028 & 0.021 &
0.022 & 0.013 & 0.018 & 0.018 & 0.027 \\ \hline
\end{tabular}
\end{center}
\end{table*}

In the following subsections, each of the 11 cool giants will be
discussed individually. For each star, calibration details
can be found in the Appendix. These, in conjunction with the general
calibration specifications discussed in Paper~II, gives us an idea on
the calibration accuracy.
If other AOT01 observations {\footnote{Each observation is determined
uniquely by its observation
number (8 digits), in which the first three digits represent the revolution
number. The observing data can be calculated from the revolution
number which is the number of days after 17 November
1995.}} are available, they are compared with each
other in the Appendix in order to assess the calibration precision of
ISO-SWS. Knowing the accuracy and the precision of the observational
data, several stellar parameters (and their error bar) are determined
from the AOT01 observation and the agreement between observational and
theoretical spectra is discussed. Furthermore, the deduced stellar
parameters are evaluated with respect to published stellar
parameters. Tables with a summary of the different published results,
together with a short description of the used methods and/or data and
of the assumed and deduced parameters by the different authors quoted
in next sections, can be found in the Appendix.


\subsection{${\delta}$ Dra{\rm: AOT01, speed 4, revolution 206}}\label{disdeldra}

\subsubsection{Comparison between the ISO-SWS spectrum and the synthetic
spectrum (Fig.\ \ref{deldra})}

\begin{figure}[h!]
\begin{center}
\resizebox{0.5\textwidth}{!}{\rotatebox{90}{\includegraphics{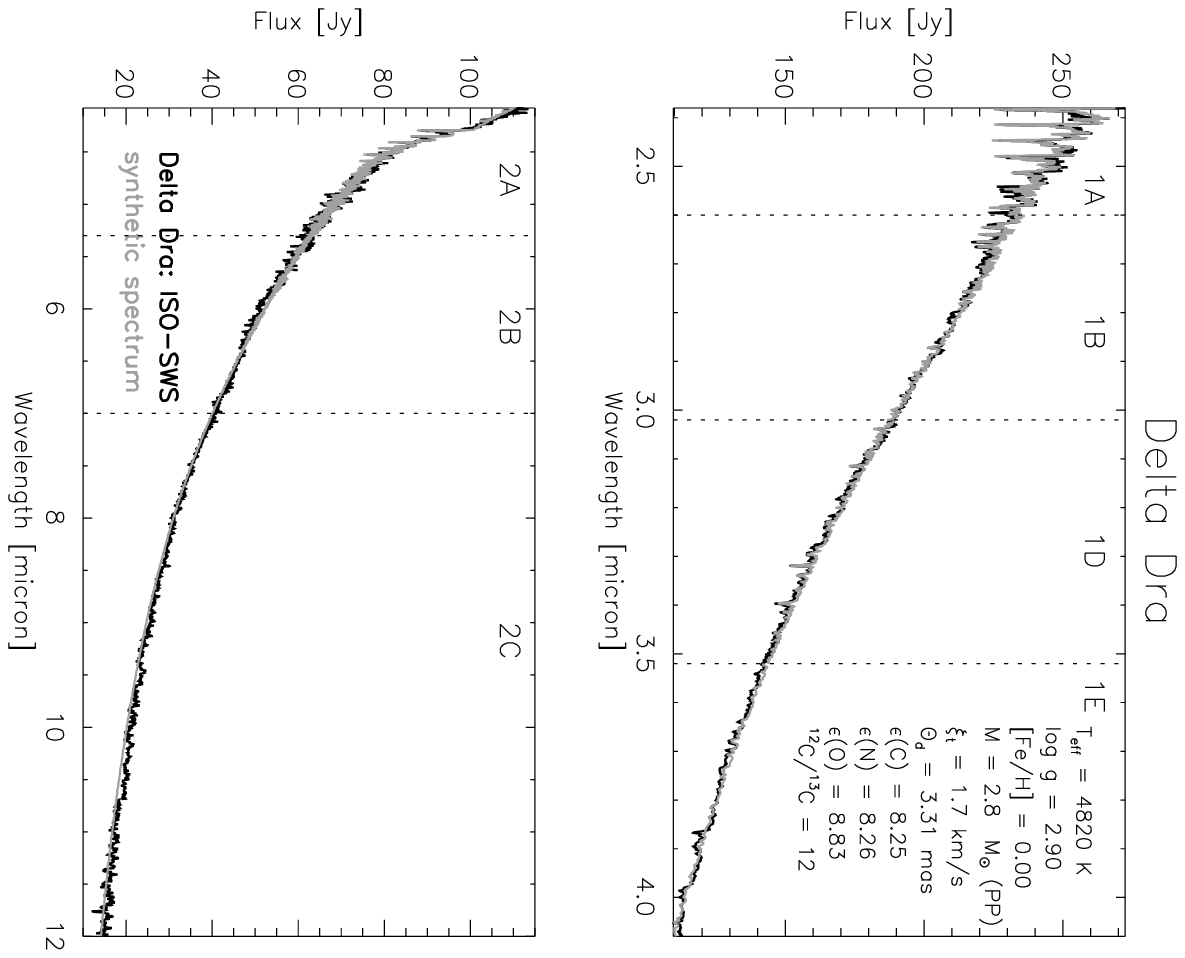}}}
\caption{\label{deldra} Comparison between band 1 and band 2 of
the ISO-SWS data of $\delta$ Dra (black) and the synthetic
spectrum (grey) with stellar parameters \teff\ = 4820\,K, $\log$ g =
2.70, M = 2.2\,\Msun, [Fe/H] = 0.00, \vt\ = 2.0\,\kms, \cc\ = 12,
$\varepsilon$(C) = 8.15, $\varepsilon$(N) = 8.26, $\varepsilon$(O)
= 8.93 and \ad\ = 3.30\,mas.
A coloured version of this plot is available in the Appendix as Fig.\
\ref{deldracol}.}
\end{center}
\end{figure}

Some of the atomic features identified in Fig.\ 3 in
\citetalias{Decin2000b} can still be recognised in $\delta$ Dra,
e.g.\ around 2.55 \mic, 2.63 \mic, 3.3 \mic, 3.4 \mic\ and 3.85
\mic. The second most abundant molecule in nearly all cool stars
is CO, because of its very high dissociation energy (11.09 eV),
and because of the high cosmic abundance of carbon and oxygen.
From $\delta$ Dra on, the first-overtone CO band dominates the
spectrum till 2.8 \mic. The problems with these CO lines are
described in point 2 in \citetalias{Decin2000b}. In band 1B, 1D
and 1E the OH $\Delta v = 1$ lines become visible. Problems with
memory effects \citepalias{Decin2000b} make it impossible to
improve the data in band 2.

The carbon abundance is deduced from the CO lines in band 1A
(where the standard deviation is larger than for the other
sub-bands in band 1), the N-abundance from  the CN lines which are
almost invisible, and the O-abundance from OH lines which are also
still very weak; the overall uncertainty on the abundance of
carbon, nitrogen and oxygen is estimated to be 0.25\,dex. Since
the chemical equilibrium determines the shape of the synthetic
spectrum, too inaccurate a nitrogen abundance would result in high
$\beta$-values. So, although the CN and NH lines remain also very
weak for the cooler giants, a wrong nitrogen abundance can be
traced.

\subsubsection{Comparison with other published stellar parameters
\rm{(see Table \ref{litddra} in the Appendix)}}\label{compparddra}

The temperatures given by the different authors quoted in Table
\ref{litddra} all agree quite
well. Only the temperature given by \citet{Faucherre1983A&A...120..263F} and
\citet{Gustafsson1974A&A....34...99G} are somewhat lower than the other values.
\citet{Gustafsson1974A&A....34...99G} have
established the temperature scale with the aid of a new
theoretical calibration of $R-I$. This colour was computed from
scaled solar model atmospheres, with the line blocking considered in a
statistical way. The use of scaled solar model atmospheres led to
underestimates of \teff, an effect which is also noticeable for the
\teff-value of $\xi$ Dra in Table \ref{litkdra}.

There is a rather large difference in published gravities, being
either around 2.9\,dex or around 2.1\,dex. Only
\citet{Gratton1982MNRAS.201..807G} and
\citet{Gratton1983MNRAS.202..231G} have tabulated a value of
$\log$ g = 2.1, a result which was obtained from a curve of growth
technique. Also the gravity value given by
\citet{Burnashev1983IzKry..67...13B} --- who has used narrow-band
photometric colours in the visible part of the electromagnetic
spectrum --- is quite discrepant from the more common listed value
of $\log$ g = 2.9. We note that also for other stars in our
sample, quite some gravity values do not correspond with the ones
given by \citet{Burnashev1983IzKry..67...13B}. Possible reasons
include inaccurate spectrophotometric data or inaccurate formulae
to convert the colour indices to the stellar parameter values.
Using a gravity of $\log$ g = 2.10, the best fit was obtained with
a
--- low --- carbon abundance of $\varepsilon$(C) = 8.00
(Fig.\ \ref{deldra210}). The discrepancies are more pronounced than in
Fig.\ \ref{deldra}, while the $\beta$-values are larger than for $\log$ g =
2.70 and $\varepsilon$(C) = 8.15.

\begin{figure}[h!]
\begin{center}
\resizebox{0.5\textwidth}{!}{\rotatebox{90}{\includegraphics{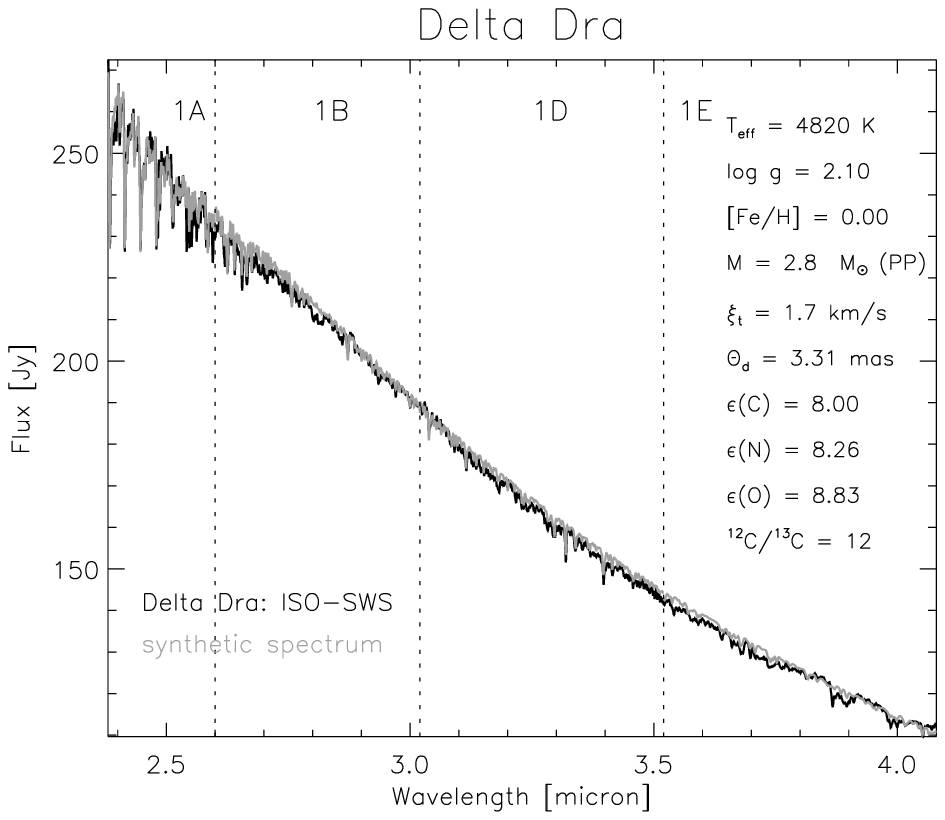}}}
\caption{\label{deldra210} Comparison between band 1 of
the ISO-SWS data (rev.\ 206) of $\delta$ Dra (black) and the synthetic
spectrum (grey) with stellar parameters \teff\ = 4820\,K, $\log$ g =
2.10, M = 2.8\,\Msun, [Fe/H] = 0.00, \vt\ = 1.7\,\kms, \cc\ = 12,
$\varepsilon$(C) = 8.00, $\varepsilon$(N) = 8.26, $\varepsilon$(O)
= 8.83 and \ad\ = 3.31\,mas.
A coloured version of this plot is available in the Appendix as Fig.\
\ref{deldra210col}.}
\end{center}
\end{figure}

Other indirect angular diameter determinations were not found.
\citet{Gratton1982MNRAS.201..807G} have derived the radius from the
visual surface brightness and the temperature, and so have obtained the
mass M$_g$ of $\delta$ Dra. \citet{Faucherre1983A&A...120..263F}
measured the angular diameter with the two-telescope
stellar interferometer (I2T) at CERGA. By using D$ = 36 \pm 8$\,pc
\citep{Jenkins1952QB813.J45......},
a radius R$ = 15 \pm 4$\,\Rsun\ was obtained. Although different
methods and/or data are used, the three values for the
radius listed in Table \ref{litddra} are in good agreement.

 From
the 10\,\% absolute flux level accuracy, the uncertainty on the
angular diameter (\ad\ = 3.30\,mas) is estimated to be 0.20\,mas. This
yields a gravity-inferred mass of M$_g = 2.18 \pm 1.04$\,\Msun. This
value for M$_g$ is higher than the gravity-inferred mass deduced by
\citet{Gratton1982MNRAS.201..807G}, due to their use of a lower
gravity value.


\subsection{${\xi}$ Dra{\rm: AOT01, speed 3, revolution 314}}\label{disksidra}

\subsubsection{Comparison between the ISO-SWS spectrum and the synthetic spectrum
(Fig.\ \ref{ksidra})}

\begin{figure}[h!]
\begin{center}
\resizebox{0.5\textwidth}{!}{\rotatebox{90}{\includegraphics{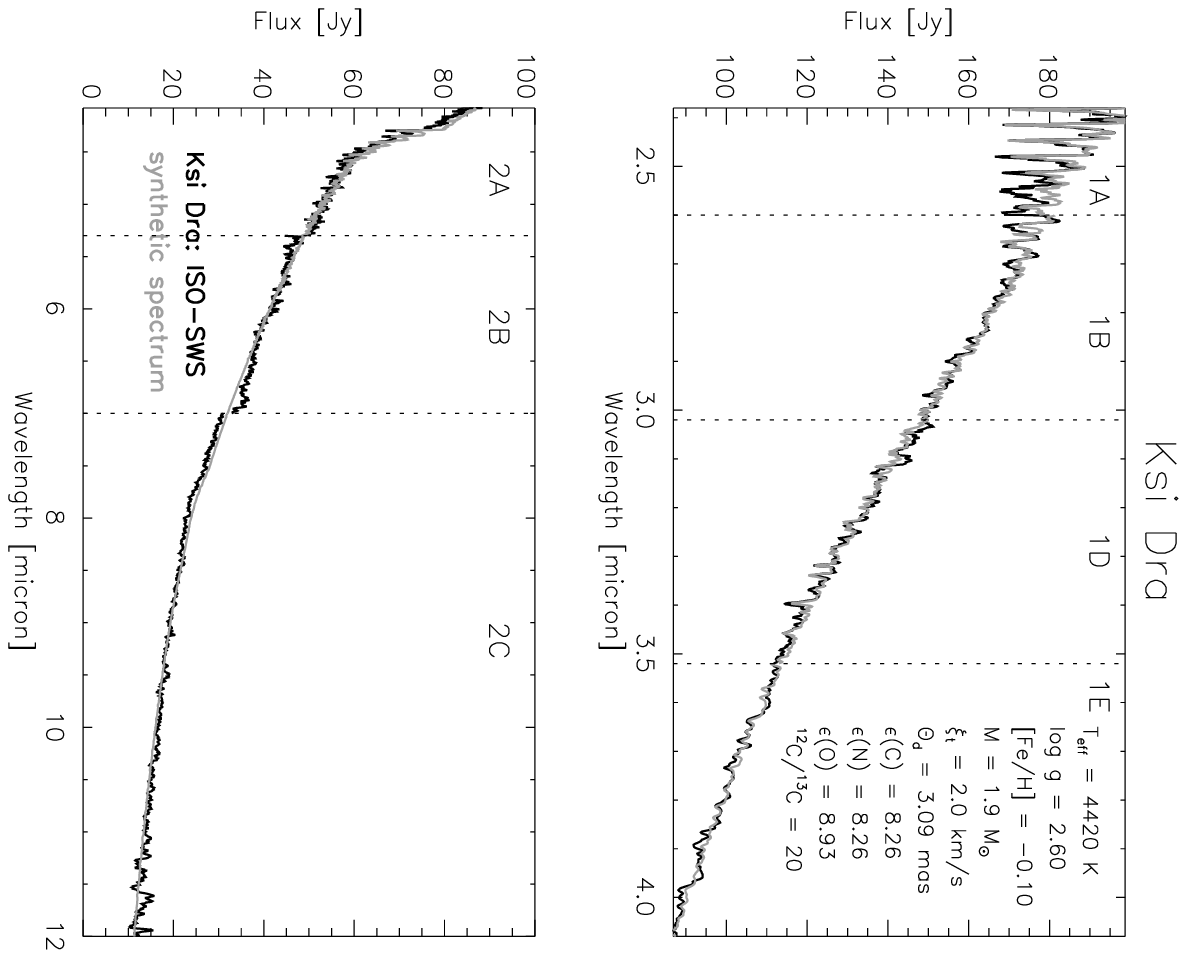}}}
\caption{\label{ksidra} Comparison between band 1 and band 2 of
the ISO-SWS data (rev.\ 314) of $\xi$ Dra (black) and the synthetic spectrum
(grey) with stellar parameters \teff\ = 4440\,K, $\log$ g = 2.40, M
= 1.2\,\Msun, [Fe/H] = $0.10$, \vt\ = 2.0\,\kms, \cc\ = 20,
$\varepsilon$(C) = 8.00, $\varepsilon$(N) = 8.26, $\varepsilon$(O)
= 8.93 and \ad\ = 3.11\,mas. A coloured version of this plot is
available in the Appendix as Fig.\ \ref{ksidracol}.}
\end{center}
\end{figure}

The larger pointing offset in the y-direction and the high noise
level reduce the reliability of the shape of the sub-bands.
Especially the shape of the data at the end of band 1A, in band 1B
and in band 1D seems unreliable. As a result, the uncertainties on
the fundamental stellar parameters are enhanced. Fortunately,
the post-He spectrum does not display these problems. Using then both
the speed-3 observation and the post-He data, the stellar parameters
were estimated. Similar discrepancies as seen in Fig.\ \ref{deldra} of $\delta$
Dra are also arising in the comparison between the observational
data of $\xi$ Dra and its synthetic spectrum. Some examples caused
by atomic transitions may be seen around 2.545 \mic, 2.595 \mic\
and 2.623 \mic.  The first-overtone CO lines are clearly visible
around 2.4 \mic. As described in \citetalias{Decin2000b}, the
strongest CO lines were predicted as somewhat too strong when a
resolution of 1500 is used. The carbon, nitrogen and oxygen
abundance, together with the \cc\ ratio are an indication that
$\xi$ Dra has already gone through the first dredge-up (see Sect.\
\ref{cc}). At 4.005
\mic, the 2-0 band head of the SiO first-overtone band emerges.

\begin{figure}[h!]
\begin{center}
\resizebox{0.5\textwidth}{!}{\rotatebox{90}{\includegraphics{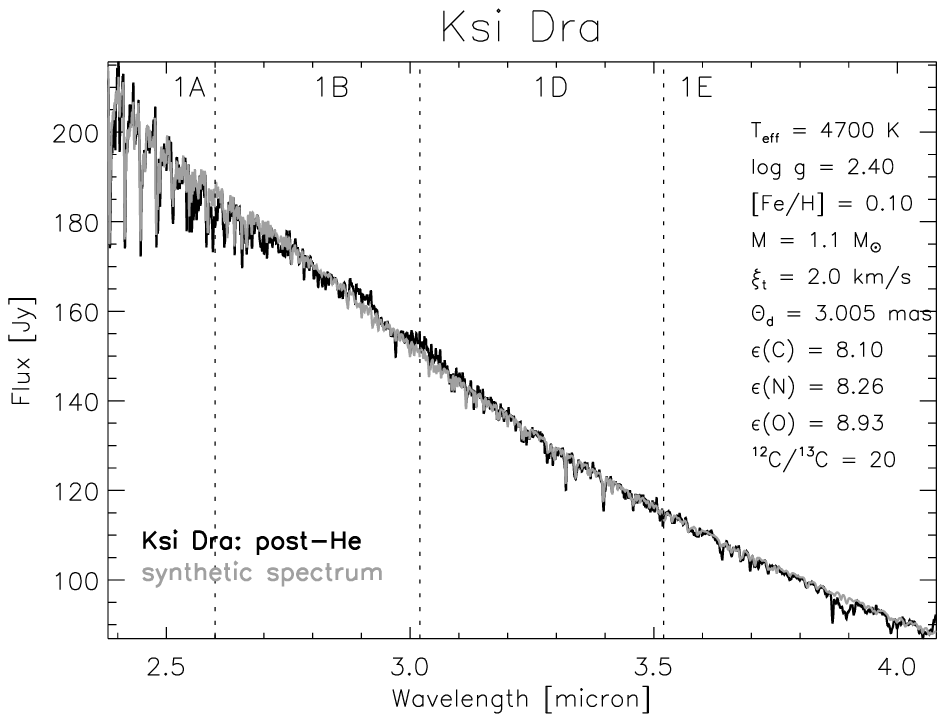}}}
\caption{\label{ksidra2} Comparison between band 1 of the
post-helium ISO-SWS data of $\xi$ Dra (black) and the synthetic
spectrum (grey) with stellar parameters \teff\ = 4440\,K, $\log$ g =
2.40, M = 1.2\,\Msun, [Fe/H] = $0.10$, \vt\ = 2.0\,\kms, \cc\ = 20,
$\varepsilon$(C) = 8.00, $\varepsilon$(N) = 8.26, $\varepsilon$(O)
= 8.93 and \ad\ = 3.125\,mas. A coloured version of this plot is
available in the Appendix as Fig.\ \ref{ksidra2col}.}
\end{center}
\end{figure}

The post-helium observation is also compared with a synthetic
spectrum generated with the same stellar parameters (but other \ad\
due to the higher absolute flux in the post-He observation) as in Fig.\
\ref{ksidra} at an appropriate resolution (Fig.\ \ref{ksidra2}).
The corresponding $\beta$-values are $\beta_{\mathrm{1A}} =
0.089$, $\beta_{\mathrm{1B}} = 0.105$, $\beta_{\mathrm{1D}} =
0.087$ and $\beta_{\mathrm{1E}} = 0.048$. The larger
$\beta$-values are due to the very strong Mg-Si-Al-Fe
feature around 2.55 \mic\ in band 1A \citepalias[see Fig.\ 3
in][]{Decin2000b} and the still ongoing calibration of the post-He
data. The spectrum resembles the spectrum of $\delta$ Dra.

\subsubsection{Comparison with other published stellar parameters \rm{(see Table \ref{litkdra} in the Appendix)}}

As for $\delta$ Dra, the temperature given by
\citet{Gustafsson1974A&A....34...99G} is lower than other
published values (see the comment given in Sect.\ \ref{compparddra}).
The different values quoted for the logarithm of the gravity are in
close agreement, with the maximum difference being
0.30\,dex and our deduced gravity value being close to the mean.
The lowest value was found by
\citet{Brown1989ApJS...71..293B} by using the K-line absolute
magnitude $M_V(K)$ and the highest value was obtained by
\citet{McWilliam1990ApJS...74.1075M} by using the well-known
relation between g, \teff, L and a mass value inferred from
evolutionary tracks.

The angular diameter deduced from the speed-3 observation tallies
very well with the IRFM angular diameter value given by
\citet{Blackwell1991A&A...245..567B} --- which is a revision of
\citet{Blackwell1990A&A...232..396B}.
Also \citet{Gratton1982MNRAS.201..807G} deduced a gravity-inferred
mass M$_g$, their value being somewhat higher than ours due to their use
of a higher radius value.

For the first time, the abundances of C, N and O are determined. Since
the spectroscopic analysis was done on the basis of a speed-3
observation and a post-He speed-4 observation, the error bar on the
abundances is estimated to be $\sim 0.30$\,dex.


\subsection{${\alpha}$ Boo{\rm: AOT01, speed 4, revolution 452}}\label{disaboo}

\subsubsection{Comparison between the ISO-SWS spectrum and the synthetic spectrum
(Fig.\ \ref{aboo})}

\begin{figure}[h!]
\begin{center}
\resizebox{0.5\textwidth}{!}{\rotatebox{90}{\includegraphics{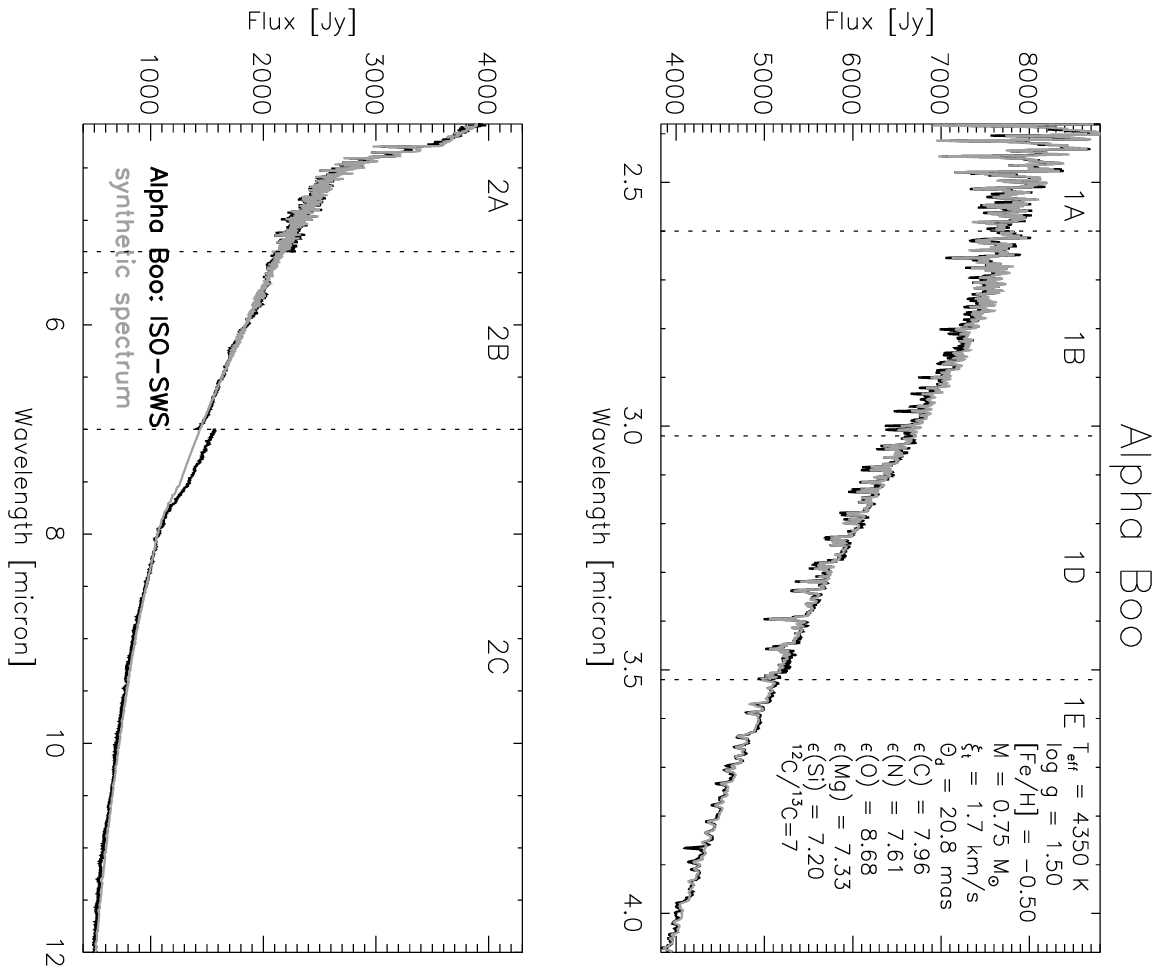}}}
\caption{\label{aboo} Comparison between band 1 and band 2 of the
ISO-SWS data (rev.\ 452) of $\alpha$ Boo (black) and the synthetic spectrum
(grey) with stellar parameters \teff\ = 4320\,K, $\log$ g = 1.50, M
= 0.75\,\Msun, [Fe/H] = $-0.50$, \vt\ = 1.7\,\kms, \cc\ = 7,
$\varepsilon$(C) = 7.96, $\varepsilon$(N) = 7.61, $\varepsilon$(O)
= 8.68, $\varepsilon$(Mg) = 7.33, $\varepsilon$(Si) = 7.20 and
\ad\ = 20.72\,mas. A coloured version of this plot is available in the
Appendix as Fig.\ \ref{aboocol}.}
\end{center}
\end{figure}

The higher standard deviation in band 1A and possible uncertainties
in the temperature structure in the outer layers determine the
maximal acceptable $\beta$-value in band 1A. The OH lines are predicted as
somewhat too weak \citepalias{Decin2000b}. From $\alpha$ Boo
on, the OH lines determine the spectral signature of band 1D. In
band 1E, one can see atomic features around 3.85 \mic\ \citepalias[see
Fig.\ 3 in][]{Decin2000b} and the SiO lines around 4 \mic. At the
beginning of this project, the first-overtone SiO lines were
always predicted as too strong. At that time, the line list of
\citet{Tipping1981} was used. An extensive theoretical
study \citep{Drira1997A&A...319..720D} and a comparison with observed SiO
lines \citep{Langhoff1993, Tsuji1994A&A...289..469T}
revealed that the Einstein coefficients based on the
results of \citet{Tipping1981} are about a factor two
larger for the 2-0 overtone band than the currently accepted value.
This explained the overestimate of the strength of the SiO
lines. Presently, the more accurate SiO line list of
\citet{Langhoff1993} is used, resulting
in a good agreement between observed and calculated spectra. The
fundamental bands of SiO and CO are visible in band 2, although no
quantitative statements can be made due to the problems with the
memory effects. The memory effects in band 2 are quite strong due
to the high flux level of $\alpha$ Boo in the infrared.

The high-quality data made it possible to pin down the atmospheric
parameters (see Paper~II). Where the first set of input parameters
contained as gravity $\log$ g = 1.75, the ISO-SWS (and FTS-KP) spectra
gave indications that this value could be refined to $\log$ g = 1.50.

\subsubsection{Comparison with other published stellar parameters
\rm{(see Table \ref{litaboo} in the Appendix)}}

The different temperature values deduced for $\alpha$ Boo vary around
4300\,K, with the lowest value being $4060$\,K
\citep{Scargle1979ApJ...228..838S} and the highest value $4628$\,K
\citep{Dyck1998AJ....116..981D}. A quite often quoted value is the one
of \citet{Frisk1982MNRAS.199..471F}, \teff = 4375\,K,  who used
photometry in the 0.4 -- 2.2\,\mic\ wavelength range in order to
determine the effective temperature of Arcturus.

Concerning the gravity, one notices that some publications give
quite a different gravity w.r.t. our deduced value (with a
difference $> 0.6$\,dex compared to $\log$ g = 1.50 deduced from
the ISO-SWS and FTS-KP spectra). The large wavelength range of
ISO-SWS, with the presence of many molecules, has proven to be
very useful for the determination of the gravity for the cool
giants.  Our deduced  value of $\log$ g = 1.50 is also
corroborated by the excellent agreement between the
FTS-KP-spectrum and the synthetic spectrum of $\alpha$ Boo
discussed in Paper~II. Using, e.g., a gravity of $\log$ g = 1.75,
leads to a gravity-inferred mass of M$_g = 1.30$\,\Msun. The
evolutionary tracks of \citet{Fagotto1994A&AS..105...29F} give
also support to a mass value around 0.8\,\Msun\ (for \teff =
4300\,K and L = 196\,\Lsun), suggesting the lower gravity value of
$\log$ g = 1.50.

All obtained values for the metallicity mention $ {\rm{[Fe/H]}}
\approx -0.50$. For the microturbulent velocity, a larger range
occurs, with a minimum of 1.5\,\kms\
\citep{Fernandez-Villacanas1990AJ.....99.1961F,
Aoki1997A&A...328..175A} and a maximum of 2.5\,\kms\
\citep{Gratton1985A&A...148..105G}. From studies on granulations,
it is clear that the use of a single parameter to represent the
non-thermal velocity field is highly phenomenological and thus, it
is not surprising that large differences in deduced values for the
microturbulence are obtained.

For the carbon abundance, most of the studies give [C/Fe] being solar
or somewhat sub-solar ($\le 0.005$), with the only exception being
\citet{Altas1987Ap&SS.134...85A}, who deduced an abnormal high carbon
abundance using the Griffin atlas \citep{Griffin1968} and the model
derived by  \citet{Mackle1975A&AS...19..303M}. Also for the
nitrogen abundance, the value mentioned by
\citet{Altas1987Ap&SS.134...85A} exceeds the other deduced values,
although here the spread is higher than for the
carbon abundance. We obtained $\varepsilon$(O) using OH-lines. Our
deduced value agrees (within the error bar) with the ones
obtained by  \citet{Lambert1981ApJ...248..228L},
\citet{Kjaergaard1982A&A...115..145K},
\citet{Altas1987Ap&SS.134...85A} and
\citet{Peterson1993ApJ...404..333P}. Only the value obtained by
\citet{Mackle1975A&AS...19..303M} is substantially lower than our
value. Within the error bars, our \cc\ ratio agrees with other
published values.

Angular diameter, luminosity and radius values are in good agreement
with other values found in the literature.

Quite a few authors have estimated the gravity-inferred mass, e.g.\
\citet{Mackle1975A&AS...19..303M}, \citet{Gratton1985A&A...148..105G},
\citet{Harris1988ApJ...325..768H}, \citet{Judge1991ApJ...371..357J},
\citet{Peterson1993ApJ...404..333P}. Differences in assumed/ adopted/ deduced
gravity, radius, angular diameter or parallax value explain the
observed differences in M$_g$.

In general, the atmospheric parameters we
derived for $\alpha$ Boo agree with the ones found by \citet{Peterson1993ApJ...404..333P}.
They have obtained the stellar parameters by computing a
model atmosphere which reproduces the observed flux distribution
of the Griffin atlas \citep{Griffin1968} of $\alpha$ Boo. Accurate
model atmospheres, opacities, line lists, ... are indispensable
for a good assessment of the stellar parameters. Minor
inaccuracies in photometric calibration formulae, in broadening
parameters, in the assumptions made in the theoretical modelling
computations, ... may lead to statistically relevant departures of
the derived parameters. One should be careful when ascribing
an atmospheric parameter based on the spectral type \citep[like done by,
e.g., ][]{Cohen1996AJ....112.2274C}. This is already obvious when comparing
the parameters of $\alpha$ Boo and $\xi$ Dra, which are both K2
giants.


\subsection{${\alpha}$ Tuc{\rm: AOT01, speed 4, revolution 866}}\label{disatuc}

\subsubsection{Comparison between the ISO-SWS spectrum and the synthetic spectrum
(Fig.\ \ref{atuc})}

\begin{figure}[h!]
\begin{center}
\resizebox{0.5\textwidth}{!}{\rotatebox{90}{\includegraphics{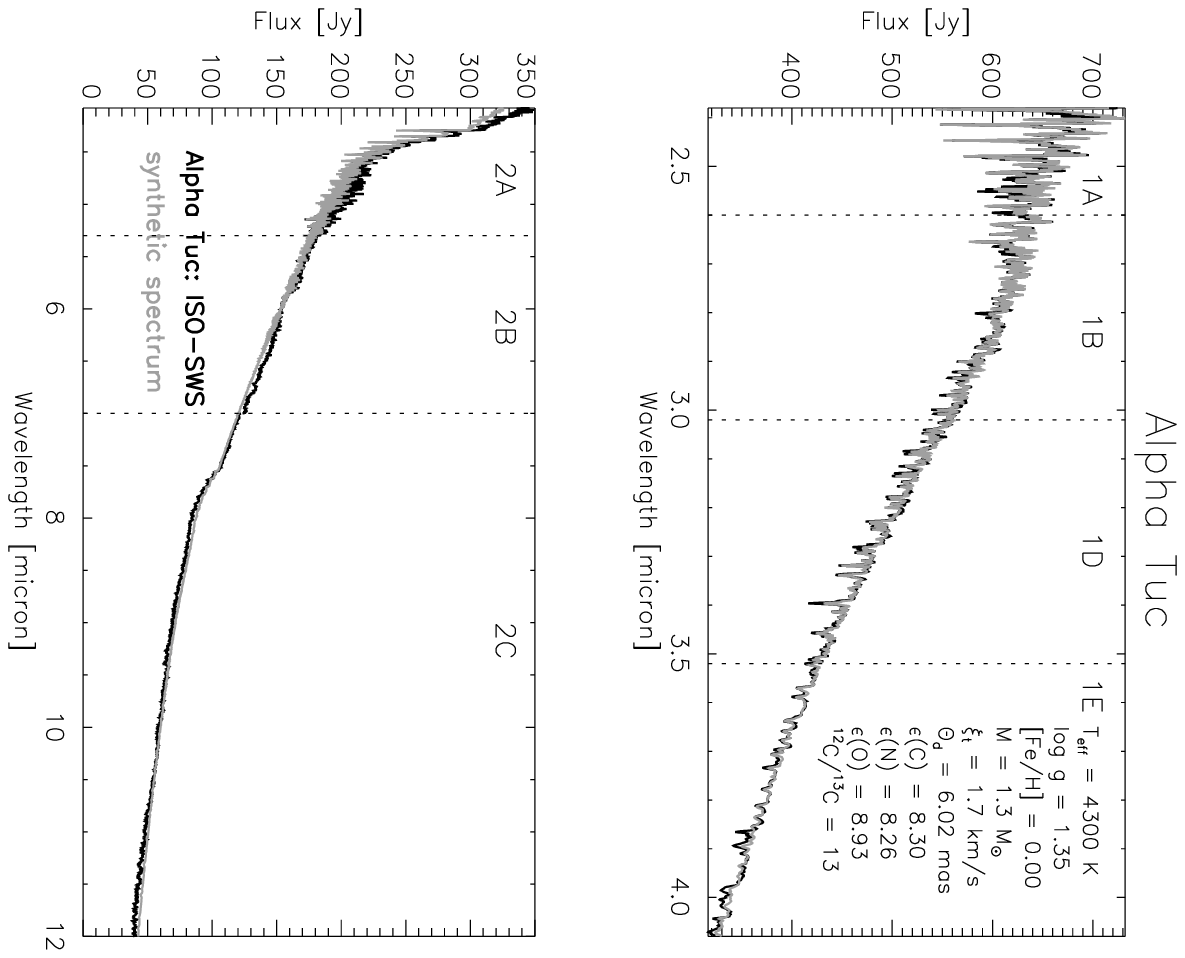}}}
\caption{\label{atuc} Comparison between band 1 and band 2 of the
ISO-SWS data (rev.\ 866) of $\alpha$ Tuc (black) and the synthetic spectrum
(grey) with stellar parameters \teff\ = 4300\,K, $\log$ g = 1.35, M
= 1.3\,\Msun, [Fe/H] = 0.00, \vt\ = 1.7\,\kms, \cc\ = 13,
$\varepsilon$(C) = 8.30, $\varepsilon$(N) = 8.26, $\varepsilon$(O)
= 8.93 and \ad\ = 6.02\,mas. A coloured version of this plot is
available in the Appendix as Fig.\ \ref{atuccol}.}
\end{center}
\end{figure}

The somewhat higher $\beta$-value in band 1A is partially due to a
uncertainties in the temperature structure in the outer layers
\citepalias{Decin2000b}
and partially due to the higher standard deviation in
band 1A. The atomic features around 3.85 \mic\ and 3.98 \mic\ remain visible
\citepalias[see Fig.\ 3 in ][]{Decin2000b}. The OH features are somewhat more
pronounced than for $\alpha$ Boo. The SiO fundamental and
first-overtone band, together with the CO fundamental band, arise
in band 2.

\subsubsection{Comparison with other published stellar parameters
\rm{(see Table \ref{litatuc} in the Appendix)}}

Only a few published values for the different stellar parameters are
available. The temperature derived from the ISO-SWS spectrum and the
$V-K$ colour is somewhat higher than the effective temperature derived from the
($V-R$) colour index \citep{Basri1979ApJ...234.1023B,
Pasquini1992A&A...266..340P} and from other broad-band photometry
\citep{Glebocki1988A&A...189..199G}. Using the Johnson $B$, $V$, $R$
and $I$ colour measurements from \citet{Mendoza1969BOTT....5...57M} as
listed by Simbad and the colour-temperature relations of
\citet{McWilliam1990ApJS...74.1075M}, an effective temperature of
4080\,K is obtained for $\alpha$ Tuc. The used $K$ magnitude of
$-0.09$\,mag listed by
\citet{Wamsteker1981A&A....97..329W} could not be checked against
other literature values, since no other values were found.

There is however a large difference between
the two listed gravities. \citet{Glebocki1988A&A...189..199G} have
estimated the mass from the mass-function  and the position of the
primary on the evolutionary track
calculations. The radius R was estimated from the bolometric
magnitude and the effective temperature deduced from infrared
broad-band photometry. \citet{Pasquini1992A&A...266..340P} have not
taken the binary character of $\alpha$ Tuc
into account, but have deduced the gravity from \Mbol\ and a mass
inferred from the comparison with evolutionary tracks. Due to a raw
division of the stellar masses into four sub-groups, their gravity is
the parameter containing the major uncertainties, which explains the
difference with our first-guess value for $\log$ g, derived from the
same \Mbol\ value, but using another M$_{hr}$ value. From the
ISO-SWS spectrum and the corresponding $\beta$-values, we could
draw the conclusion that a gravity of 1.90 is too high.

Our deduced angular diameter of $6.02 \pm 0.37$\,mas corresponds with the value
given by \citet{Basri1979ApJ...234.1023B}, who have used a relation which
links the colour $V-R$ to the angular diameter. These two values
are higher than the value mentioned by \citet{Stencel1980ApJS...44..383S}. Both
\citet{Basri1979ApJ...234.1023B} and \citet{Stencel1980ApJS...44..383S} have
used the scaling laws described by \citet{Barnes1976MNRAS.174..489B} and \citet{Barnes1976MNRAS.174..503B} to convert fluxes
observed on the Earth to fluxes at the stellar surface. Since this scaling law
is based on the photometric colour ($V-R$), an uncertainty in ($V-R$) can induce
a significant error. The photometric colour ($V-R$) of $\alpha$ Tuc has been
quoted as uncertain in both papers (($V-R$) = 0.9 in
\citet{Stencel1980ApJS...44..383S} and
($V-R$) = 1.04 in \citet{Basri1979ApJ...234.1023B}), explaining the large
difference in \ad.

The high-quality ISO-SWS spectra of $\alpha$ Tuc gave us the
possibility to determine for the first time the C, N and O abundances
and the \cc-ratio for this target.


\subsection{${\beta}$ UMi{\rm: AOT01, speed 4, revolution 182}}\label{disbumi}

\subsubsection{Comparison between the ISO-SWS spectrum and the synthetic spectrum
(Fig.\ \ref{bumi})}

\begin{figure}[h!]
\begin{center}
\resizebox{0.5\textwidth}{!}{\rotatebox{90}{\includegraphics{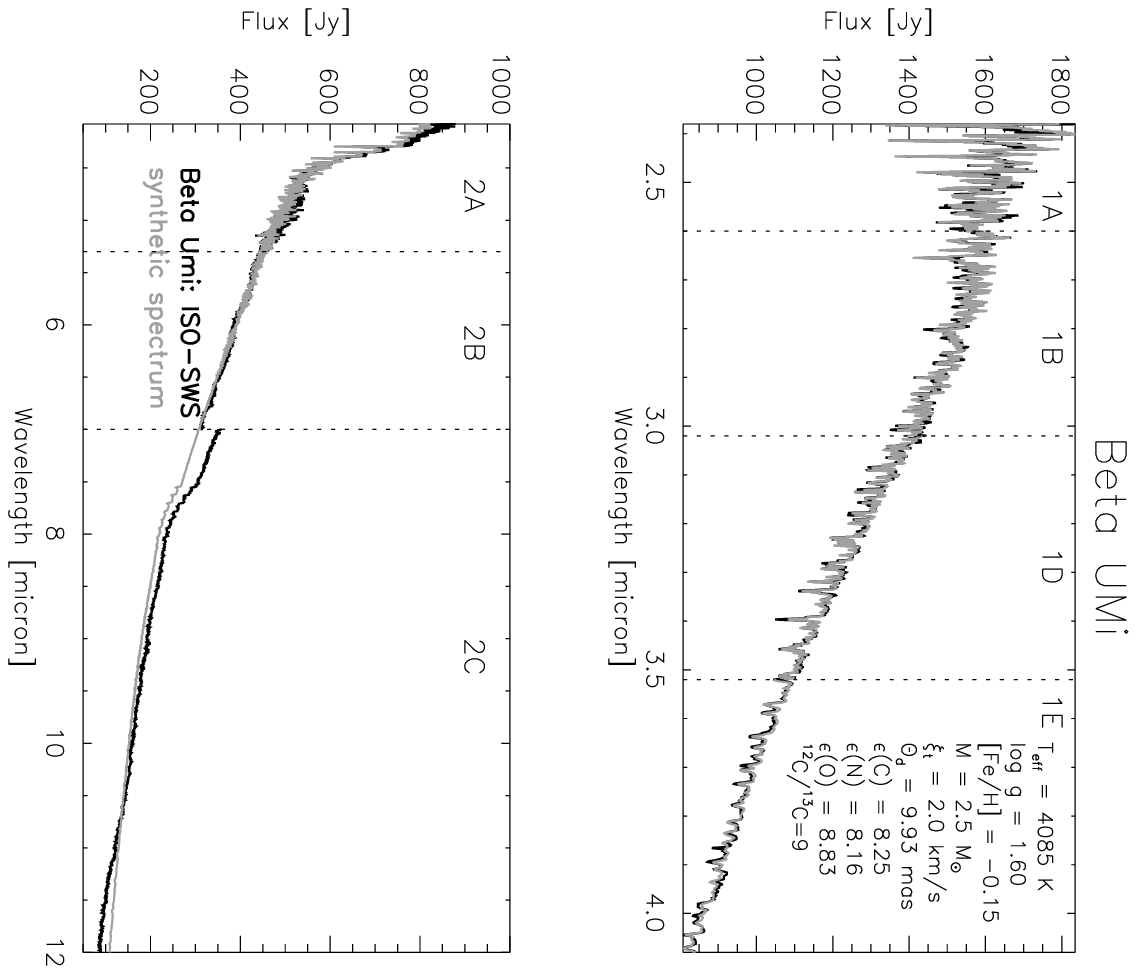}}}
\caption{\label{bumi} Comparison between band 1 and band 2 of the
ISO-SWS data (rev.\ 182) of $\beta$ UMi (black) and the synthetic spectrum
(grey) with stellar parameters \teff\ = 4085\,K, $\log$ g = 1.60, M
= 2.50\,\Msun, [Fe/H] = $-0.15$, \vt\ = 2.0\,\kms, \cc\ = 9, $\varepsilon$(C) =
8.25, $\varepsilon$(N) = 8.16, $\varepsilon$(O) = 8.83 and \ad\ =
9.93\,mas. A coloured version of this plot is available in the
Appendix as Fig.\ \ref{bumicol}.}
\end{center}
\end{figure}

Band 1A has the largest $\beta$-value. For band 1A, the same
reasons as for the previous stars can be quoted. The discrepancy
between 3.91 and 3.97~\mic\ causes the somewhat larger
$\beta_{\mathrm{1E}}$-value. One can notify an analogous
discrepancy in Fig.\ \ref{bumivers} in the Appendix. The pointing offset
($|$dy$| > 1''$), together with a somewhat
larger standard deviation after rebinning for the speed-4
observation in that wavelength range, render this discrepancy not significant.
The OH spectral features dominate band 1D and band 1E, while the
first-overtone band of SiO is clearly visible around 4 \mic. The
CO fundamental and SiO fundamental and first-overtone bands remain
visible in band 2.

\subsubsection{Comparison with other published stellar parameters \rm{(see Table \ref{litbumi} in the Appendix)}}\label{compparbumi}

The published temperature values all agree very well, the
only exception being the value derived by \citet{Lambert1981ApJ...248..228L}
and used by some other authors. As quoted by \citet{Ries1981PhDT.........4R},
\citet{Harris1988ApJ...325..768H} and \citet{Luck1995AJ....110.2968L}, the
effective temperature and gravity derived by
\citet{Lambert1981ApJ...248..228L} are too high and should be lowered
by $\sim 240$\,K and $\sim 0.40$\,dex respectively. Taking this remark
into account and knowing that the gravity values quoted by
\citet{Burnashev1983IzKry..67...13B} differ often with other
values (see Sect.\ \ref{compparddra}), our deduced gravity is
in good agreement, although being at the lower end, with other deduced
gravity values. The metallicity and
microturbulent velocity agree with the other values listed in Table
\ref{litbumi}. The \cc\ ratio derived from the ISO-SWS spectrum
is somewhat lower than other published values, but the difference
is not significant due to the large standard deviation
and the other problems discussed in \citetalias{Decin2000b} for band 1A.
The angular diameter from the SWS-spectrum lies in between the values
quoted by \citet{Judge1991ApJ...371..357J} and
\citet{Stencel1980ApJS...44..383S}. Being somewhat larger than the value
given by \citet{Judge1991ApJ...371..357J}, our deduced values
of R, M$_g$ and L are also higher than those listed by
\citet{Judge1991ApJ...371..357J}.


\subsection{${\gamma}$ Dra{\rm: AOT01, speed 4, revolution 377}}\label{disgdra}

\subsubsection{Comparison between the ISO-SWS spectrum and the synthetic spectrum
(Fig.\ \ref{gamdra})}

\begin{figure}[h!]
\begin{center}
\resizebox{0.5\textwidth}{!}{\rotatebox{90}{\includegraphics{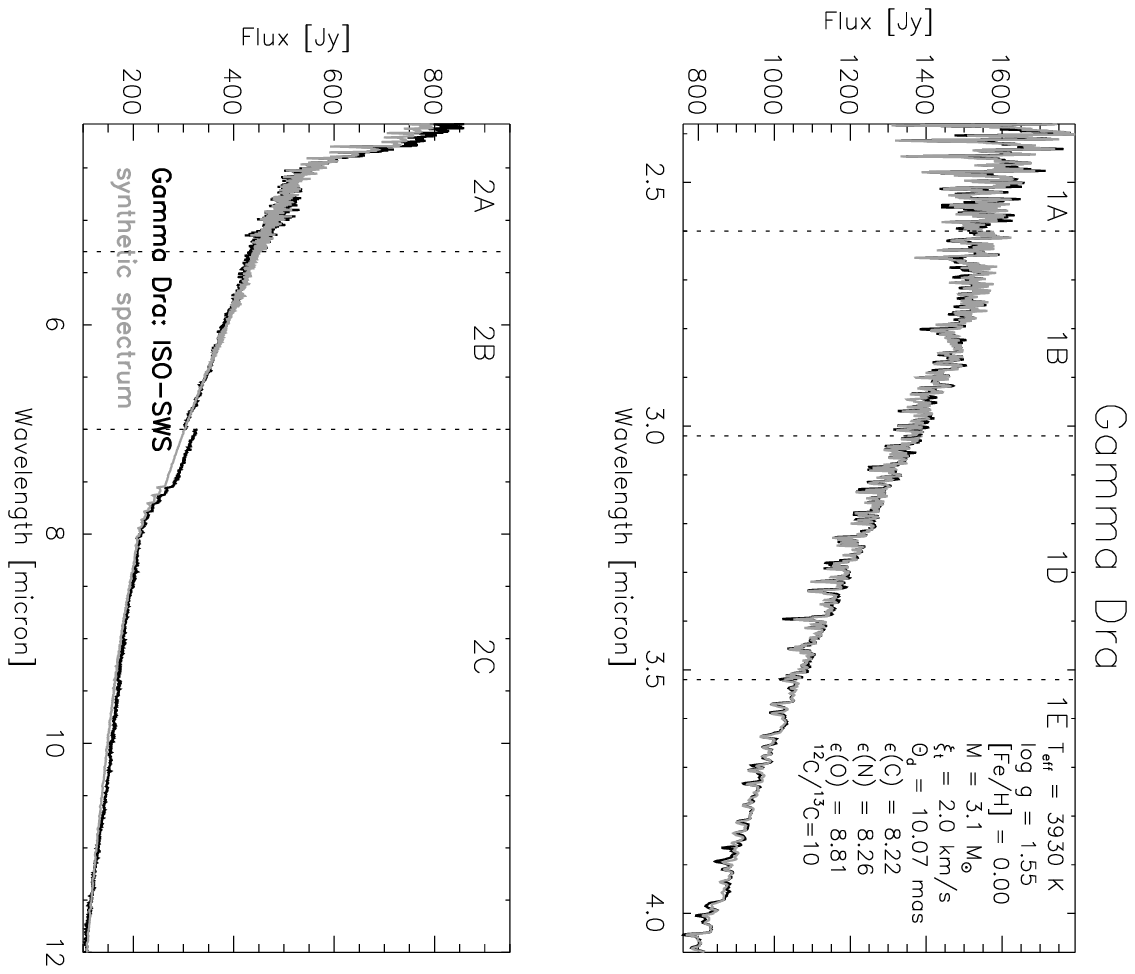}}}
\caption{\label{gamdra} Comparison between band 1 and band 2 of
the ISO-SWS data (rev.\ 377) of $\gamma$ Dra (black) and the synthetic
spectrum (grey) with stellar parameters \teff\ = 3960\,K, $\log$ g =
1.30, M = 1.7\,\Msun, [Fe/H] = 0.00, \vt\ = 2.0\,\kms, \cc\ = 10,
$\varepsilon$(C) = 8.15, $\varepsilon$(N) = 8.26, $\varepsilon$(O)
= 8.93 and \ad\ = 9.98\,mas. A coloured version of this plot is
available in the Appendix as Fig.\ \ref{gamdracol}.}
\end{center}
\end{figure}

It was difficult to decide between $\log$ g $\approx 1.20 $ and $\log$
g $\approx 1.55$ for $\gamma$ Dra, since the Kolmogorov-Smirnov test ---
applied in the region from 2.8 - 3.2 \mic, where the
synthetic spectrum is quite sensitive to the gravity --- yields almost
the same $\beta$-value. Since a gravity of $\log$ g = 1.30 was
obtained from $K$, BC$_K$, $\pi$, \teff($V-K$) and M$_{hr}$, this
gravity was chosen to represent the `true' gravity of the star.
The small difference between the two $\beta$-values is reflected
in the higher uncertainty of $\log$ g in Table
\ref{parameters}. The other comments for $\gamma$ Dra are similar to the ones
for $\alpha$ Boo.

\subsubsection{Comparison with other published stellar parameters}

Different methods like IRFM, colour indices, interferometry, the
Engelke function \citep{Engelke1992AJ....104.1248E}, ... almost
all yield  the same effective
temperature for $\gamma$ Dra. The value given by
\citet{Tomkin1975ApJ...199..436T} is quite low, which may be due to inaccurate
colour indices or calibration. We note that
\citet{Faucherre1983A&A...120..263F}, who derived effective
temperatures from interferometric measurements of angular diameters,
underestimated the effective temperature of $\delta$ Dra compared to
our value, but for
$\gamma$ Dra obtained a value slightly higher than ours.  As
for $\alpha$ Boo, the gravity deduced by \citet{Cohen1996AJ....112.2274C} from
the spectral type seems to be erroneous.

The reference quoted by \citet{Harris1984ApJ...285..674H} for the
determination of the atmospheric parameters could not be traced
back. The only two other deduced values for the gravity listed in Table
\ref{litgamdra} are $\log$ g $\simeq 1.55$ and $\log$ g $\simeq
1.20$. As described in previous paragraph, the $\beta$-values from
the Kolmogorov-Smirnov test for these two gravity values are quite alike.

Our deduced value for the angular diameter \ad\ = $9.98 \pm
0.63$\,mas is in close agreement with the values obtained by using the
IRFM method \citep{Blackwell1977MNRAS.180..177B,
Leggett1986A&A...159..217L, Bell1989MNRAS.236..653B} or by using the
Engelke function as was done by
\citet{Cohen1996AJ....112.2274C}. While our results and the values
adopted by \citet{Robinson1998ApJ...503..396R} for the stellar radius
are based on the Hipparcos' parallax and the deduced (or adopted)
angular diameter, \citet{Leggett1986A&A...159..217L} used the
parallax value given by \citet{Jenkins1963GCTP..C......0J} ($\pi = 17
\pm 6$\,mas). Consequently, their deduced radius (and luminosity) is
higher than ours.

Only \citet{Harris1988ApJ...325..768H} and
\citet{Robinson1998ApJ...503..396R} have estimated a gravity-inferred
mass from adopted $g$ and R values. The value of
\citet{Robinson1998ApJ...503..396R} is higher than our
derived M$_g$ value due to the use of a higher gravity, while the
value obtained by
\citet{Harris1988ApJ...325..768H} (M = 0.9\,\Msun) is much lower. This
low value was discarded by \citet{Harris1988ApJ...325..768H},
who instead adopted M = 2.0\,\Msun.

C, N and O abundances agree fairly well with the ones obtained by
\citet{Lambert1981ApJ...248..228L}.


\subsection{${\alpha}$ Tau{\rm: AOT01, speed 4, revolution 636}}\label{disatau}

$\alpha$ Tau has been discussed in \citetalias{Decin2000A&A...364..137D}.


\subsection{H Sco{\rm: AOT01, speed 4, revolution 847}}\label{dishsco}

\subsubsection{Comparison between the ISO-SWS spectrum and the synthetic spectrum
(Fig.\ \ref{hd149})}

\begin{figure}[h!]
\begin{center}
\resizebox{0.5\textwidth}{!}{\rotatebox{90}{\includegraphics{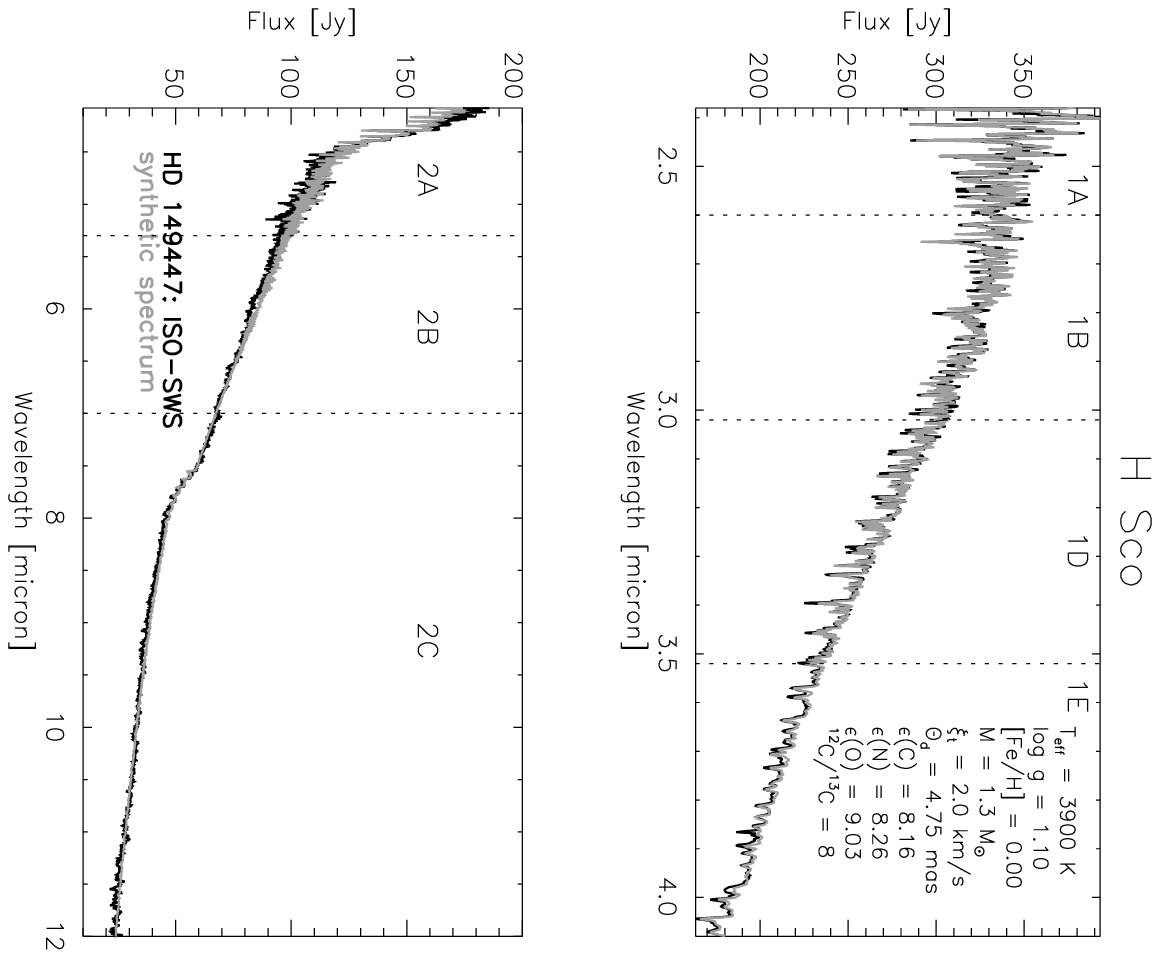}}}
\caption{\label{hd149} Comparison between band 1 and band 2 of the
ISO-SWS data (rev.\ 847) of H Sco (black) and the synthetic spectrum (grey)
with stellar parameters \teff\ = 3900\,K, $\log$ g = 1.30, M =
2.0\,\Msun, [Fe/H] = 0.00, \vt\ = 2.0\,\kms, \cc\ = 8, $\varepsilon$(C) =
8.20, $\varepsilon$(N) = 8.26, $\varepsilon$(O) = 8.83 and \ad\ =
4.73\,mas. A coloured version of this plot is available in the
Appendix as Fig.\ \ref{hd149col}.}
\end{center}
\end{figure}

The use of the initial values for the effective temperature and
gravity of \teff\,=\,3750\,K and $\log$ g\,=\,1.00 resulted in
synthetic spectrum being very discrepant from the ISO-SWS
spectrum. From the slope of the ISO-SWS data between 2.8 and
4.08\,\mic, the effective temperature was found to be higher by $\sim
150$\,K. The logarithm of the gravity had to be increased with
0.3\,dex. The used $K$ magnitude to derive the initial values of
\teff\ and $\log$ g was quoted to be uncertain
\citep{Price1968AJ.....73..431P}, which may result in initial values
being quite off.

From H Sco on, the spectra are dominated by water lines. It
has been calculated that at a temperature of about 3900~K the
concentrations of water and OH are equal; below this temperature,
water is the dominant oxygen-containing molecule as well as the
dominant opacity source \citep{Wallace1995Sci...268.1155W}.

\subsubsection{Comparison with other published stellar parameters}\label{lithd}

HD~149447 is classified by Simbad as a K6~III
giant. \citet{Perrin1998A&A...331..619P} estimated the effective
temperature for a K5III giant (with
$V-K = 3.52$) as being \teff$\ = 3980$\,K and for an M0~III giant
(with $V-K = 3.78$) as \teff$\ = 3895$\,K. \citet{Gray1992} has, however,
listed somewhat lower temperatures, i.e.\ \teff\ = 3915\,K for a K5~III
giant and \teff\ = 3726\,K for an M0~III giant. Using the Johnson
$B$, $V$, $R$ and $I$ colour measurements listed by
\citet{Johnson1966CoLPL...4...99J} and the colour-temperature relations of
\citet{McWilliam1990ApJS...74.1075M}, we derived an effective
temperature \teff $\approx 3900$\,K. These \teff\ values support our
\teff\ value estimated from the ISO-SWS spectrum of H Sco.

The quantities $\log$ g, mass, \vt, [Fe/H], CNO-abundances,
\cc-ratio, \ad, L and R are here derived for the first time for this
stellar source.

\subsection{${\beta}$ And{\rm: AOT01, speed 4, revolution 795}}\label{disband}

\subsubsection{Comparison between the ISO-SWS spectrum and the synthetic spectrum
(Fig.\ \ref{band})}

\begin{figure}[h!]
\begin{center}
\resizebox{0.5\textwidth}{!}{\rotatebox{90}{\includegraphics{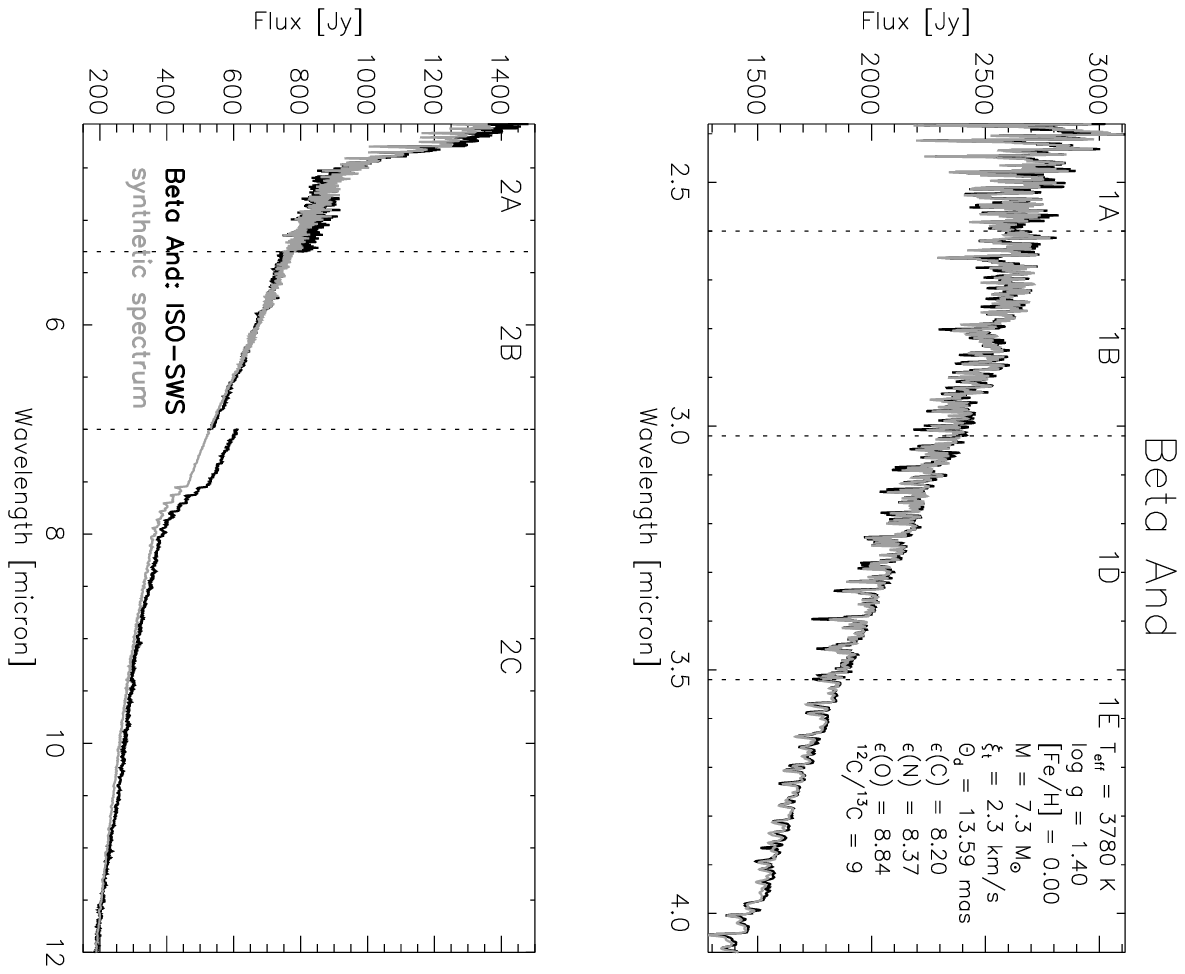}}}
\caption{\label{band} Comparison between band 1 and band 2 of the
ISO-SWS data (rev.\ 795) of $\beta$ And (black) and the synthetic spectrum
(grey) with stellar parameters \teff\ = 3880\,K, $\log$ g = 0.95, M
= 2.5\,\Msun, [Fe/H] = 0.00, \vt\ = 2.0\,\kms, \cc\ = 9, $\varepsilon$(C)
= 8.12, $\varepsilon$(N) = 8.37, $\varepsilon$(O) = 9.08 and \ad\
= 13.30\,mas. A coloured version of this plot is available in the
Appendix as Fig.\ \ref{bandcol}.}
\end{center}
\end{figure}

Just like the K-giants, this `normal' M-giant displays the
molecular features and the discrepancies summarised in Paper~I.
However, H$_2$O, rather than OH, has now become the main absorber
in the wavelength region from 2.8 \mic\ to 4 \mic, although the OH
lines remain very well visible.

\subsubsection{Comparison with other published stellar parameters
\rm{(see Table \ref{band} in the Appendix)}}

As for $\gamma$ Dra and $\alpha$ Tau, the effective temperatures
given by \citet{Tomkin1975ApJ...199..436T, Linsky1978ApJ...220..619L,
Scargle1979ApJ...228..838S} are quite low compared to the other values
derived from different methods. The interferometric value by
\citet{Dyck1998AJ....116..981D} is the highest \teff\ value in Table
\ref{litband}.

From $K$, BC$_K$, $\pi$, \teff($V-K$) and M$_{hr}$ an initial
input gravity of $\log$ g = 0.95 was derived. Using this gravity
value, \teff\ = 3880\,K, \vt\ = 2.0 \kms, [Fe/H] = 0.00,
$\varepsilon$(C) = 8.12, $\varepsilon$(N) = 8.37, $\varepsilon$(O)
= 9.08 and \cc\ = 9, a good agreement with the observed ISO-SWS
spectrum was obtained. This gravity value of $\log$ g = 0.95 is,
however, much lower than most of the other quoted values in Table
\ref{litband}. Examining the other gravity values, we see that
\begin{itemize}
\item{\citet{Clegg1979ApJ...234..188C} have derived $\log$ g from an
{\it assumed} mass of 1\,\Msun, a \teff\ value derived from
spectral type and colours and have estimated a luminosity value using the
HR diagram of \citet{Scalo1976ApJ...206..474S}. Our deduced M$_{hr}$
of 2.2\,\Msun would result in a higher luminosity and thus in a lower
gravity.}
\item{Both \citet{Smith1985ApJ...294..326S} and
\citet{Lazaro1991MNRAS.249...62L} have estimated the effective
temperature from the $V-K$ colour and the luminosity
from M$_V$($K$) obtained via the Wilson-Bappu effect and
BC$_K$. Comparing, however, these M$_V$($K$) values from
\citet{Wilson1976ApJ...205..823W} with M$_V$ values obtained using the
$V$ colour given by Simbad and the Hipparcos' parallax, we note that
these M$_V$($K$) are {\it systematically} higher than our derived M$_V$
values (with a mean difference of 0.73\,mag). Using another
sample of cool giants, \citet{Tsuji1981A&A....99...48T} compared the
results from the Wilson-Bappu effect with those by
\citet{Eggen1973PASP...85..379E}, showing that the results from the
Wilson-Bappu effect are systematically $\sim 0.5$\,mag fainter. Hence,
\citet{Tsuji1981A&A....99...48T} decided to use the mean of the two
values. Using too high an \Mbol\ value (by $\sim 0.73$\,mag) results in too
high a $\log$ g value (by $\sim 0.3$\,dex). Moreover, taking these
questionable values for the luminosity into account, the mass values
estimated by \citet{Smith1985ApJ...294..326S} and
\citet{Lazaro1991MNRAS.249...62L} using evolutionary tracks are
meaningless. In addition, \citet{Lazaro1991MNRAS.249...62L} have
estimated stellar parameters
per spectral type, and so have adopted a mean mass of
1.5\,\Msun. Although having the same spectral type, stellar objects
may differ significantly in stellar parameters.}
\item{\citet{Judge1991ApJ...371..357J} have estimated the gravity from
photospheric line studies, but these results could not be traced
back.}
\end{itemize} Thus, we may conclude that the other quoted $\log$ g
values in Table \ref{litband} are either too high or are meaningless.

The derived angular diameter tallies with other indirectly
determined angular diameters given in Table \ref{litband}. The use of
old parallax values by \citet{Faucherre1983A&A...120..263F,
Volk1989AJ.....98.1918V, Judge1991ApJ...371..357J} --- which may
be almost a factor $3$ higher than the Hipparcos' parallax ---
resulted, however, in a radius value much higher than
any literature values. As a consequence, our deduced luminosity
exceeds the other quoted values.
Since our deduced gravity is at the lower end of the other quoted
gravity values, our deduced C-abundance is also somewhat lower (see
Paper~I). The abundances of nitrogen and oxygen agree with other quoted values.


\subsection{${\alpha}$ Cet{\rm: AOT01, speed 4, revolution 797}}\label{disacet}

\subsubsection{Comparison between the ISO-SWS spectrum and the synthetic spectrum
(Fig.\ \ref{acet})}

\begin{figure}[h!]
\begin{center}
\resizebox{0.5\textwidth}{!}{\rotatebox{90}{\includegraphics{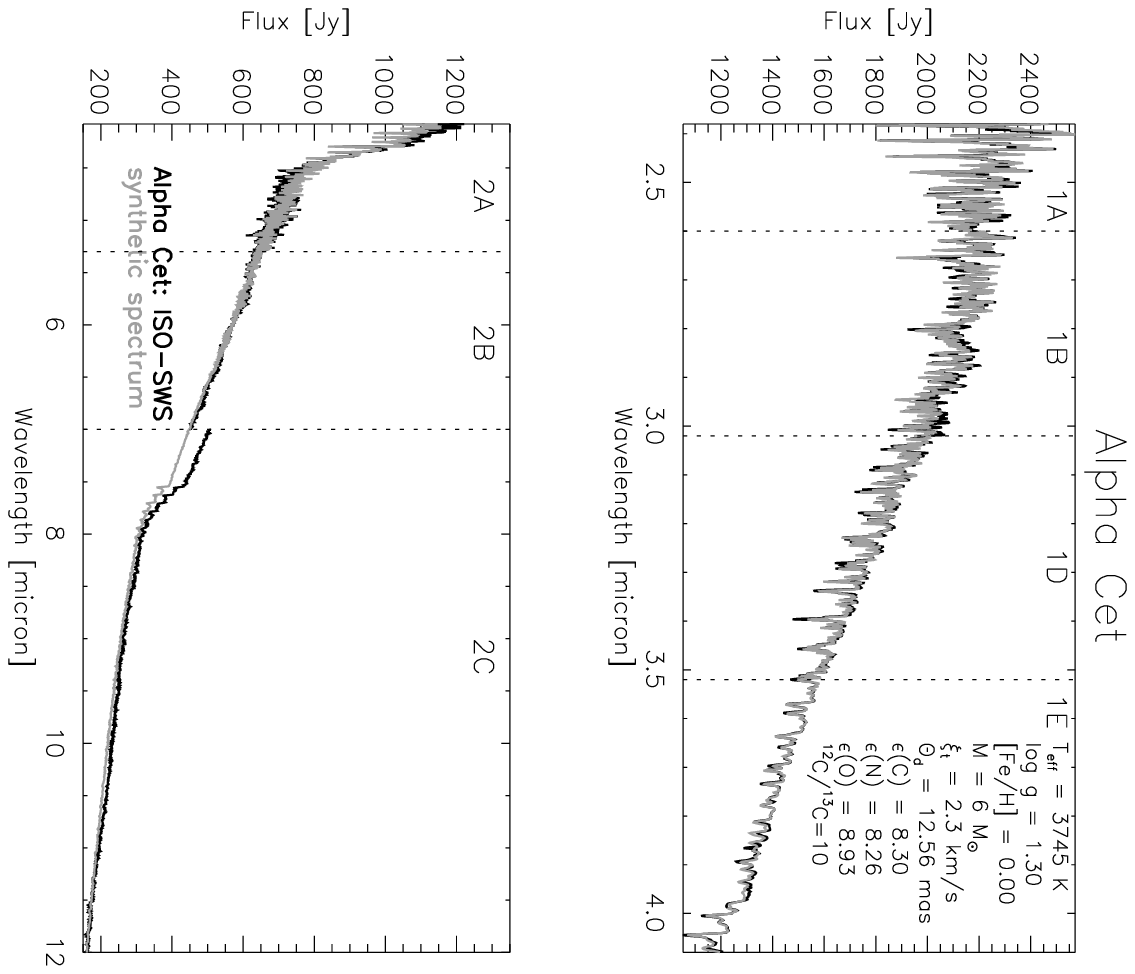}}}
\caption{\label{acet} Comparison between band 1 and band 2 of the
ISO-SWS data (rev.\ 797) of $\alpha$ Cet (black) and the synthetic spectrum
(grey) with stellar parameters \teff\ = 3740\,K, $\log$ g = 0.95, M
= 2.7\,\Msun, [Fe/H] = 0.00, \vt\ = 2.3\,\kms, \cc\ = 10, $\varepsilon$(C)
= 8.20, $\varepsilon$(N) = 8.26, $\varepsilon$(O) = 8.93 and \ad\
= 12.52\,mas. A coloured version of this plot is available in the
Appendix as Fig.\ \ref{acetcol}.}
\end{center}
\end{figure}

For this cool giant, the spectrum resulting from the
computations with a standard stellar model, i.e.\ without  a
stellar wind, shows a very good agreement
with the ISO-SWS spectrum of $\alpha$ Cet. The
discrepancies described in Paper~I occur also for
$\alpha$ Cet. The small atomic feature around 3.85\,\mic\ remains
marginally visible.

\subsubsection{Comparison with other published stellar parameters
\rm{(see Table \ref{litacet} in the Appendix)}}

\citet{Napiwotzki1993A&A...268..653N} quoted that the
resulting IRFM temperatures are too low by 1.6 -- 2.8\,\%, but the
effective temperature found by \citet{Tsuji1981A&A....99...48T} is
$\sim 160$\,K higher than the effective temperature deduced from the
$V-K$ colour and confirmed by the ISO-SWS data. Using improved H$^-$ opacity,
\citet{Blackwell1991A&A...245..567B} derived an effective temperature
from the IRFM being 3745\,K, which matches our derived temperature value.

Concerning the quoted gravity values in Table \ref{litacet}, the same
comments can be given as for $\beta$ And. The use of too low a
luminosity value results in too high a gravity value
\citep{Clegg1979ApJ...234..188C, Lazaro1991MNRAS.249...62L,
Tsuji1986A&A...156....8T}. Moreover, assuming a stellar mass
(1\,\Msun\ by \citet{Clegg1979ApJ...234..188C}, 3\,\Msun\ by
\citet{Tsuji1986A&A...156....8T}, 1.5\,\Msun\ by
\citet{Lazaro1991MNRAS.249...62L}) based on the spectral type, can
introduce significant errors on the gravity. The obtained value by
\citet{Judge1991ApJ...371..357J} could not be traced
back. \citet{Cohen1996AJ....112.2274C} used a statistical relation
between spectral type and gravity which may result in inaccurate
gravity values taking into account the divergent stellar parameters
for different targets having the same spectral type (cfr.\ e.g.\
$\alpha$ Cet and $\beta$ Peg). \citet{Burnashev1983IzKry..67...13B}
has derived gravity values from narrow-band photometric colours. As
can be seen from Table \ref{litacet}, the obtained values differ
largely (see also the remarks given in Sect.\ \ref{compparddra} and
\ref{compparbumi}). Hence, once more, we may conclude that the other
quoted gravity values in Table \ref{litacet} are not accurate.

Our angular diameter deduced from the ISO-SWS spectrum agrees quite
well with other angular diameters deduced using indirect methods
\citep[e.g.\ ][]{Blackwell1991A&A...245..567B,
Judge1991ApJ...371..357J}. \citet{Judge1991ApJ...371..357J} used a
somewhat lower parallax value quoted by
\citet{Jenkins1952QB813.J45......} (which differs by a factor $\sim 1.2$ with
the Hipparcos' parallax), resulting in a higher radius and luminosity
value than ours.

Our deduced carbon abundance differs somewhat from the values
deduced by \citet{Tsuji1991A&A...245..203T}, who used
second-overtone CO-lines. These values are higher than the values
inferred from first-overtone CO-lines by
\citet{Tsuji1986A&A...156....8T} and
\citet{Lazaro1991MNRAS.249...62L}.
\citet{Tsuji1991A&A...245..203T} discussed this problem in
conjunction with the smaller \vt, deduced from the
second-overtone, than from the first-overtone CO lines. Probably,
the origin of the discrepancy between the carbon abundances
derived from the more or less saturated lines of the CO
first-overtone and those from the less saturated second-overtone
lines can, at least partly, be explained by an over-simplification
of the treatment of the microturbulent velocity, for which in
classical models a depth-independent, isotropic Gaussian model is
taken. Even in classical abundance analysis, an error of only
0.5\,\kms\ in \vt\ produces an error as large as 0.3\,dex in the
derived abundances. Since line asymmetries in cool giant stars are
as large as 0.5\,\kms, a poor modelling of the turbulence may also
produce similar or even larger errors on the abundances derived
from saturated lines. Our carbon abundance is, however, deduced
from first-overtone lines. Its higher value w.r.t.\ the value
given by \citet{Tsuji1986A&A...156....8T} may be explained by our
lower \vt. Moreover, when using the parameter values as deduced by
\citet{Tsuji1981A&A....99...48T,
Tsuji1986A&A...156....8T,Tsuji1991A&A...245..203T} a bad
correspondence is seen between the observed and synthetic spectra:
when using $\varepsilon$(C) = 8.06 the CO ($\Delta v = 2$) lines
are predicted as being far too weak, while they are predicted as
being far too strong for $\varepsilon$(C) = 8.50. It should also
be noted that the agreement in band 1B was very bad.


\subsection{${\beta}$ Peg{\rm: AOT01, speed 4, revolution 551}}\label{disbpeg}

\subsubsection{Comparison between the ISO-SWS spectrum and the synthetic spectrum
(Fig.\ \ref{bpeg})}

\begin{figure}[h!]
\begin{center}
\resizebox{0.5\textwidth}{!}{\rotatebox{90}{\includegraphics{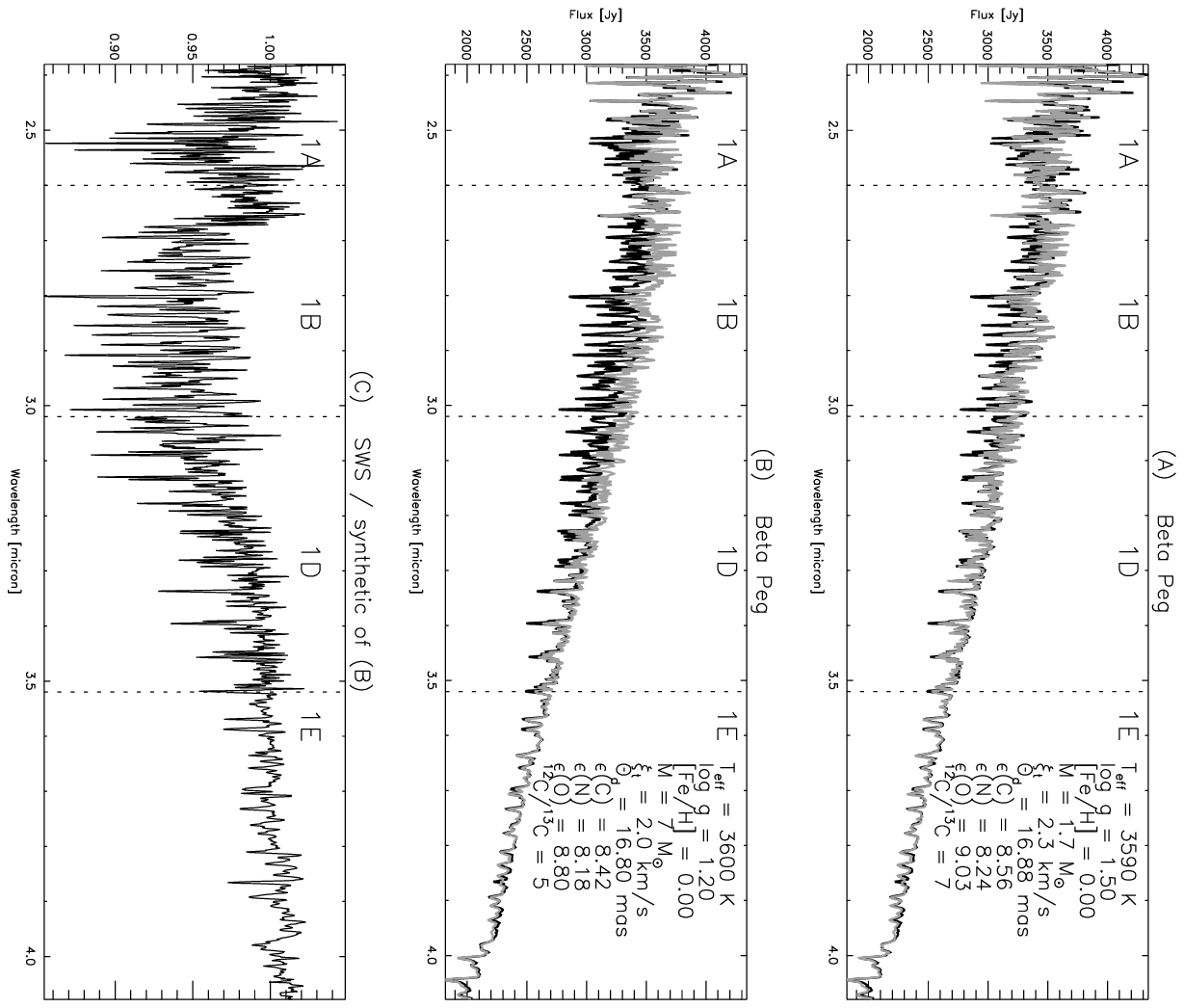}}}
\caption{\label{bpeg} {\underline{Panel (A):}} Comparison between band 1 of the
ISO-SWS data (rev.\ 551) of $\beta$ Peg (black) and the synthetic spectrum
(grey) with stellar parameters \teff\ = 3590\,K, $\log$ g = 1.50, M
= 1.7\,\Msun, [Fe/H] = 0.00, \vt\ = 2.3\,\kms, \cc\ = 7, $\varepsilon$(C)
= 8.56, $\varepsilon$(N) = 8.24, $\varepsilon$(O) = 9.03 and \ad\
= 16.88\,mas.
{\underline{Panel (B):}} Comparison between band 1 of the
ISO-SWS data (rev.\ 551) of $\beta$ Peg (black) and the synthetic spectrum
(grey) with stellar parameters \teff\ = 3600\,K, $\log$ g = 0.65, M
= 2\,\Msun, [Fe/H] = 0.00, \vt\ = 2.0\,\kms, \cc\ = 5, $\varepsilon$(C)
= 8.20, $\varepsilon$(N) = 8.18, $\varepsilon$(O) = 8.93 and \ad\
= 16.60\,mas.
{\underline{Panel (C):}} ISO-SWS observational data (rev.\ 551) of $\beta$ Peg
divided by the synthetic spectrum with the same parameters as in panel
(B). A coloured version of this plot is available in the Appendix as
Fig.\ \ref{bpegcol}.}
\end{center}
\end{figure}

By using the classical models and synthetic spectra as calculated by
the MARCS- and Turbospectrum codes, it turned out impossible to
determine stellar parameters from the ISO-SWS data of $\beta$ Peg: in
the range 2.6 -- 3.2\,\mic, certain lines were always predicted as
far too weak (see, e.g., Fig.\ \ref{bpeg}, where the synthetic spectra
for 2 different sets of parameters are plotted).

Already from the early eighties, it was clear that the
near-infrared absorption lines of CO and H$_2$O molecules yield
insight into the structure of the inner parts of an expanding
envelope \citep{Tielens1983ApJ...271..702T}. The picture emanating
from these data was that three different regions are to be
recognised:\\ (a) an extending pulsating photosphere terminated
by\\ (b) a stationary layer which starts \\ (c) the cool expanding
envelope.

In panel (C) in Fig.\ \ref{bpeg} --- where the SWS
observational data of $\beta$ Peg are divided by the synthetic
spectrum specified in panel (B) --- a clear pattern is
appearing. Comparing this pattern with typical molecular absorption
patterns as displayed e.g.\ in Fig.\ 3.5 in \citet{Decinthesis} or in
Fig.\ 4 in \citet{Decin2000A&A...364..137D}, we see that --- at least
--- H$_2$O is necessary to explain this excess absorption.

The discovery of such excess absorption/emission features visible
in the ISO-SWS data was for the first time discussed by
\citet{Tsuji1997A&A...320L...1T}. They concluded that these features could be
interpreted as due to an excess absorption originating in a `warm
molecular envelope'. These layers are clearly distinct from the
much cooler circumstellar envelopes (CSE) already known from radio
observations. At that time, it was however still not clear how
such a molecule forming region could be developed.

\begin{figure}[h!]
\begin{center}
\resizebox{0.5\textwidth}{!}{\rotatebox{90}{\includegraphics{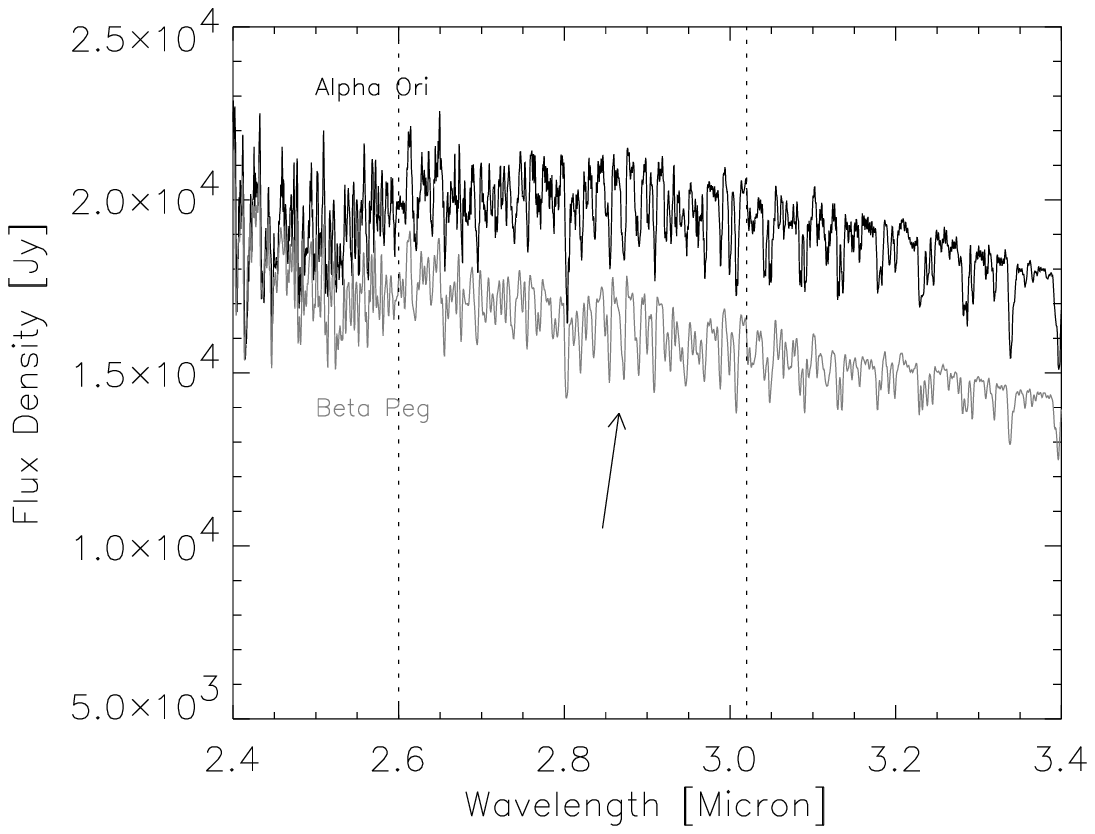}}}
\caption{\label{aoriversbpeg}$\alpha$ Ori and $\beta$ Peg are
compared in the wavelength region from 2.4 to 3.4 \mic. The flux
values of $\beta$ Peg are multiplied by a factor 5. The
'problematic' OH/H$_2$O lines are indicated by the arrow.}
\end{center}
\end{figure}

A similar structure of features was found from ISO observations in
quite some other stars (Fig.\ \ref{aoriversbpeg}), among them
$\alpha$ Ori \citep{Cami1997fiso.work..159C,
Justtanont1998A&A...330L..17J, deJong1999,
Yamamura1999A&A...341L...9Y, Ryde1999A&A...341..579R}. Together
with $\mu$ Cep, $\alpha$ Ori is one of the two objects whose CSE
kinetic temperature distribution  has been analysed
\citep{Rodgers1991ApJ...382..606R}. Indications were found by
other methods and different authors that a hydrostatic model is
not able to model these stars correctly \citep[e.g.\
][]{Hinkle1982ApJ...252..697H, Aringer1999A&A...342..799A}. Such
discrepancies, found by both  models developed by Tsuji and
collaborators and models developed by the Uppsala group, may be
too large to be relaxed by improving the model of the photosphere.
Dynamical models for the outer layers of oxygen-rich AGB-stars
reveal, on the other hand, a layered structure of the molecules
and much higher column densities than predicted by hydrostatic
models \citep{Woitke1999A&A...348L..17W, Hofner2001}. This is a
natural consequence of the shock waves occurring in these models.
The overall amount of levitated gas found in the models exceeds
the lower density limits of CO, H$_2$O, CO$_2$ and SO$_2$ inferred
from the ISO-SWS observations. Therefore, the levitation of the
outer atmospheres of AGB-stars by pulsations  seems to be a
possible mechanism to form these `warm molecular layers'. Other
effects, such as dust formation and chemical non-equilibrium, are
probably also involved.

Thus, it is an established fact that a better understanding of the
complicated structure of cool stars --- and more precisely their
spectral features around 3\,\mic\ --- will only be reached once
dynamical modelling of stars, including good opacity calculations,
NLTE and a CSE, are fully developed. One should therefore be very
careful in the interpretation of results based on current
modelling.

In order to determine $\varepsilon$(C), the \cc-ratio, \ad, L, R and
M$_g$ we have adopted as \teff\  and $\log $ g the values obtained in
Sect.\ \ref{firstguess}. Furthermore, we adopted [Fe/H] = 0.00, \vt\ =
2.0 \kms, $\varepsilon$(N) = 8.18 and $\varepsilon$(O) = 8.93 assuming
that $\beta$ Peg has already gone through the first dredge-up. For the
computation of the upper and lower limit of the stellar mass, $\Delta
\log$ g = 0.40\,dex was assumed.

\subsubsection{Comparison with other published stellar parameters
\rm{(see Table \ref{litbpeg} in the Appendix)}}

As for $\alpha$ Boo and $\beta$ And, the highest effective
temperature value for $\beta$ Peg in Table \ref{litbpeg}
is given by the interferometric value of \citet{Dyck1998AJ....116..981D}. The
other \teff\ values do agree quite well.

A discussion on the other quoted gravity values in Table \ref{litbpeg}
can be found in the corresponding sections for $\beta$ And and
$\alpha$ Cet. As for $\alpha$ Cet, \citet{Tsuji1986A&A...156....8T}
has determined \Mbol\ as being the mean of the value given by the
Wilson-Bappu relation \citep{Wilson1976ApJ...205..823W} (\Mbol\ =
$-1.9$) and the result found by \citet{Eggen1973PASP...85..379E} (\Mbol\
= $-3.2$). Since this mean value is higher than our value for \Mbol\ by
$0.79$\,mag (see Table \ref{firstlogg}, \Mbol = $-3.34$), the obtained
value for $\log$ g is higher by
$-0.3$\,dex. \citet{Tsuji1991A&A...245..203T} quoted, however, a lower
gravity value, being $\log$ g = 0.5\,dex, but the reason for this
decrease is not mentioned.

The microturbulence was determined by various authors. They have
used different kinds of spectral lines, such as the SiO first-overtone
lines \citep{Lambert1987Ap&SS.133..369L}, the CO first-overtone lines
\citep{Tsuji1986A&A...156....8T}, the CO second-overtone lines
\citep{Tsuji1991A&A...245..203T}, CN lines \citep{Aoki1997A&A...328..175A},
Fe~I, Ni~I and Ti~I lines \citep{Smith1985ApJ...294..326S}. Unfortunately, these
studies all yield different values.

Also for $\beta$ Peg, a large difference is visible between the carbon
abundance deduced from the CO ($\Delta v = 2$) lines and from the CO
($\Delta v = 3$) lines as obtained by
\citet{Tsuji1986A&A...156....8T, Tsuji1991A&A...245..203T}.
\citet{Smith1990ApJS...72..387S} have used both CO ($\Delta v = 2$) and CO
($\Delta v = 3$)  lines. They are the only ones having deduced
$\varepsilon$(C), $\varepsilon$(N) and $\varepsilon$(O) simultaneously based
on  high-resolution spectra of OH ($\Delta v=2$),
CO ($\Delta v = 3$), CO ($\Delta v = 2$) and  CN ($\Delta v = 2$)
lines.
Using the assumed stellar parameters as described in previous section,
$\varepsilon$(C), \ad,
L, R, M$_g$ and the \cc-ratio are deduced from the ISO-SWS spectrum of
$\beta$ Peg. We assigned a quite high error bar on the
derived $\varepsilon$(C) and \cc-ratio, since these ratios were
deduced only from the first
$^{12}$CO and $^{13}$CO features before 2.45\,\mic, where the spectrum
is less disturbed by the excess absorption of H$_2$O. As for $\alpha$
Cet, the \cc-ratio obtained by
\citet{Lazaro1991MNRAS.249...62L} is higher than the value deduced in
this and other studies.

Our obtained angular diameter agrees well
with the angular diameter obtained most recent by
\citet{Blackwell1991A&A...245..567B} using IRFM. Also the luminosity and the
radius correspond with other quoted values.

\section{Discussion on the derived stellar
parameters}\label{discussion}

\subsection{Temperatures derived from ${V-K}$ colours}
\label{colours}

\begin{table*}
\caption{\label{colour2}Effective temperatures derived from the
(\teff, $V-K$) relationship as given by
\citet{Ridgway1980ApJ...235..126R} (RJWW80),
\citet{DiBenedetto1993A&A...270..315D,DiBenedetto1998A&A...339..858D}
(diBen9398) and
\citet{Bessell1998A&A...333..231B} (BCP98) are tabulated in columns 3,
4, and 5 respectively. In column 6, $T_{\rm{eff}}(K_0, {\rm BC}_K,
\theta_{\rm{D}})$ is listed and in the last column the effective
temperature deduced from the ISO-SWS spectra is given.}
\begin{center}
\begin{tabular}[b]{lcccccc}\hline
\rule[0mm]{0mm}{5mm} & & \teff($(V-K)_0$) &
\teff($(V-K)_0$) & \teff($(V-K)_0$) &  & \\
\rule[-3mm]{0mm}{3mm}\raisebox{1.5ex}[0pt]{name} &
\raisebox{1.5ex}[0pt]{$(V-K)_0$} &
RJWW80 & diBen9398 & BCP98 &
\raisebox{1.5ex}[0pt]{$T_{\rm{eff}}(K_0, {\rm BC}_K, \theta_{\rm{D}})$} &
\raisebox{1.5ex}[0pt]{\teff(ISO)} \\
\hline \hline
\rule[0mm]{0mm}{5mm}$\delta$ Dra & 2.26 & 4846 & $4784 \pm 48$ & 4828 & $4858 \pm 157$ & $4820 \pm 140$ \\
$\xi$ Dra & 2.70 & 4437 & $4412 \pm 44$ & 4457 & $4457 \pm 166$ & $4440 \pm 300$ \\
$\alpha$ Boo & 2.88 & 4321 & $4280 \pm 43$ & 4331 & $4226 \pm
139$ & $4320 \pm 140$ \\
$\alpha$ Tuc & 2.91 & 4302 & $4264 \pm 43$ & 4312 & $4052 \pm 133$ & $4300 \pm 140$ \\
$\beta$ UMi & 3.25 & 4101 & $4061 \pm 41$ & 4111 &
$3963 \pm 132$ & $4085 \pm 140$ \\
$\gamma$ Dra & 3.54 & 3973 & $3926 \pm 39$ & 3970 &
$3978 \pm 134$ & $3960 \pm 140$ \\
$\alpha$ Tau & 3.67 & 3931 & $3874 \pm 39$ & 3915 &
$3831 \pm 129$ & $3850 \pm 140$ \\
H Sco & 4.08 & 3793 & $3688 \pm 98$ & 3764 & $4112 \pm 139$ & $3900 \pm 140$ \\
$\beta$ And & 3.82 & 3845 & $3722 \pm 95$ &  3856 & $3790 \pm 127$ &
$3880 \pm 140$ \\
$\alpha$ Cet & 4.07 & 3796 & $3689 \pm 98$ & 3768 & $3679 \pm 124$ &
$3740 \pm 140$ \\
\rule[-3mm]{0mm}{3mm}$\beta$ Peg & 4.61 & 3648 & $3622 \pm 102$ & 3615
& $3554 \pm 147$ &
($3600$) \\ \hline
\end{tabular}
\end{center}
\end{table*}

It is instructive to compare the ISO-SWS \teff, mostly sensitive to
the slope of the spectrum between 2.8 and 4\,\mic, to \teff\ derived
by other means. As already said in Sect.\ \ref{firstguess}, one can
calculate \teff\ directly from $V-K$ using (semi-) empirical (\teff,
$(V-K)_0$) temperature scales. Such calibrations are e.g.\  given
by \citet{Ridgway1980ApJ...235..126R}, \citet{DiBenedetto1993A&A...270..315D,
DiBenedetto1998A&A...339..858D} and
\citet{Bessell1998A&A...333..231B} (see Sect.\
\ref{firstguess}). \citet{Ridgway1980ApJ...235..126R} have
established an empirical effective temperature calibration by using
angular diameters from lunar occultation in combination with infrared
photometry. More recently, Michelson interferometry has been producing
much more precise radii for the K and M stars
\citep{DiBenedetto1993A&A...270..315D,
DiBenedetto1998A&A...339..858D}.

\teff\ may also be derived from m$_{\rm{bol}}$ and the angular
diameter deduced from the ISO-spectra. It is easily shown that:
$\log$(\teff)$ =  2.720 - 0.5
\log(\theta_{\rm{D}}/1000) - {\rm{m_{bol}}}/10$, with \ad\ in mas and assuming
that M$_{\mathrm{bol},\odot} = 4.74$
\citep{Bessell1998A&A...333..231B}. From m$_{\rm{bol}}$($K_0$, BC$_K((V-K)_0)$)
in Table \ref{firstlogg} and \ad\ from Table \ref{parameters},
T$_{\rm{eff}}(K_0,{\rm BC}_K, \theta_{\rm{D}})$ is calculated.
This T$_{\rm{eff}}(K_0,
{\rm BC}_K, \theta_{\rm{D}})$ is listed in  column 6 in Table \ref{colour2},
together with \teff\ derived from the three \teff($V-K$) calibrations
discussed above.
The comparison to \teff(IS0) (Fig.\ \ref{VminK}) shows that these
methods do agree within the $\sim5\,\%$ level, with the exception
being H Sco (see further discussion)! Only for a few stars,
\teff(ISO) has been adjusted slightly w.r.t.\ the first guess
value \teff($(V-K)_0$) derived from
\citet{Bessell1998A&A...333..231B}. This good general agreement in
\teff\ proves that \teff\ as derived from ISO-SWS spectra is
compatible with \teff\ derived from photometry. Since all the
\teff($(V-K)_0$) calibrations are based on almost the same
principle, it is normal that all the \teff($(V-K)_0$) values are
clustered together w.r.t.\ T$_{\rm{eff}}(K_0,{\rm BC}_K,
\theta_{\rm{D}})$ , as is seen in the lowermost plot of Fig.\
\ref{VminK}. For H Sco and $\alpha$ Tuc, the derived
T$_{\rm{eff}}(K_0,{\rm BC}_K, \theta_{\rm{D}})$ is quite
discrepant w.r.t.\ the other \teff-values. Concentrating on the
most discrepant star, H Sco, this too high a
T$_{\rm{eff}}(K_0,{\rm BC}_K, \theta_{\rm{D}})$ value can be
either due to too low an m$_{\rm{bol}}$ or \ad\ value. The used
$K$ magnitude of 0.00\,mag is from
\citet{Price1968AJ.....73..431P}, who quoted this value as being
uncertain. We note that an error of +0.4\,mag in $K$ ($K =
0.4$\,mag), inducing a 0.11\,mag decrease in BC$_K$ around 4000\,K
so that m$_{\rm{bol}}$ = 2.99\,mag, leads to an error in \teff of
$-275$\,K (T$_{\rm{eff}}(K_0,{\rm BC}_K, \theta_{\rm{D}}) =
3837$\,K; $-367$\,K for the increase in $K$, $+92$\,K for in the
decrease in BC$_K$). Simultaneously, this increase in $K$, through
a decrease of $V-K$ by 0.4\,mag, induces an increase of about
75\,K in the \teff\ deduced from \teff($(V-K)_0$) calibrations.
Concerning the bolometric correction BC$_K$ we have used the
relation found by \citet{Bessell1998A&A...333..231B}. As can be
seen from Fig.\ 20 in \citet{Bessell1998A&A...333..231B}, the
BC$_K$ values of \citet{Bessell1984PASP...96..247B} are
systematically somewhat higher in the region $2.3 < V-K <
4.5$\,mag, with a maximum difference around $V-K = 3$\,mag and
being almost 0 for $V-K \approx 4$. So, uncertainties in BC$_K$
can not explain the observed differences. For H Sco, only 1
ISO-SWS observation is available. Inspecting the literature values
for the angular diameter for the other cool stars in the sample,
we see that \ad(ISO) is not systematically higher or lower than
the other quoted values. Taking the worst-case scenario, in which
the absolute-flux level of the ISO-SWS data is too low by $10\,\%$
(a value deduced from the comparison between the different
observations of one target, as done in the Appendix) and using the
same \teff(ISO) value of 3900\,K, the angular diameter of H Sco
would increase from 4.73\,mas to 4.96\,mas inducing a decrease in
T$_{\rm{eff}}(K_0,{\rm BC}_K, \theta_{\rm{D}})$ from 4112\,K to
4016\,K, still different by $\sim 250$\,K from the
\teff($(V-K)_0$) values. Moreover, as discussed in Sect.\
\ref{lithd}, we derive an effective temperature of
\teff\,=\,3900\,K using the Johnson $B$, $V$, $R$ and $I$ colour
measurements of \citet{Johnson1966CoLPL...4...99J} and the
colour-temperature relation of
\citet{McWilliam1990ApJS...74.1075M}. This value of 3900\,K ---
being $\sim 150$\,K higher than the initial guess value based on
$K = 0.00$\,mag --- is supported by the slope of the ISO-SWS
spectrum of H Sco. We may conclude that a least the (high)
uncertainty in the $K$ magnitude can explain the larger deviation
of T$_{\rm{eff}}(K_0,{\rm BC}_K, \theta_{\rm{D}})$  w.r.t.\ the
other \teff\ values.

Taking a (more typical) error of $-0.1$\,mag in the $K$ magnitude of
$\alpha$ Tuc,  would also shift this target in within the $3\,\%$
level of correspondence in the lowermost plot of Fig.\ \ref{VminK}.

Taking all these remarks into account, we may conclude that
uncertainties in the $K$ magnitude or in \ad\ may cause H Sco and
$\alpha$ Tuc to deviate from the 3\,\% level of correspondence of the
different obtained \teff-values. Moreover, this high level of
correspondence proves the consistency between $V-K$, BC$_K$, \teff\
and \ad\ derived from IR and optical data.

\begin{figure}[h!]
\begin{center}
\resizebox{0.5\textwidth}{!}{\rotatebox{90}{\includegraphics{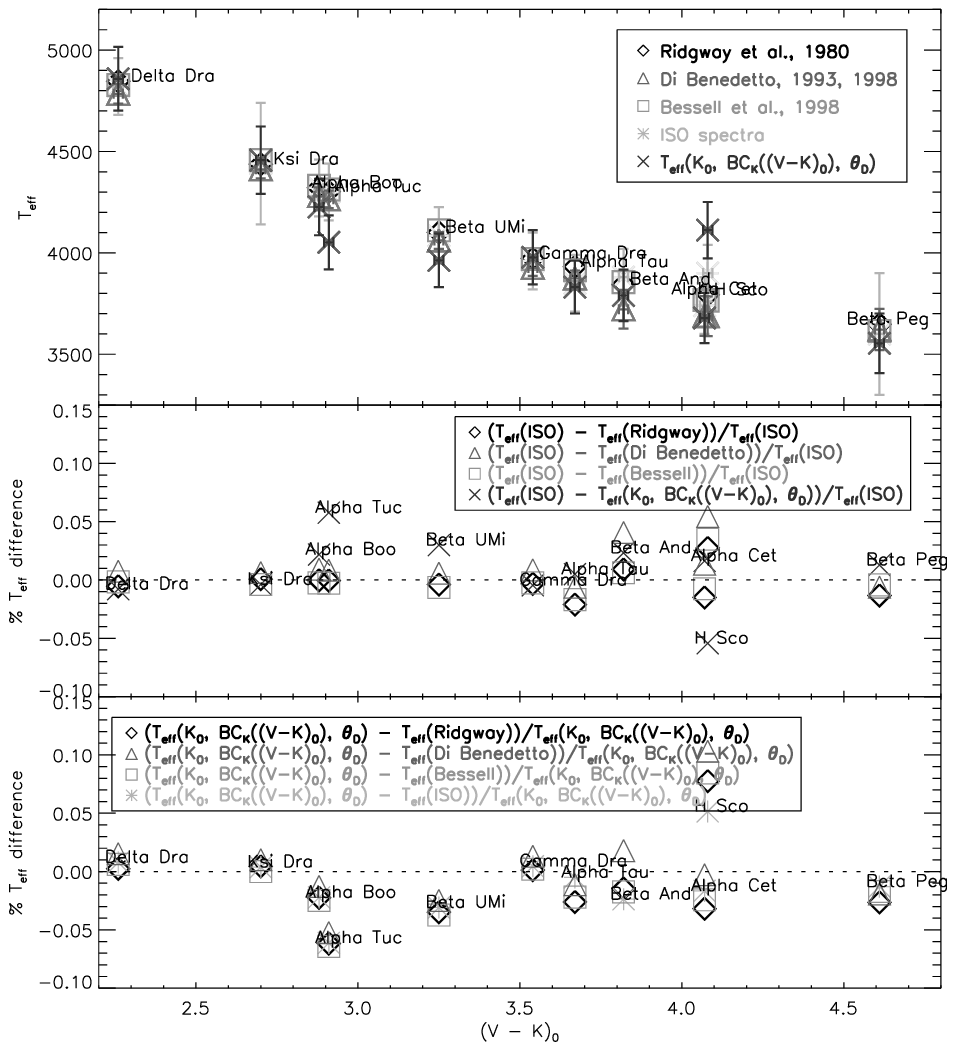}}}
\caption{\label{VminK}Comparison between \teff($V-K$) derived from
the relationship given by \citet{Ridgway1980ApJ...235..126R},
\citet{DiBenedetto1993A&A...270..315D},
\citet{DiBenedetto1998A&A...339..858D} or
\citet{Bessell1998A&A...333..231B}, T$_{\rm{eff}}(K_0, {\rm BC}_K, \pi)$
and the effective temperature deduced from the ISO-SWS spectra. A
coloured version of this plot is available in the Appendix as Fig.\
\ref{VminKcol}.}
\end{center}
\end{figure}


\subsection{\cc\ ratio}\label{cc}

Classical models (i.e.\ those where stellar convective regions are
instantly mixed, where no transport of matter occurs in the
radiative regions and without rotation at any depth)
predict that the
carbon isotopic ratio \cc\ in low-mass stars (M $\la 2$\,\Msun)
declines from 90 to about 20 -- 30 along the red giant
branch \citep[e.g.][]{Iben1967ApJ...147..624I,
Dearborn1976ApJ...203..455D,Charbonnel1998A&A...332..204C}. This
pattern is, however, not seen
for our program stars (see Fig.\ \ref{figcc}). Observations of CNO
elements in a large number of evolved stars \citep[for an overview,
see ][]{Sneden1991IAUS..145..235S} permitted already to confirm the
qualitative occurrence of the first dredge-up, but quantitative
disagreements appeared between the abundance pattern of real
low-mass stars and of `model' stars: in a
large fraction of giants the observed \cc\ and
$^{12}{\rm{C}}/^{14}{\rm{N}}$ ratios appeared to be substantially
lower than the predicted post-dredge-up ratios. Therefore, different
processes involving an extra-mixing process have been proposed to
explain the deviations from standard theoretical predictions
\citep[for an overview, see ][]{Charbonnel1994A&A...282..811C,
Charbonnel1998A&A...332..204C}. Observational results of
\citet{Charbonnel1998A&A...332..204C} confirmed that the extra-mixing
process becomes efficient on the RGB only when the low-mass stars
reach the so-called luminosity function bump, i.e.\ the evolutionary
point where the hydrogen burning shell crosses the chemical
discontinuity created by the convective envelope at its maximum
extent, tending to indicate that the nature of this extra-mixing
process on the RGB is related to {\it rotation}.

\begin{figure}[h!]
\begin{center}
\resizebox{.48\textwidth}{!}{\rotatebox{90}{\includegraphics{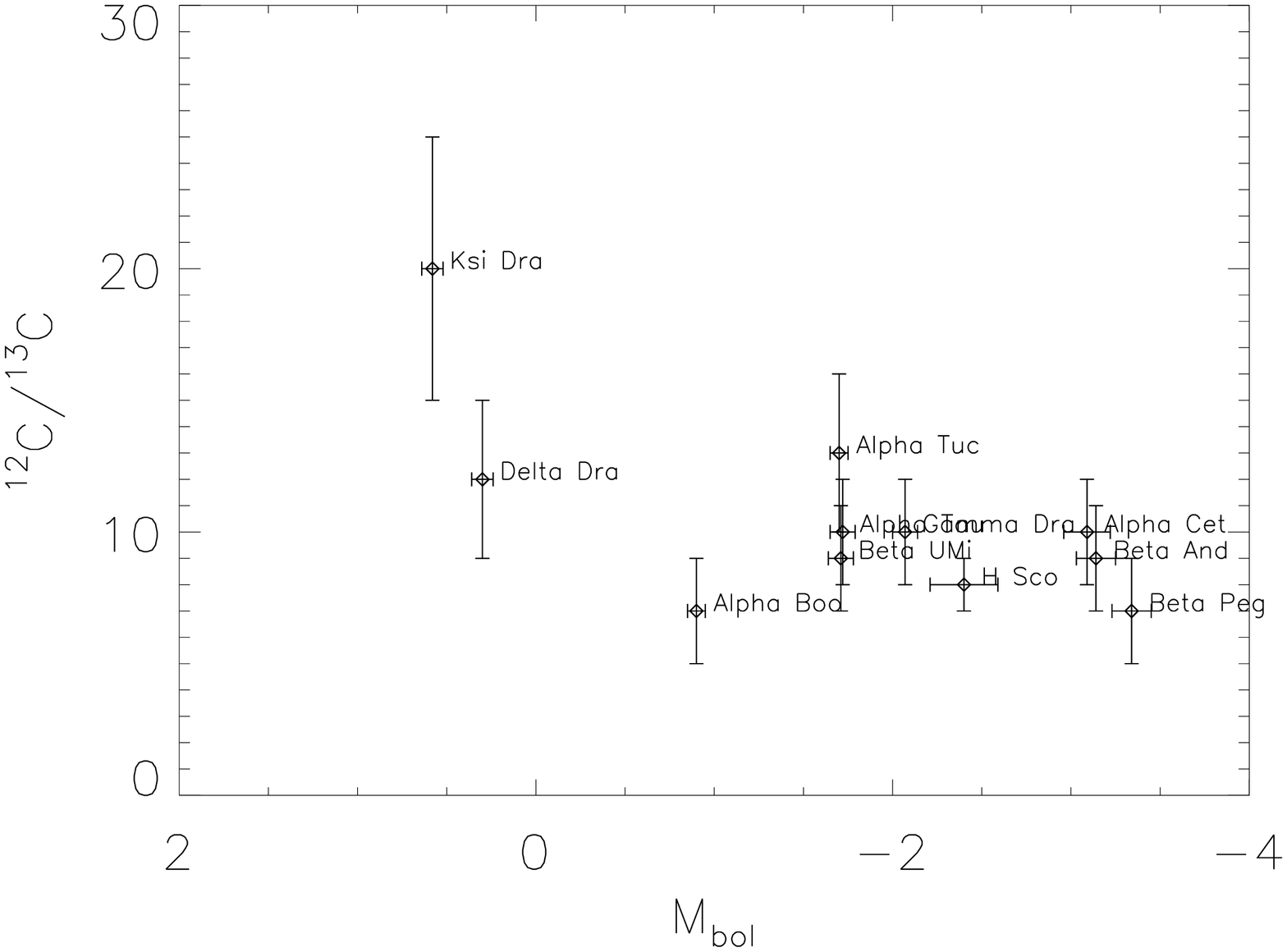}}}
\caption{\label{figcc}The ratio of \cc\ is plotted against the
bolometric magnitude of the program stars. The absolute magnitudes
are tabulated in Table \ref{firstlogg}. This figure should be
compared with Fig.\ 1 in \citet{Charbonnel1998A&A...332..204C}.}
\end{center}
\end{figure}

Our observational results are summarised in Fig.\ \ref{figcc},
where we show the observed dependence of the \cc\  ratio on
M$_{\rm{bol}}$. The same conclusions can be drawn  as by
\citet{Charbonnel1998A&A...332..204C}, namely (i) the least
luminous star in our sample, $\xi$ Dra, shows a limit on the \cc\
ratio in agreement with standard predictions, and (ii) for stars
with M$_{\rm{bol}} < 0.9$, the observed isotopic ratio drops to
values between 7 and 13, well below the standard predicted
post-dilution ratio. As quoted by
\citet{Charbonnel1998A&A...332..204C}, it is interesting to note
that the RGB luminosity function bump of 47 Tuc (with
${\rm{[Fe/H]}} \simeq -0.65$) is at ${\rm{M_{bol}}} = 1.05 \pm
0.20$, i.e.\ precisely where the disagreement between standard
predictions and observed \cc\ ratios appears. Our observational
results, in which for the first time the \cc-ratio of $\delta$
Dra, $\alpha$ Tuc and H Sco are included, are thus a confirmation
of the claims of \citet{Charbonnel1998A&A...332..204C}.

Moreover it has to be noted that for more massive stars (M $\ge$
1.7 -- 2.3\,\Msun, depending on the
metallicity) the evolutionary scenario is different: the
hydrogen-burning shell never reaches the region of constant
molecular weight that has been homogenised by the convective
envelope during the first dredge-up. Under these conditions, the
extra-mixing process does not occur in these stars and their
abundance pattern should resemble the pattern predicted by
standard stellar theory. Taking into account the (large) error
bars on the derived M$_g$ values or on the assumed M$_{\rm{hr}}$ values,
and assuming that the cool giants in our
sample are still on the RGB --- see also the discussion in the
previous section --- and that no other extra-mixing process occurs,
the low values of the \cc\ ratios are indicative for low-mass stars (M
$\la$ 2\,\Msun) in which this extra-mixing process occurs.

\section{Conclusions} \label{conclusions}

In this paper, the atmospheric parameters of 11 giants have been
determined by using the method described in
\citetalias{Decin2000A&A...364..137D} and \citetalias{Decin2000b}. A
confrontation with other published stellar parameters, other ISO-SWS
data and the synthetic spectra always showed very `positive' results
both for the theoretical modelling and for the calibration of the
ISO-SWS detectors. The very small discrepancies still remaining in band
1 are at the 1 -- 2\,\% level for the giants, proving not only that
the calibration of the (high-flux) sources in this band is better than
$\sim 2$\,\%, but also that the description of cool-star atmospheres
and molecular linelists is already quite accurate at
the ISO-SWS spectral resolution ($R \sim 1000$).

In the discussion of the deduced parameters, we demonstrated that
\teff(ISO) is consistent with other \teff($(V-K)_0$) or
\teff($K_0$, BC$_K$, \ad(ISO)) values in within the few percent
level, proving once more that the used MARCS-models to derive the
quantities from the ISO-SWS spectra are very trustworthy and that
the flux calibration of ISO-SWS is indeed reassuring.

The obtained \cc\ ratios are compatible with other sources \citep[e.g.\
][]{Charbonnel1995ApJ...453L..41C}.

Like already pointed out in \citetalias{Decin2000b}, these
synthetic spectra will now be used for the --- latest --- OLP10
calibration of ISO-SWS. Also other consortia constructing
instruments for ground-based telescopes --- i.e.\ MIDI (\,=\, the
Mid-Infrared Interferometric instrument for the VLTI with first
observations foreseen in 2001) and new satellites --- i.e. SIRTF,
the Space InfraRed Telescope Facility, with launch foreseen in
December 2001, and Herschel, a far-infrared submillimeter
telescopes with launch foreseen in 2007 --- will use this type of
study and/or these SEDs for determining the required sensitivity
of their instruments. A confrontation with existing SEDs, like the
ones of M.\ Cohen, is therefore planned in a forthcoming paper of
this series, in which also the new atomic linelist constructed by
Sauval \citep{Sauval2000} will be implemented.

\begin{acknowledgements}LD acknowledges support from the Science Foundation of
Flanders. This research has made use of the SIMBAD
database, operated
at CDS, Strasbourg, France and of the VALD database, operated at Vienna,
Austria. It is a pleasure to thank the referees, J.\ Hron and F.\
Kupka, for their careful reading of the manuscript and for their
valuable suggestions.
\end{acknowledgements}


\aareferences
